\documentclass[fleqn,usenatbib]{mnras}
\usepackage{graphicx}	
\usepackage{amsmath}	
\usepackage{amssymb}	
\usepackage{comment}

\usepackage{newtxtext,newtxmath}

\usepackage[T1]{fontenc}

\DeclareRobustCommand{\VAN}[3]{#2}
\let\VANthebibliography\thebibliography
\def\thebibliography{\DeclareRobustCommand{\VAN}[3]{##3}\VANthebibliography}

\usepackage{soul}
\def\msun{{\rm M_{\odot}}}
\newcommand\rsun{{\rm R_{\odot}}}

\def\be{\begin{equation}}
\def\ee{\end{equation}}

\def\del#1{{}}

\newcommand{\bref}[1]{{#1}}

\newcommand\mj{{\,{\rm M}_{\rm Jup}}}

\title[HMYSO outbursts]{Episodic accretion and mergers during growth of massive protostars}

\author[Elbakyan et al.]{
Vardan G. Elbakyan$^{1,4}$\thanks{vardan.elbakyan@leicester.ac.uk}, Sergei Nayakshin$^{1}$, Dominique M.-A. Meyer$^{2}$, Eduard I. Vorobyov$^{3,4}$
\\
$^{1}$Department of Physics and Astronomy, University of Leicester, Leicester LE1 7RH, UK\\
$^{2}$Institut f{\"u}r Physik und Astronomie, Universit{\"a}t Potsdam, Karl-Liebknecht-Strasse 24/25, 14476 Potsdam, Germany\\
$^{3}$University of Vienna, Department of Astrophysics, Vienna, 1180, Austria\\
$^{4}$Institute of Astronomy, Russian Academy of Sciences, 48 Pyatnitskaya St., Moscow, 119017, Russia
}

\date{Accepted XXX. Received YYY; in original form ZZZ}

\pubyear{2022}

\begin{document}
\label{firstpage}
\pagerange{\pageref{firstpage}--\pageref{lastpage}}
\maketitle

\begin{abstract}
3D simulations of high mass young stellar object (HMYSO) growth show that their circumstellar discs fragment onto multiple self-gravitating objects. Accretion of these by HMYSO may explain episodic accretion bursts discovered recently. We post-process results of a previous 3D simulation of a HMYSO disc with a 1D code that resolves the disc and object dynamics down to the stellar surface. We find that burst-like deposition of material into the inner disc seen in 3D simulations by itself does not always signify powerful accretion bursts. Only high density post-collapse clumps crossing the inner computational boundary may result in observable bursts. The rich physics of the inner disc has a significant impact on the expected accretion bursts: (1) In the standard turbulent viscosity discs, migrating objects can stall at a migration trap at the distance of a few au from the star. However, in discs powered by magnetised winds, the objects
are able to cross the trap and produce bursts akin to those observed so far. 
(2) Migrating objects may interact with and modify the thermal (hydrogen ionisation) instability of the inner disc, which can be responsible for longer duration and lower luminosity bursts in HMYSOs. 
(3) If the central star is bloated to a fraction of an au by a previous episode of high accretion rate, or if the migrating object is particularly dense, a merger rather than a disc-mediated accretion burst results; (4) Object disruption bursts may be super-Eddington, leading to episodic feedback on HMYSO surroundings via powerful outflows.


\end{abstract}




\section{Introduction} 

Due to their rarity, formation of high-mass stars is less well understood than that of their lower mass counterparts. Nevertheless, it is becoming clear that, similarly to low mass stars, high-mass stars grow via an accretion disc that transports material from the collapsing molecular cloud onto the central protostar \citep[see review by][]{2018Motte}. Circumstellar discs around high-mass young stellar objects (HMYSOs) can reach a few hundreds of au in size and have masses of several 10 \bref{per cent} of their host stars \citep{2014Tan}, potentially rendering them gravitationally unstable \citep{2016Kratter}. Long term evolution of such discs is a matter of intense ongoing research.

Numerical simulations show that massive self-gravitating discs form spiral density arms \citep{2007Krumholz, 2011Kuiper} which are prone to fragmentation -- formation of self-bound gas fragments  \citep{2006KratterMatzner, 2018MeyerKuiper, 2019Ahmadi}. What becomes of the fragments depends on a number of intricate detail, \bref{such as internal structure of the fragment, its efficiency of mass accretion, and interaction with other structures in the disc} \citep{2004MayerL,2017NayakshinDesert} and is a subject of ongoing research. Regardless of all the physical and numerical uncertainties, the fact that most stars, and especially the massive ones,  are found in binary systems \citep{2011SanaEvans} tells us that gravitational fragmentation of early circumstellar discs is an extremely ubiquitous process.

One outcome of disc fragmentation is fragment survival on long period orbits \citep[e.g.,][]{2019ChonHosokawa}. The fragments may mature into planets, brown dwarfs, or stellar companions  \citep{2008Kratter,2009StamatellosWhitworth,2013Vorobyov,2016Kratter,2017Boss,2020OlivaKuiper}. 

Numerical simulations of fragmenting discs usually show formation of multiple rather than single fragments \citep[e.g.,][]{2004MayerL,2018VorobyovElbakyan,2011ChaNayakshin}. The most common evolutionary pathway for the fragments in these simulations is their inward migration on a timescale as short as $\sim 10^4$ years \citep[e.g.,][]{2010BoleyEtal,2011BaruteauMeru}. A number of authors studied high luminosity flares resulting due to accretion of the fragments onto the surface of the central star  \citep{2005VorobyovBasu, 2006VorobyovBasu, 2011MachidaInutsuka, 2016Hosokawa, 2017MeyerVorobyov, 2018VorobyovElbakyan, 2018ZhaoCaselli}, which may be related to the well-known FU Ori outbursts of low mass stars \citep{1996HartmannKenyon, 2014AudardAbraham, 2022Fischer}. This process is also interesting because accretion of gas onto the central star at very high rates modifies its evolution towards the main sequence \citep{2012DunhamVorobyov, 2012BaraffeVorobyov, 2017VorobyovElbakyan}. For massive stars specifically, accretion outbursts are key to radiative and outflow feedback on the immediate environment from which the stars grow \citep{2008McKeeTan, 2009Krumholz}, as well as a influencing parameter of their pre-main-sequence 
evolutionary path \citep{meyer_mnras_484_2019}. 

Discs around high-mass protostars have been observed  recently \citep{2015Johnston, 2015Zinchenko, 2016Ilee,  2017Kraus,liu_apj_904_2020}, with what appears to be an ongoing disc fragmentation in some cases \citep{2018Ilee, 2020Johnston, 2021Suri}. Further, recent observations of high luminosity outbursts from a few HMYSOs show that massive YSOs do indeed exhibit accretion outbursts \citep{2017Caratti, 2018Hunter, 2021Stecklum, 2021Chen}. This provides a solid observational motivation to simulate fragment formation in gravitationally unstable discs of HMYSOs and their eventual accretion onto the central star. 

Unfortunately, multidimensional simulations of circumstellar discs are numerically expensive, especially in the innermost disc regions where the orbital timescale is very short. Therefore, 3D simulations are usually conducted over relatively short timescales (a few hundred outer disc orbits) \citep[e.g.][]{2020Zhu} or use a large inner boundary ($\sim$20~\bref{au}) for the disc \citep[e.g.][]{2017FlockNelson, 2018MeyerKuiper, 2020OlivaKuiper}. For the innermost ($\sim$20~\bref{au}) region of the disc, the so-called “sink cell” approach is used. \bref{The sink cell is usually considered as a passive boundary with zero gradients. It is fixed in position within the grid and dynamically inactive. Sink cells are usually used in grid-based codes with a general non-Cartesian (e.g., spherical, cylindrical, polar) grid to deal with the singularity at the centre \citep[e.g.,][]{1982BossBlack}. It} is assumed that the material that passes through the inner boundary \bref{of active disc into the sink cell} lands onto the protostar directly \citep[but see][]{2019VorobyovSkliarevskii}.  In reality, the material passing through the inner boundary of the computational domain has to go through the inner accretion disc before landing onto the star. This process is not instantaneous, as the viscous time at tens of au can be as long as $\sim 10^4$ years. Various instabilities (e.g., thermal, magnetorotational) are known to operate in the inner disc region \citep{1994BellLin, 2001Armitage}. This could lead to a significant accretion rate variations onto the central star \citep[e.g.,][]{2010ZhuHartmann, 2014BaeHartmann, 2014OhtaniVorobyov, 2020Kadam}, even when the rate at which the matter is deposited into the inner disc is constant \citep[e.g.,][]{2001Armitage}. Furthermore, the properties of the inner disc also dictate how the fragments born in the outer disc migrate through it towards the star, potentially even reversing the direction of their migration \citep{2018VorobyovElbakyan, 2019VorobyovSkliarevskii, 2021Guilera}. Thus, it is important to follow the evolution of the circumstellar discs down to the central star surface. 

One dimensional (1D) numerical models of circumstellar discs are currently the only practical way of achieving this goal.  In our recent paper, \citet[][hereafter Paper I]{2021ElbakyanNayakshin}, we used the 1D approach to compare the properties of the bursts on HMYSOs triggered in the inner disc by different mechanisms: thermal instability (TI), magnetorotational instability (MRI), and planet disruption. We found that the luminous accretion bursts observed from HMYSOs are unlikely to result from one of these two instabilities in the inner disc, but may be a result of tidal disruption of a gaseous protoplanet. On the other hand, the 1D approach is especially unsuitable for the model in which the bursts are due to tidal disruptions of a protoplanet or a more massive object. These objects are born by disc fragmentation in the spiral density areas. This process is inherently non-axisymmetric and cannot be modelled from first principles in 1D. As a stop-gap measure, in Paper I we assumed a constant mass deposition rate into the disc interspersed with an occasional fragment injection into it.

In this paper, we make the first attempt to combine the advantages of the 3D and 1D numerical techniques to study episodic accretion in HMYSOs on scales from the stellar surface to thousands of au.  \cite{2019MeyerVorobyovElbakyan} modelled \bref{the} collapse of a massive rotating molecular cloud core, resulting in formation of the circumstellar disc on scales from 0.1~pc to 20~\bref{au}, with the \bref{modified version of 3D} hydrodynamics code PLUTO \citep{2007Mignone}\bref{, which considers the radiation transport, self-gravity \citep{2010Kuiper, 2011Kuiper} and stellar evolution \citep{2013KuiperYorke}}.
We design a post-processing procedure in which the capture of material by the PLUTO inner "sink cell" is treated as an injection of matter into the inner disc that we model here with an improved version of the 1D code used in Paper I. In this way we are able to model the viscous evolution of a circumstellar disc along with the dynamics of an embedded object inside the region unresolved in the 3D study. Our main goal is to investigate how the physics of the innermost disc region with an embedded object in it may affect the character of HMYSO accretion outbursts.

The paper is organized as follows. In Section~\ref{sec:model} we  introduce the 1D and 3D hydrodynamical models used in this paper, and the connection between them. In Section~\ref{sec:clump_wout} we consider the outcome of  diffuse matter or pre-collapse clump deposition into the inner disc during a major accretion episode in 3D simulations. This section is important in setting the scene for, and contrasting this scenario to, the rest of the paper where a high density post-collapse object is injected into the inner disc. In Section~\ref{sect:obj_migration}  we study injection of high mass post-collapse objects  ($M_{\rm obj}>100\mj$) into the inner disc during the same high accretion rate episode. In Section~\ref{sec:Low_M_both_models} we present the results for a less intense accretion burst in 3D simulations, in which less massive post-collapse objects are injected into the 1D disc. We also investigate how the nature of the angular momentum transfer -- via disc turbulent viscosity or MHD driven winds -- affect our results. Section~\ref{sec:future} is devoted to model caveats and a limited parameter space exploration. In Section~\ref{sec:discuss}, we provide our main conclusions.

\section{Methods}\label{sec:model}

\subsection{3D simulations of the region $r\geq 20$~au}\label{sec:3D}

\begin{figure}
\begin{centering}
\includegraphics[width=1\columnwidth]{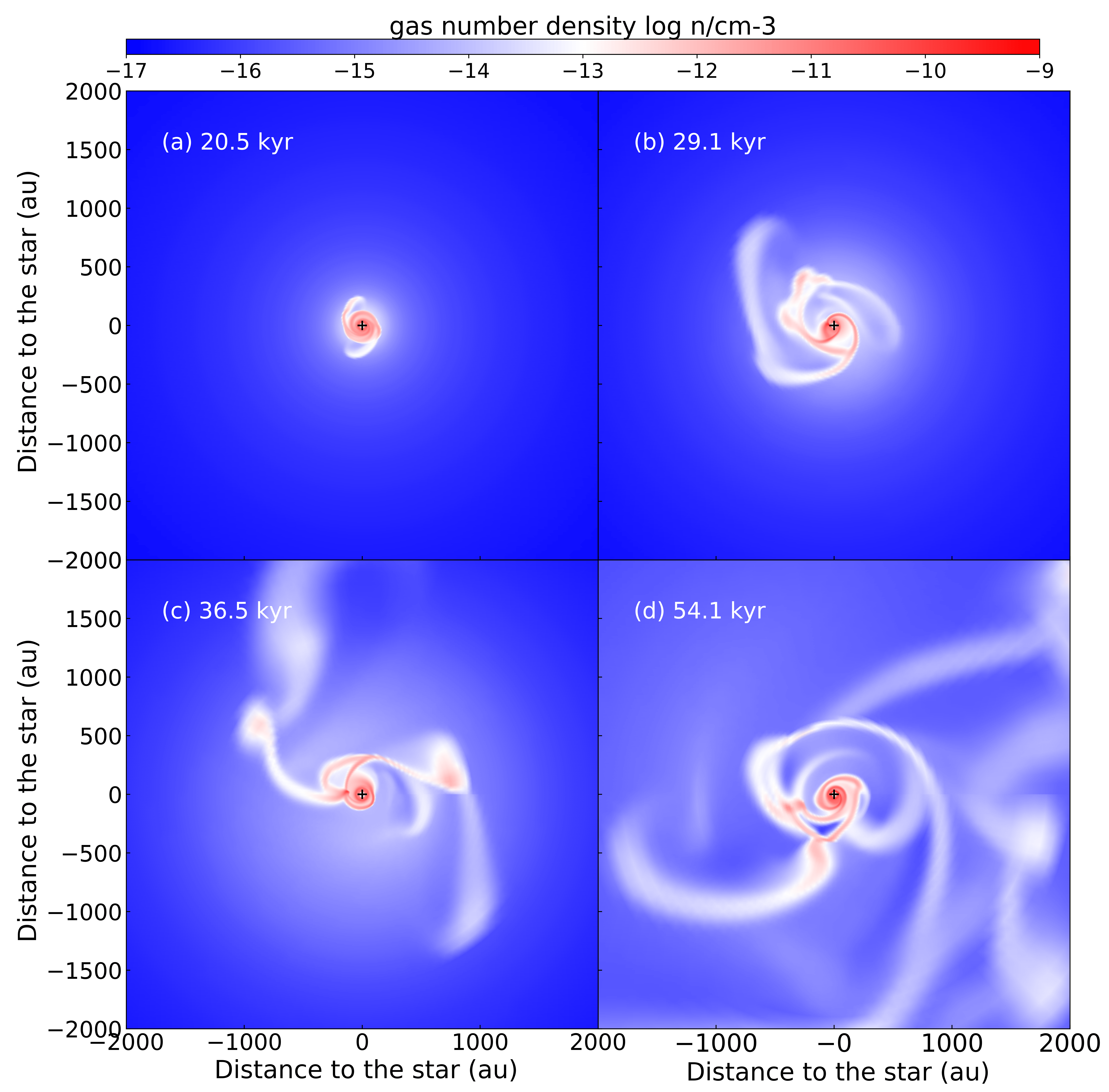}
\par\end{centering}
\caption{\label{fig:000} Mid-plane gas number density (in log particle per cm$^{3}$) of the accretion disc in the 3D simulation \textit{Run-long-5\%} from \citet{2019MeyerVorobyovElbakyan} at four different times (shown in the legend in kyr). The black cross marks the position of the central sink cell. In this paper, we use the accretion rate from this simulation to investigate the evolution of the inner 20 \bref{au} disc unresolved in the 3D simulation. \bref{The Gouraud shading interpolation technique is used for plotting the figure.}}
\end{figure}

In Figure~\ref{fig:000} we present several snapshots of the circumstellar disc mid-plane gas number density maps taken in the model \textit{Run-long-5\%} from \citet{2019MeyerVorobyovElbakyan}. The simulation starts with a solid-body rotating \bref{molecular cloud} of mass 100$M_{\odot}$ with a uniform temperature of 10K \bref{and with a spherically symmetric density distribution profile $\rho(r)\propto r^{-3/2}$. The inner boundary of the cloud constitutes a semi-permeable sink cell, which is fixed at the radial distance $r=20$~au at the origin of the computational domain. The outer radius of the cloud is located at $r=0.1$~pc.}
The spatial grid is made of $N_{\rm r}=138$, $N_{\rm \theta}=21$, and $N_{\rm \phi} =138$  zones in the spherical polar coordinate system \bref{$[20~\rm{au},0.1~\rm{pc}]\times[0,\pi/2]\times[0,2\pi]$. The grid is spaced logarithmically in radial direction, as cosine in $\theta$ \citep{2015OrmelShiKuiper}, and uniformly in $\phi$}.  As the cloud collapses, the central protostar grows in mass, and a massive disc forms. The early circumstellar disc is not azimuthally symmetric, with spiral arms fragmenting on numerous gas fragments. Some fragments migrate outwards, but most migrate inwards rapidly by losing angular momentum due to gravitational torques on them from the disc. When massive gaseous clumps enter the inner computational boundary set at 20~\bref{au} from the central star (which is shown with the black cross), bursts of high mass accretion onto the star result. 
\bref{The stellar properties (e.g. stellar radius and photospheric luminosity) are calculated using the pre-calculated protostellar evolutionary tracks of \citet{2009HosokawaOmukai}. }

It is important to stress for what is following that simulations such as that presented in this section remain numerically challenging. \bref{The} disc fragmentation process, and the associated mass and angular momentum transport within it, depend on numerical resolution achieved in the simulation  \citep{2016Hosokawa}. \citet{2013MachidaDoi} showed that numerical codes need to have spatial resolution down to $\sim (0.01-0.1)$~\bref{au} to capture disc fragmentation with numerical convergence. \bref{The more recent study by \citet{2020OlivaKuiper} found that the number of fragments that are present in the disc at a given time converges for the spatial resolution $\lesssim~1$~au at $r=30$~au.} The simulations by \citet{2019MeyerVorobyovElbakyan} have sub-\bref{au} spatial resolution in the inner disc, but not in the outer $\bref{r} > 1000$~\bref{au} disc. The resolution study by \bref{\citet{2018MeyerKuiper}} showed that the characteristics and the number of fragments formed in the disc do vary with the spatial resolution of the model (see their fig. 10). In these simulations, fragments formed in the outer $\bref{r} > 1000$~\bref{au} disc have spatial dimensions of order 100~\bref{au}. During their inward migration, the size of the fragments often shrinks due to their cooling to a few tens of au. When exploring the fate of fragments entering the inner 20 \bref{au} region, we shall keep in mind (see \S \ref{sec:1D-3D}) the fact that the internal structure of the fragments has not yet been fully resolved in \bref{the} 3D simulations due to numerical limitations.

\bref{
We use the PLUTO code \citep{2007Mignone} to solve the hydrodynamic equations of mass, momentum, and energy transport for the gas 
\begin{equation}
	   \frac{\partial \rho}{\partial t}  + \mathbf{\nabla}  \cdot (\rho\mathbf{v}) =   0,
\label{eq:euler1}
\end{equation}
\begin{equation}
	   \frac{\partial \rho \mathbf{v} }{\partial t}  + 
           \mathbf{\nabla} \cdot ( \rho  \mathbf{v} \otimes \mathbf{v})  + 
           \mathbf{\nabla}p 			      =   \mathbf{f},
\label{eq:euler2}
\end{equation}
\begin{equation}
	  \frac{\partial E }{\partial t}   + 
	  \mathbf{\nabla} \cdot ((E+p) \mathbf{v})  =	   
	  \mathbf{v} \cdot \mathbf{f} ,
\label{eq:euler3}
\end{equation}
where $\rho$ is the gas density,  $\mathbf{v}$ is the gas velocity, $p=(\gamma-1)E_{\rm int}$ is the thermal pressure, $E  = E_{\rm int} + \rho \mathbf{v}^{2}/2$ represents the total (internal plus kinetic) energy, and $\gamma=5/3$ is the adiabatic index. The force density vector $\mathbf{v}$ is defined as
\begin{equation}
	  \mathbf{ f } = -\rho \mathbf{\nabla} \Phi_{\rm tot} 
			- \frac{\rho \kappa_{\rm R}(T_\star)}{c}\mathbf{F_{\star}}
			- \frac{\rho \kappa_{\rm R}(T)}{c}\mathbf{F},
\label{eq:f}
\end{equation}
where $\Phi_{\rm tot}$ is the  total gravitational potential, $\kappa_{\rm R}$ is the Rossland opacity, c is the speed of light, $\mathbf{F}$ is the radiation flux and $\mathbf{F_{\star}}$ is the stellar radiation flux.
}

\bref{
The radiation transfer is calculated by solving the equation of radiation transport for the thermal radiation energy density $E_{\rm R}$ in the flux-limited diffusion approach treated within the gray approximation~\citep{2013Kolb} using the scheme from \citet{2010Kuiper}
\begin{equation}
	  \frac{\partial }{\partial t} \Big( \frac{ E_{\rm R} }{ f_{\rm c}  }  \Big)  + 
	  \mathbf{\nabla} \cdot \mathbf{F}  =	   
	  -\mathbf{\nabla} \cdot \mathbf{F_{\star}},
\label{eq:rad1}
\end{equation}
where $f_{\rm c}=1/( c_{\rm v} \rho/4 a T^{3} + 1 )$ with the calorific capacity $c_{\rm v}$ and the radiation constant $a$. The radiation flux is defined as $\mathbf{F} = -D \mathbf{\nabla} E_{\rm R}$,  where $D$ is the diffusion constant. The radial dependence of stellar flux is calculated as
\begin{equation}
	  \mathbf{F_{\star}}(r)  = \mathbf{F_{\star}}( R_{\star}) \left( \frac{ R_{\star} }{ r } \right)^{2} e^{-\tau(r)},
\label{eq:rad2}
\end{equation}
with the optical depth of the medium $\tau(r)$, estimated with both constant gas opacity and dust opacities from~\citet{1993Laor}. Self-gravity is included by solving the Poisson equation using the PETSC library \citep{petsc-efficient} for the total gravitational potential of gas and the stellar gravitational contribution. We neglect turbulent viscosity by assuming that the most efficient angular momentum transport mechanism consists of the gravitational torque generated once a self-gravitating disc has formed after collapse \citep[see also][]{2009VorobyovBasu, 2016Hosokawa}. We refer the reader interested further in the 3D numerical model to \citet{2018MeyerKuiper}.
}

\subsection{1D model of the region $r\leq 20$~au}\label{sec:1D}

We use a 1D azimuthally symmetric and vertically averaged  viscously evolving \citet{1973ShakuraSunyaev} disc model to solve system dynamics inside 20 \bref{au}. Our code stems from \cite{2012NayakshinLodato} with several modifications. Our viscous disc equations  additionally include the \bref{effects of the} tidal torque arising from the migrating object, the mass deposition from the object into the disc, and mass deposition from outer regions into the disc:
\begin{equation}
\begin{split}
    \frac{\partial\Sigma}{\partial t} = \frac{3}{r} \frac{\partial}{\partial r} \left[ r^{1/2} \frac{\partial}{\partial r} \left(r^{1/2}\nu \Sigma\right) \right] - \frac{1}{r} \frac{\partial}{\partial r} \left(2\Omega^{-1} \lambda \Sigma\right) + \\ 
    \frac{\dot{M}_{\mathrm{ obj}}}{2\pi r}\delta (r - r_{\mathrm{obj}}) + \frac{\dot{M}_{\mathrm{dep}}}{2\pi^{3/2} r \sigma} \exp\left[-\left(\frac{r-r_{\mathrm{dep}}}{\sigma}\right)^2\right],
\end{split}
\label{dSigma_dt}
\end{equation}
where $\Sigma$ is the gas surface density, $\Omega$ is the angular velocity at radial distance $r$, $\nu=\alpha_{\rm eff} c_{\rm{s}} H$ is the kinematic viscosity of the disc, $c_{\rm{s}}$ is the gas sound speed in the disc midplane, and $H$ is the vertical scale height of the disc. The external mass deposition into the inner disc, $\dot M_{\rm dep}$, is discussed in \S \ref{sec:1D-3D}.

In our model, $\alpha_{\rm eff}$ is the sum of constant $\alpha=10^{-2}$ and the self-gravity term $\alpha_{\rm sg}$, which depends on disc temperature and surface density and is defined as
\begin{equation}
   \alpha_{\mathrm sg} = 
   \begin{cases}
        10^{-2}\left[\left(\frac{1.5}{Q}\right)^2-1\right], & \text{ if } Q<1.5 \\
        0, & \text{ if } Q\geq1.5
   \end{cases},
   \label{alpha_sg}
\end{equation}
where $Q=c_{\rm s}\Omega/(\pi G \Sigma)$ is the Toomre parameter. We use a logarithmic radial grid with 200 zones. In \cite{2012NayakshinLodato}, the equation of state for the disc assumed an ideal gas pressure dominated form for completely ionised hydrogen. We now use \bref{an} equation of state for a H/He mixture that includes Hydrogen molecule rotational and vibrational degrees of freedom, its dissociation, and Hydrogen atom ionisation. Radiation pressure is included.

The mass loss rate of the object, $\dot{M}_{\rm{ obj}}$, is computed as in \citet{2012NayakshinLodato}. In this model, the object is surrounded by an exponential atmosphere and the mass loss proceeds through the Lagrange L1 point. The mass loss rate is exponentially small while object radius, $R_{\rm obj}$, is significantly smaller than its Hill radius, $R_{\rm Hill}$, but increases very rapidly when $R_{\rm obj} > R_{\rm Hill}$. The mass lost by the object through the L1 point is deposited at the corresponding circularisation point, $r_{\rm obj}=a_{\rm obj}(1-R_{\rm Hill}/a_{\rm obj})^4$, where $a_{\rm obj}$ is the radial distance of the object to the central star.

The specific tidal torque of the object onto the disc, $\lambda$, is defined as 
\begin{equation}
    \label{eq:pl_torque}
    \lambda = \lambda_{\mathrm{I}}(1-f_{\mathrm{II}}) + \lambda_{\mathrm{II}}f_{\mathrm{II}},
\end{equation}
where $\lambda_{\mathrm{I}}$ and $\lambda_{\mathrm{II}}$ are the normalized specific torques for Type I and Type II migration, respectively, and $0 \leq f_{\mathrm{II}} \leq 1$ is a function that  controls the transition between Type I and Type II migration regimes. Specifically, when the object migrates via pure Type I regime, $f_{\rm II}=0$; in the Type II regime, $f_{\rm II}=1$. To ensure a smooth transition between the regimes, we define $f_{\mathrm{II}}$ as
\begin{equation}
    \label{eq:fII}
    f_{\mathrm{II}} = \mathrm{min}(1,\mathrm{exp}[-(C_{\mathrm{obj}}-1)^2]),
\end{equation}
where $C_{\rm obj}$ is the \cite{2006Crida} gap opening parameter, given by
\begin{equation}
    C_{\mathrm{obj}} = \frac{3}{4}\frac{H_{\mathrm{obj}}}{R_{\mathrm{Hill}}} + \frac{50\alpha_{\mathrm{eff}}}{q}\left(\frac{H_{\mathrm{obj}}}{a_{\mathrm{obj}}}\right)^2,
\end{equation}
and $H_{\rm obj}$ is the vertical scale height of the disc at the position of the object, $q$ is the object to star mass ratio, $q = M_{\rm obj}/M_*$. The expressions for the normalized specific torque for Type I and Type II migration are presented in Appendix~\ref{sect:app}.



 The migration rate of the object is found via
\begin{equation}
    \frac{\mathrm{d}a_{\mathrm{obj}}}{\mathrm{d}t} = \frac{-a_{\mathrm{obj}}}{t_{\mathrm{migr}}} = a_{\mathrm{obj}}\frac{2\Gamma_{\mathrm{tot}}}{J_{\mathrm{obj}}}
\end{equation}
where $J_{\rm obj}=M_{\rm obj}\sqrt{GM_*a_{\rm obj}}$ is the angular momentum of the object and $\Gamma_{\rm tot}$ is the total torque acting on the object due to its interaction with the disc. In the Type II migration regime, i.e., when $C_{\rm obj} \leq 1$, $M_{\rm obj} \Gamma_{\rm tot}$ is found by integrating the object torque on the disc and changing the sign in front. This ensures angular momentum conservation in the object-disc angular momentum exchange.

For a locally isothermal disc, \citet{2002TanakaWard} derived the total disc torque acting on the object as
\begin{equation}
    \Gamma_{\mathrm{tot}} = - f_{\mathrm{migr}}\Gamma_{\mathrm{0}},
    \label{Gtot0}
\end{equation}
with
\begin{equation}
   \Gamma_{\mathrm{0}} = (qa_{\mathrm{obj}}/H)^2\Sigma_{\mathrm{obj}}a_{\mathrm{obj}}^4\Omega_{\mathrm{obj}}^2,
   \label{G0}
\end{equation}
where $f_{\mathrm{migr}}$ is a dimensionless factor that depends on the disc properties, and index "obj" indicates that the value is calculated at the radial position of the object. \citet{2019FletcherNayakshin} found that the \citet{2002TanakaWard} expression with $f_{\mathrm{migr}}=1$ approximates planet migration in massive self-gravitating discs well.


Equations~(\ref{Gtot0})-(\ref{G0}) were used in Paper I. However, the inner disc may not be cooling efficiently, and the innermost region can be far from $Q\sim 1$ regime and Tanaka's formulation of the torque may not be accurate. In this paper, following \citet{2011Paardekooper} we define the total disc torque on the object in the Type I regime as the sum of the Lindblad torque and the corotation torques,
\begin{equation}
   \Gamma_{\mathrm{tot}} = \Gamma_{\mathrm{L}}+\Gamma_{\mathrm{c}}.
   \label{eq:gamma_tot}
\end{equation}
The expressions for $\Gamma_{\mathrm{tot}}$ are given in  \citet{2011Paardekooper}.



\subsection{Connection of the 3D disc model with the 1D circumstellar disc}\label{sec:1D-3D}

\begin{figure}
\begin{centering}
\includegraphics[width=1\columnwidth]{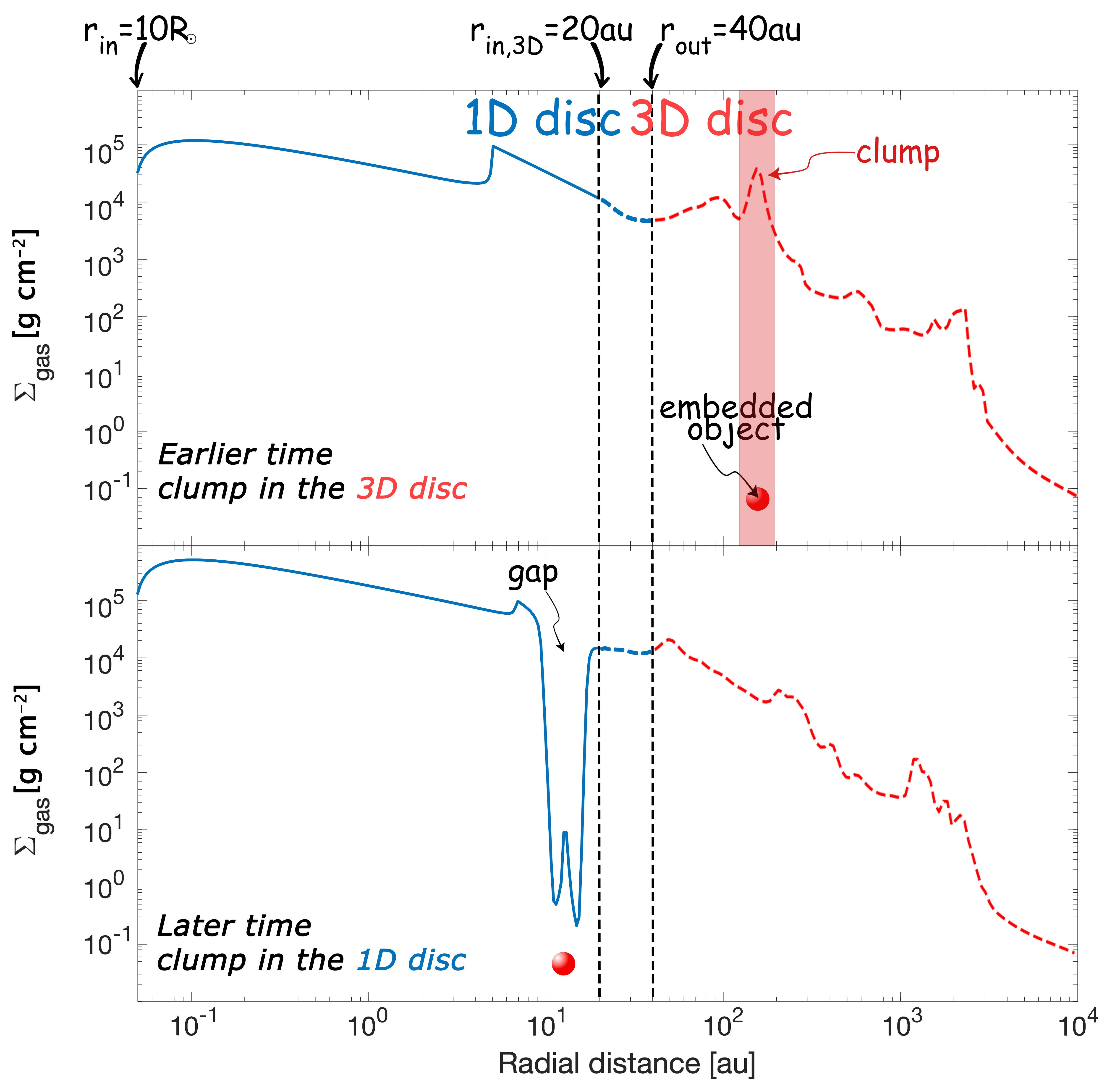}
\par\end{centering}
\caption{ A cartoon illustration of the connection between our 1D and 3D models. Surface density profile of the 1D disc (blue line) and 3D disc (red dashed line) at the time before ({\bf top panel}) and after ({\bf bottom panel}) a clump crosses the boundary. The peak in the disc surface density pointed out with an arrow is the clump. The shaded area around the peak represents the approximate linear size of the clump. The dense post-collapse embedded object, shown with a red circle, is not to scale. As the clump crosses the inner edge of 3D risk, $r_{\rm in, 3D}$, its mass, excluding the mass of embedded object, is added to the disc as described in \S \ref{sec:1D-3D}. The embedded object migrates inside the 1D disc and may be tidally disrupted in the sub-\bref{au} disc. Note that this figure is for illustration purposes only and does not imply that migrating objects always open a gap inside the 1D disc.
\label{fig:00}}
\end{figure}

In Figure~\ref{fig:00} we depicted in cartoon form the geometrical connection of our 1D disc with the 3D disc simulations. The material from the 3D disc is added to the 1D disc in a Gaussian ring with a centroid radius $r_{\rm dep}=20$~\bref{au}, and dispersion $\sigma=0.05r_{\rm dep}$. The radial distance of the inner edge of 3D disc is $r_{\rm in,3D}=20$~\bref{au}. As we only post-process the 3D data, the 3D and 1D models are not interacting with each other entirely self-consistently in our setup; e.g., there is no angular momentum transfer due to viscous torques between the 1D and 3D discs. This does not significantly impact our results in terms of gas flow, as the most interesting region of the disc is far inside the boundary between the 1D and 3D discs. However, we find that embedded object dynamics is sensitive to this transition region. This is to be expected, as the speed and direction of the object migration depends on the balance of torques acting on it from the disc regions interior and exterior to it. Ideally, we need continuous time-resolved 3D disc data close to the inner disc boundary. However, such data are memory expensive; only infrequent data dumps are available.


To mitigate this problem, we extend our 1D disc to $r_{\rm out}=40$~\bref{au}, thus mimicking a part of the outer disc. 
The value of 40~\bref{au} is chosen by trial and error as  most optimal.  The outer boundary condition for the 1D model is reflective. That is, matter is allowed to flow from the 3D disc into the 1D one through the $\dot M_{\rm dep}$ term in eq. \ref{dSigma_dt}, but not allowed to cross the outer 1D boundary in the opposite direction into the 3D domain. At the inner boundary, $r_{\rm in}$=0.0465~au~$=10 \rsun$,  we make the standard assumption of zero torque \citep[e.g.,][]{2015Armitage} and assume that the material crossing the inner edge of the 1D disc is being directly accreted onto the central star.


For the $\dot M_{\rm dep}$ term in Eq.~\ref{dSigma_dt} we use the tabulated gas accretion rate history of the 3D model \textit{Run-long-5\%} from \citet{2019MeyerVorobyovElbakyan}. This gas was assumed to land directly onto the star in the former paper, but here we solve for its dynamics with the 1D code. There are, however, physical uncertainties that we need to address. In the 3D simulations, a "clump" is defined as a significant spike in the gas flow rate history through the inner boundary. \bref{We separate the bursts caused by a passage of gaseous clumps from regular long-term variability caused by perturbations from large-scale spiral arms using the algorithm described in \S \ref{sect:low_mass_objects}.} As noted in the Introduction, resolving the innermost regions of gas clumps in 3D simulations remains challenging. It is likely that current 3D disc models have not yet reached convergence in the clump centres. Since we aim to model the clumps in very hot and dense regions close to the star, we need this missing information. In its absence, we devise a scheme to bracket the physical uncertainties in the clump structure.
During a massive clump passage through the inner boundary of the 3D disc, we divide the mass input into the "diffuse gas" mass injection and the embedded object.

As is well known in star formation, {\em spherical} collapse of a molecular cloud initially results in formation of "first cores", objects with mass of a few Jupiter masses and the size of a few au \citep{1969Larson}. These objects are dominated by molecular hydrogen. As they gain mass and contract, they heat up; eventually, when reaching the central temperature of $\sim 2000$~K, they collapse dynamically due to dissociation of molecular hydrogen, forming second cores \citep{2000MasunagaInutsuka, 2020Bhandare}. The second cores have central densities of at least $10^{-4}$ g/cm$^3$ and \bref{temperatures of the order higher than $10^4$~K}, and they can be called proper seeds of protostars.

\citet{2018MeyerKuiper} has found that the central temperature of some of their clumps does reach the threshold value for molecular hydrogen dissociation ($\sim$2000~K) during their inward migration. As a result, these clumps would likely go through  a dynamical collapse in their centres, forming a second core there. However, due to clump rotation and their location inside a massive hot disc and interactions with other clumps, it is not clear whether all of the clump mass would collapse onto the second core \citep[see also][]{2019VorobyovElbakyan}. It is possible that all the clumps formed in 3D simulations of \citet{2018MeyerKuiper} would be hot enough to contain second cores if it was numerically feasible to resolve their centres better. \bref{In the highest-resolution run of \citet{2020OlivaKuiper} 19 fragments live longer than 200 yr out of which 10 experience hydrogen dissociation and form second cores. Thus, the assumption of the embedded second core object formation inside the inward migrating clump is justified.}

Given these numerical uncertainties of 3D simulations, when a gas clump passes through the inner boundary of the 3D disc, we divide the clump into a dense embedded object and a diffuse envelope. The diffuse envelope is deposited into the inner 1D disc as described above through the $\dot M_{\rm dep}$ mass injection term in Eq.~\ref{dSigma_dt}, whereas the embedded object is inserted into the 1D disc as a one-off event as described in detail in \S \ref{sect:obj_migration}. The mass of the embedded object in this approach is a fraction of the total clump mass, and this fraction is a free parameter that we shall vary in the models below.

\subsection{The internal structure of the embedded object in the 1D model}\label{sec:internal}

A detailed modelling of object internal structure requires the use of specialised planet or stellar evolution codes \citep[e.g.,][]{2012Vazan,2016Hosokawa}; this is outside the scope of our paper. Here, we employ a toy polytropic sphere model for the internal structure of the object to explore the range of possible outcomes and formulate questions to address in future work. As shall be seen later, embedded objects in the first core (pre-\bref{second-}collapse) configuration are insufficiently dense to approach the central star close enough to yield bright and short duration accretion outbursts similar to those observed so far. Therefore, here we only consider embedded objects in the post-collapse state -- the second cores of \cite{1969Larson} -- that are dominated by the ionised hydrogen. We model them as  uniform composition polytropic spheres with a polytropic index $n =3/2$. Such objects obey a mass-radius relation of the form
\begin{equation}
    R_{\rm obj} M_{\rm obj}^{1/3} = \text{Const}\;,
    \label{R-M}
\end{equation}
where the constant on the right depends on the specific entropy of the gas in the object. When the objects are injected from the 3D outer disc into the inner 1D disc,  we specify their initial radius and mass, thus specifying the constant in Eq.~\ref{R-M}. 

Radiative cooling reduces specific entropy of the object so that it contracts with time as long as $M_{\rm obj}$ is constant. The rate of object contraction, i.e., $dR_{\rm obj}/dt$ is found by solving for the evolution of its total energy:
\begin{equation}
    \frac{dE_{\rm tot}}{dt} = -L_{\rm obj}\;,
\end{equation}
where $L_{\rm obj}$ is the total luminosity of the object, and $E_{\rm tot}$ is the total energy of a polytropic sphere with $n =3/2$, 
\begin{equation}
    E_{\mathrm{tot}} = -\frac{3}{7}\frac{GM_{\mathrm{obj}}^2}{R_{\mathrm{obj}}}\;.
\end{equation}

We further assume that the object is on the Hayashi track, which is a reasonable assumption for objects that went through the second collapse recently. The luminosity of the object would then be $ L_{\mathrm{obj}} = 4\pi R_{\mathrm{obj}}^2 \sigma_{\mathrm{b}} T_{\mathrm{Hayashi}}^4$, with the effective temperature set to $T_{\mathrm{Hayashi}} = 3000$~K. We neglect a weak dependence of  $T_{\mathrm{Hayashi}} = 3000$~K on mass of the object \citep[see, e.g.,][]{Palla12-Hayashi-Track}.

In the context of our toy model for the object, we calculate the net object luminosity as
\begin{equation}
\label{eq:Lobj}
    L_{\mathrm{obj}} = 4\pi R_{\mathrm{obj}}^2 \sigma_{\mathrm{b}} \left(T_{\mathrm{Hayashi}}^4 - T_{\mathrm{d,obj}}^4\right) \Theta,
\end{equation}
where $T_{\mathrm{d,obj}}$ is the disc midplane temperature at the position of the object. The function $\Theta$ is defined as 
\begin{equation}
\label{eq:heating_off}
    \Theta = \begin{cases}
      1\quad \text{for} \; T_{\rm d, obj} \leq T_{\rm Hayashi} \\
      0\quad \text{for} \; T_{\rm d, obj} > T_{\rm Hayashi}\;.
    \end{cases}
\end{equation} 
In this model external irradiation of the object by the hot disc surroundings slows down its radiative cooling but we do not allow the heat to flow into the object when $T_{\rm d, obj} > T_{\rm Hayashi}$. It has previously been shown that at sufficiently high $T_{\rm d, obj}$ (usually much higher than effective temperature, $T_{\rm Hayashi}$, here) the object may be completely disrupted by the thermal bath effect \citep{2012Vazan} rather than by the stellar tides that we focus on in this paper. We leave consideration of this important effect to a future study.

\section{No post-collapse embedded objects} \label{sec:clump_wout}

\subsection{Diffuse matter deposition}

\begin{figure}
\begin{centering}
\includegraphics[width=1\columnwidth]{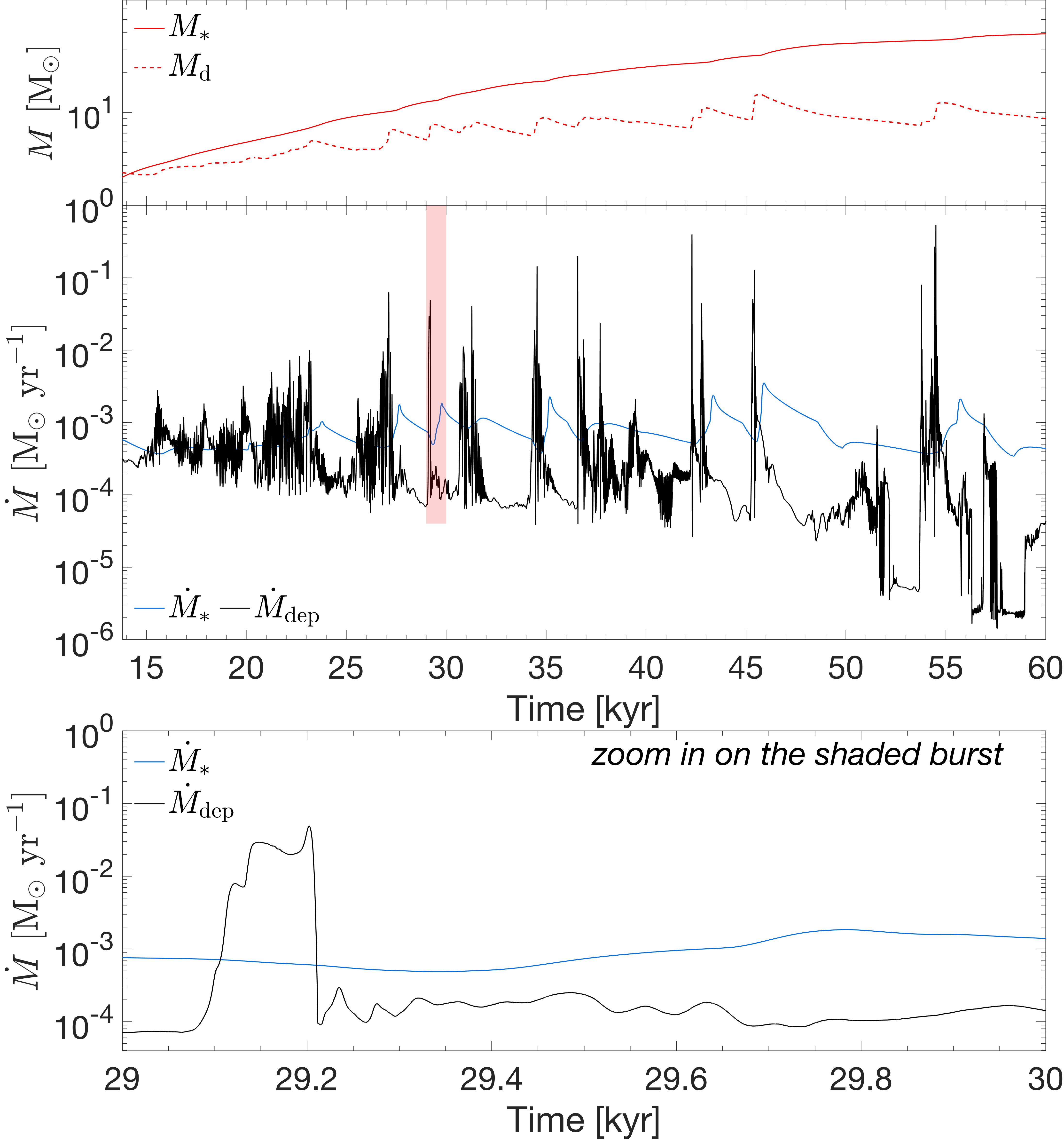}
\par\end{centering}
\caption{\label{fig:0} Comparison of accretion rate time evolution in the 3D simulations of \citet{2019MeyerVorobyovElbakyan} with that obtained here when including the 1D disc model of the sink cell region. \bref{\textbf{Top panel:} The mass of the star and 1D disc are shown with solid and dashed red lines, respectively.} \textbf{Middle panel:} The mass deposition rate onto the outer edge of 1D disc through the inner edge of 3D disc ($\dot M_{\rm dep}$, black curve) and the mass accretion rate through the inner edge of 1D disc onto the central star (blue line). The shaded area shows the region of the plot presented in the bottom panel. \textbf{Bottom panel:} Zoom-in onto one of the accretion bursts due to the passage of a massive clump (the area highlighted in the \bref{middle} panel).}
\end{figure}

As explained in the Introduction, our main aim is to understand how the presence of the inner disc within the region unresolved in 3D simulations modifies gas accretion onto the star. In our numerical method (\S \ref{sec:model}), the material passing from the 3D disc into the 1D disc region is deposited into the latter continuously (as diffuse gas) except during bright accretion flares. During the flares, the material is divided into a dense self-gravitating embedded object and the diffuse part. In this section, we study the limiting case when all the matter entering the 1D disc is in the diffuse form. This scenario is interesting in its own right, as some of the accretion rate spikes in 3D simulations may not contain dense post-collapse embedded objects (cf. \S\S \ref{sec:1D-3D}, \ref{sec:internal}). It also represents a useful reference case to contrast with the inner disc variability in simulations in which a dense embedded object is present in the disc in the sections to follow.

The black solid curve in the \bref{middle} panel of  Fig.~\ref{fig:0} is the gas accretion rate onto the central star from the 3D simulations by \cite{2019MeyerVorobyovElbakyan}. This is the rate at which the gas is recorded to pass through the $r_{\rm in,3D}=20$~\bref{au} inner boundary of their computational domain. In the context of our paper, this is the rate at which the gas is deposited from the 3D disc into the 1D disc (cf. Fig. \ref{fig:00}), that is, $\dot M_{\rm dep}$ term in Eq.~\ref{dSigma_dt}. Solving that equation without an embedded object in this section, we obtain an accretion rate through the inner boundary of the 1D disc, $r_{\rm in}=10~R_{\odot}$, which is the stellar surface. Our accretion rate onto the star, $\dot M_*$, is shown with the blue solid curve in Fig.~\ref{fig:0}. The red solid and dashed curves \bref{in the top panel} depict time evolution of the central star mass, $M_*$, and the 1D disc mass, $M_{\rm d}$, respectively. \bref{Since the diffuse matter from the 3D disc is added to the 1D disc, $M_{\rm d}$ shows sharp increase during the high $\dot{M}_{\rm dep}$ periods.} Note that $M_{\rm d}$ is comparable to the mass of the central star during the time interval shown in Fig. \ref{fig:0}. The high disc to star mass ratio is the reason for the disc being strongly self-gravitating and hatching numerous gaseous clumps. The highest of the accretion rate spikes seen in the black curve are produced when these clumps pass through  $r_{\rm in,3D}=20$~\bref{au} inner boundary of 3D simulations. In the bottom panel of Figure~\ref{fig:0} we zoom-in onto a strong $\dot{M}_{\rm dep}$ outburst occurring at $t\approx 29$~kyr and highlighted by the shaded area in the \bref{middle} panel. \bref{We choose the highlighted burst for a few reasons. It has a high amplitude (more than four magnitudes in total luminosity) and is a result of clump passage through the inner boundary of 3D disc. Also, this burst is not part of a clustered group of accretion outbursts, but rather an isolated burst. This further assures that the burst is a result of singular clump passage and not a group of clumps or a simultaneous passage of a clump and other structures in the disc.}



The main conclusion that we draw from comparing the stellar accretion rates from \cite{2019MeyerVorobyovElbakyan} and this paper, that is, the black and the blue curves, is that the inner disc acts as a low frequency variability filter. The inner disc dumps short duration bursts, whether intense or low amplitude. The physical reason for this is the fact that material entering the 1D disc close to its outer edge, $r_{\rm in,3D}=20$~\bref{au}, spends the viscous time, $\tau_{\rm visc} = r^2/\nu$,
\begin{equation}
\label{eq:t_visc}
   \tau_{\rm visc} \sim 1500 \text{ years } \; \left( \frac{r}{10 \text{ au}}\right)^{3/2}\;
   \left( \frac{M_*}{10  \msun}\right)^{-1/2} 
   \alpha_{-1}^{-1} \left( \frac{h}{0.1}\right)^{-2}
\end{equation}
diffusing towards the star. Here $h = H/r$, and $\alpha_{-1} = \alpha/0.1$, much larger than just the viscous $\alpha=0.01$ for this simulation, taking into account the fact that the disc viscosity may be dominated by the disc self-gravity (Eq.~\ref{alpha_sg}). We observe that it is not reasonable to expect $\tau_{\rm visc}$ to be shorter than about $10^3$ years at tens of au. This has the effect of delaying the outbursts (cf. the bottom panel in Fig. \ref{fig:0}), and also decreasing their amplitude correspondingly. It thus appears that for this particular 3D simulation, the inner boundary condition set at 20 \bref{au} influences the accretion rate history onto the star strongly, somewhat artificially over-predicting the amplitude of the accretion bursts onto the star. 


On the other hand, our azimuthally symmetric (1D) inner disc treatment with the outer boundary set at tens of au, probably over-estimates the degree to which the outbursts generated by gravitational instability with subsequent fragmentation of the outer disc are damped by the inner disc region. Firstly, we find that the regions outside a few au can be self-gravitating. The self-gravity viscosity prescription (Eq.~\ref{alpha_sg}) is a widely used stop-gap measure in 1D codes, but clearly this is insufficient to resolve non axisymmetric structures in the disc. Secondly, in this section, we assumed that the intense bursts in simulations by \cite{2019MeyerVorobyovElbakyan} do not contain dense embedded objects. While we are unable to remedy the former problem of our 1D approach, the latter assumption will now be relaxed for the rest of the paper.

\subsection{Pre-\bref{second-}collapse objects}\label{sec:pre-collapse}

As discussed in \S \ref{sec:1D-3D}, formation of both stars and planets begins with the birth of the first hydrostatic object called the first core \citep{1969Larson}. The distinguishing characteristic of these is their very large linear size and thus low density. First cores have central temperatures lower than $\sim 2000$~K as at higher temperatures they collapse to the second cores \citep[e.g.,][]{2000MasunagaInutsuka}. We can therefore estimate the order of magnitude of the linear size and the mean density of first cores by requiring their virial temperature to be lower than 2000 K:
\begin{equation}
    T_{\rm vir}=\frac{GM_{\rm obj}\mu m_{\rm H}}{3k_{\rm B}R_{\rm obj}}\lesssim2000~\rm{K}.
\end{equation}
The first core with the mass $M_{\rm obj}=10\mj$ and the radius $R_{\rm obj}=1$~\bref{au} will have volume density $\rho_{\rm obj} \approx 10^{-9}$~g~cm$^{-3}$, which is $3-6$ orders of magnitude lower than the densities of post-collapse second cores \citep{2000MasunagaInutsuka,2018Bhandare}.

The radial distance in the disc at which the radius of the first core will become larger than its Hill radius and the core will be tidally disrupted is 
\begin{equation}
\label{eq:rdis}
    r_{\rm dis} \approx 11 \left[\frac{M_{\rm obj}}{10\mj}\right]^{2/3} \left[\frac{M_{\rm *}}{10\msun}\right]^{1/3} \left[\frac{T_{\rm vir}}{2000\rm{K}}\right]^{-1}~\rm{au}.
\end{equation}
The first core with the fiducial parameters in this equation is tidally disrupted at radial distance $r\approx10$~\bref{au}. \bref{Less bound objects with lower $T_{\rm vir}$ will get disrupted at even larger radial distances. Indeed, disruption of weakly bound first core objects at $r\gtrsim10$~au in the disc is frequently seen in 2D \citep[e.g.,][]{2019VorobyovElbakyan} and 3D \citep[e.g,][]{2020OlivaKuiper} numerical models.} On the other hand,  disruption of an object in a disc with stellar accretion rate $\dot M_*$ will lead to an accretion outburst only if the mass of the disrupted object is larger than the local disc mass at the point of disruption. Therefore,
\begin{equation}
   M_{\rm obj}>\dot M_* t_{\rm visc}= \frac{\dot M_*}{\alpha_{\rm v} h^2 \Omega_{\rm K}}.
\end{equation}
For such observable bursts, the disruption of the object must take place at radial distance
\begin{equation}
\begin{split}
  r_{\rm dis} < \left(\frac{G M_* M_{\rm obj}^2 \alpha^2 h^4} {\dot{M}_*^2}\right)^{1/3} = 
  0.07 \left[\frac{\alpha_{\rm v}}{0.01}\right]^{2/3} \left[\frac{h}{0.1}\right]^{4/3} \times \\
  \left[\frac{M_{*}}{10\msun}\right]^{1/3} \left[\frac{M_{\rm obj}}{10\mj}\right]^{2/3} \left[\frac{\dot M_*}{10^{-4}}\right]^{-2/3}~\rm{au}.
  \end{split}
\end{equation}
Evidently, the object must be disrupted at a sub-\bref{au} distance to trigger an accretion outburst. However, as it was shown earlier  (see Eq.~\ref{eq:rdis}) a first core object is tidally disrupted at $r \gtrsim 10$~\bref{au}, before reaching sub-\bref{au} radial distance. Concluding, detectable protostar accretion burst are unlikely to be triggered by disruption of pre-collapse (first) cores.


\section{Outbursts due to disruption of post-collapse objects, high mass case, $M_{\rm obj}>100~M_{\rm Jup}$}
\label{sect:obj_migration}

Unlike the case presented in \S\ref{sec:clump_wout}, where no post-collapse embedded object is formed inside the clump, here we assume that the clump crossing the \bref{internal boundary} of the 3D disc has already passed through the phase of run-away collapse and formed an embedded object at its centre before passing through the $r_{\rm in,3D}=20$~\bref{au} \bref{boundary}.

\subsection{Fiducial case}\label{sec:fiducial}

Earlier we showed that if clumps passing from the 3D disc to the 1D disc contain no high density embedded objects, then all the high accretion rate peaks are washed out in the 1D disc region due to a long viscous time (Eq.~\ref{eq:t_visc}) with which matter passes through it. In this section, we present co-evolution of the inner disc and an embedded object carried into the inner disc by the clump responsible for the accretion outburst seen in the black curve  in the bottom panel of Fig. \ref{fig:0}. We assume specifically that the mass of the embedded object, $M_{\rm obj}$, is a 25 \bref{per cent} of the clump mass. The total mass of the latter is $2.08 M_{\odot}$; thus the mass of the embedded object in this example is $0.502\msun = 545$ Jovian masses. \bref{The total clump mass is calculated as follows: (i) the maximum value of $\dot{M}_{\rm dep}$ is found during the passage of the clump through the inner edge of 3D disc, (ii) the minimum values of $\dot{M}_{\rm dep}$ are found to the right and left sides of the maximal value, (iii) $\dot{M}_{\rm dep}$ is integrated between the minimums and gives the total mass of the clump.}  The object is inserted in the beginning of the accretion burst, as explained in detail below. It is further assumed that the initial radius of the object is equal to 70 Jovian radii ($\approx 7 \rsun$). We study evolution of embedded objects with different masses in \S\ref{sec:masses} and with different initial radii in \S\ref{sect:cooling_heating}. \bref{The choice of 25 per cent mass assumption for the embedded object as a fiducial case is based only on the desire to pick a representative value from the range of masses studied in \S\ref{sec:masses}.}

\begin{figure}
\begin{centering}
\includegraphics[width=1\columnwidth]{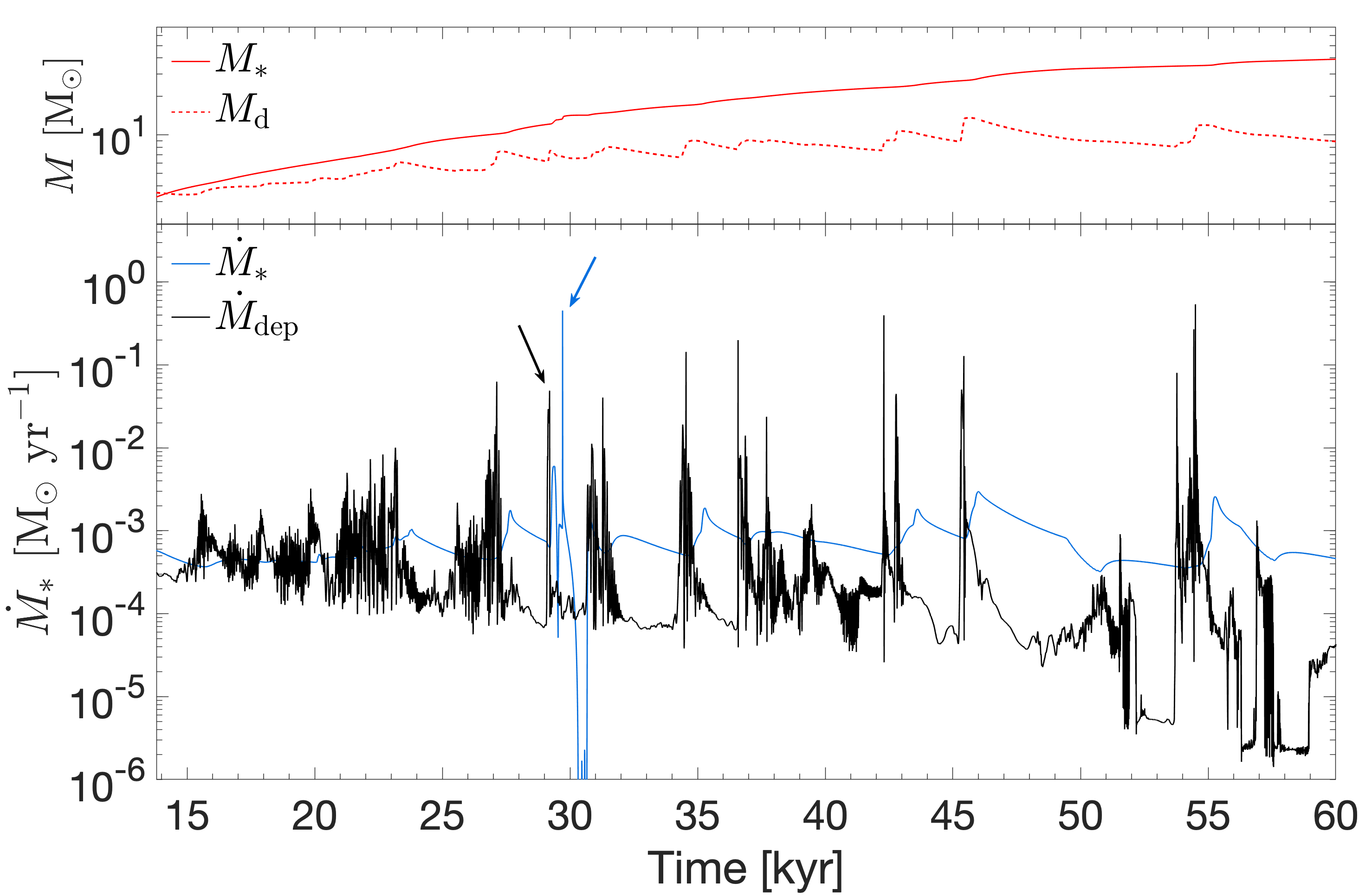}
\par\end{centering}
\caption{\label{fig:1} Similar to Figure~\ref{fig:0} but now with the embedded object inserted into the 1D disc. The black arrow shows the $\dot{M}_{\rm dep}$ burst caused by the clump passing through the inner edge of the 3D disc, while the blue arrow shows the accretion burst due to the disruption of the embedded object. See \S \ref{sect:obj_migration} for detail.}
\end{figure}

Figure~\ref{fig:1} shows the resulting accretion rate history onto the star. It is instructive to compare it to Figure~\ref{fig:0}, which is an entirely analogous calculation but without the embedded object. There is an intense spike of accretion onto the star in the blue solid curve in Fig. \ref{fig:1} at time $t\approx 30$~kyr due to tidal disruption of the object in the inner disc. Viewed on timescales of thousands of years, it appears that the main effect of the inner disc is to delay the original accretion burst (pinpointed with the black arrow) to the later burst (the blue arrow). However, on human time scales there are important differences in the accretion disc history onto the star before and after the burst, and the duration of the burst itself is strongly modified.



\begin{figure}
\begin{centering}
\includegraphics[width=1\columnwidth]{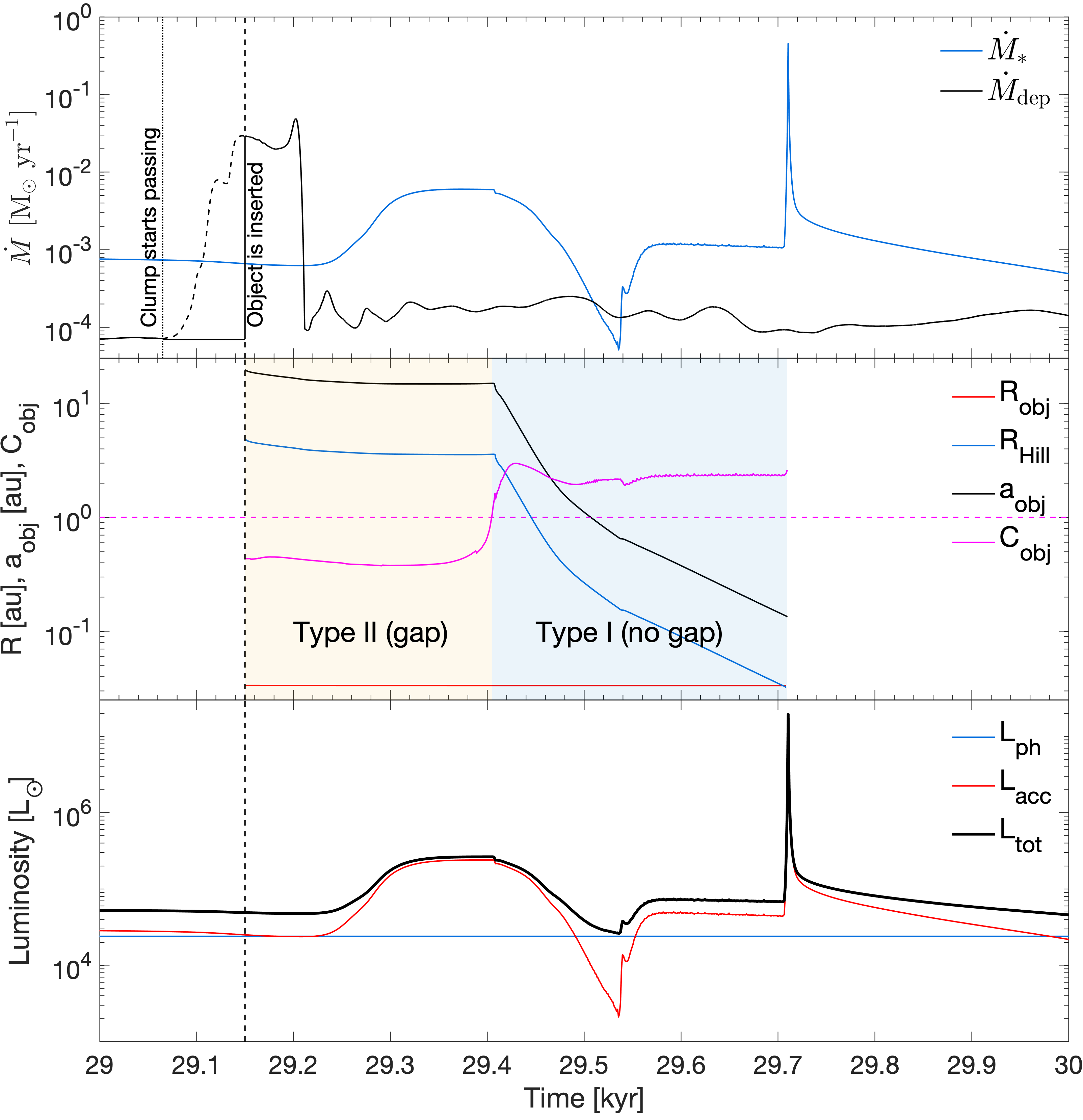}
\par\end{centering}
\caption{\label{fig:2} Disc and HMYSO properties during the embedded object disruption event. \textbf{Top panel:} Accretion rate history onto the HMYSO (blue line) and mass deposition rate onto the outer edge of 1D disc (solid black line). The vertical dashed line marks the moment of object injection into the disc. The dashed part of mass deposition rate shows the original part of the curve that was replaced with constant low deposition rate to subtract the embedded object mass from the clump mass. \textbf{Middle panel:} Radius ($R_{\rm obj}$) and Hill radius ($R_{\rm Hill}$) of the object are show, respectively, with red and blue lines. Orbital distance of the object ($a_{\rm obj}$) is shown with black line, while the gap opening parameter ($C_{\rm obj}$) is shown with solid magenta line. The horizontal dashed magenta line shows the threshold value of $C_{\rm obj}=1$ below which the object opens a gap in the disc and migrates in Type II regime, while above it no gap is opened and the object migrates in Type I regime. \textbf{Bottom panel:} Photospheric ($L_{\rm ph}$), accretion ($L_{\rm acc}$), and total ($L_{\rm tot}$) luminosity of the central HMYSO presented, respectively, with blue, red, and black lines.}
\end{figure}

In Fig. \ref{fig:2} we present detail of disc and object evolution following its insertion into the  1D disc. The top panel of Figure~\ref{fig:2} shows the accretion rate histories in the 3D (black) and the 1D calculation (blue) between 29 kyr and 30 kyr. The panel also shows the timing of object insertion in our numerical method. In 3D simulations,  the clump starts passing through the inner edge of the 3D disc at time $t=$29065~yr. Here we set $\dot{M}_{\rm dep}$ constant equal to $\dot{M}_{\rm dep}(t=29065)$ and wait until the time integral of the excess mass passing through $r_{\rm in,3D}$ reaches the mass of the embedded object. This occurs at time $t=$29150~yr when the object is inserted and released in the inner disc (the vertical dashed line in Fig. \ref{fig:2}). This prescription ensures mass conservation. \bref{A discussion of the object insertion procedure can be found in \S \ref{sect:ecc_orbit}. }



In the middle panel of Figure~\ref{fig:2} we plot the radius of the object, $R_{\rm obj}$, its Hill radius, $R_{\rm Hill}$, the radial distance of the object ($a_{\rm obj}$) to the star, and the gap opening parameter $C_{\rm obj}$. Shortly after its insertion inside the disc, the object opens a deep gap in the disc. This is expected since $C_{\rm obj}$ is less than the critical value of 1 (dashed magenta line) required for gap opening. We observe that during the next 250~years the object first migrates from 20~\bref{au} to 15~\bref{au}  in the Type II regime (yellow shaded area in the middle panel of Fig.~\ref{fig:2}). Once the parameter $C_{\rm obj}$ becomes greater than 1, the gap in the disc closes, the object switches to Type I migration regime, and starts migrating inward very rapidly, reaching  0.13~\bref{au} in just about 300~years. We focus on the migration of the object in more detail in \S\ref{sec:disc_profiles}. Given this very short migration time, the cooling of the object is ineffective and 
it shrinks only very slightly below the initial radius of 70 Jovian radii. As the object gets closer to the star, its Hill radius decreases precipitously. When $R_{\rm Hill}$ approaches and eventually drops below $R_{\rm obj}$, the object starts to lose  mass extremely rapidly, resulting in its runaway tidal disruption. This happens at radial distance $r=0.134$~\bref{au}. The mass lost from the object is deposited in the disc at about 0.1~\bref{au} from the star. As the viscous time is very short on these scales, a very powerful accretion outburst results, reaching the peak accretion rate of $4.5\times10^{-1}~M_{\odot}$~yr~$^{-1}$.


The total luminosity of the central star ($L_{\rm tot}$) is a sum of its photospheric luminosity ($L_{\rm ph}$) and the accretion luminosity ($L_{\rm acc}$). The accretion luminosity is defined here as
\begin{equation}
\label{eq:Lacc}
    L_{\mathrm{acc}} = \frac{GM_*\dot{M}_*}{R_*},
\end{equation}
where $M_*$ and $R_*$ are, respectively, the mass and the radius of the protostar. If  photospheric luminosity dominates over the accretion luminosity, then accretion outburst is unlikely to be detectable \citep{2016Elbakyan}.


The bottom panel in Figure~\ref{fig:2} shows photospheric, accretion, and total ($L_{\rm ph}+L_{\rm acc}$) luminosity of the central protostar in our 1D model. For illustration purposes here, we assume $L_{\rm ph}$ to be constant and set it to the pre-burst bolometric luminosity of $\sim20~M_{\odot}$ high-mass stellar object S255IR-NIRS3 from \citet{2017Caratti}, $L_{\rm bol}=2.4\times10^4~L_{\odot}$, and consider it constant during the burst. We use this object since it is the most studied one from a very few known outbursting HMYSOs and its mass is close to the stellar mass in our model (TBD). \bref{Also note that bolometric luminosity of a young massive star with this mass is comparable to $2.4\times10^4~L_{\odot}$.} We can see that $L_{\rm acc}$ follows time evolution of stellar accretion rate $\dot M_*$ shown in the top panel of Fig. \ref{fig:2}, and is in general higher than the photospheric luminosity. Shortly after object disruption the accretion luminosity, and the total luminosity of the protostar, rises rapidly by a factor of $\sim300$, reaching $2\times10^7~L_{\odot}$. \bref{The assumption of constant $L_{\rm ph}$ during the outburst event is a simplification of the model given that the radius of HMYSO can change dramatically, reaching a few hundred solar radii during the burst \citep[e.g.,][see also \S \ref{sec:stellar_rad}]{2009HosokawaOmukai}. Thus, for more realistic results, the evolution of HMYSOs needs to be studied self-consistently with the stellar evolution code. We plan to conduct such a study in the future.}

\begin{figure}
\begin{centering}
\includegraphics[width=1\columnwidth]{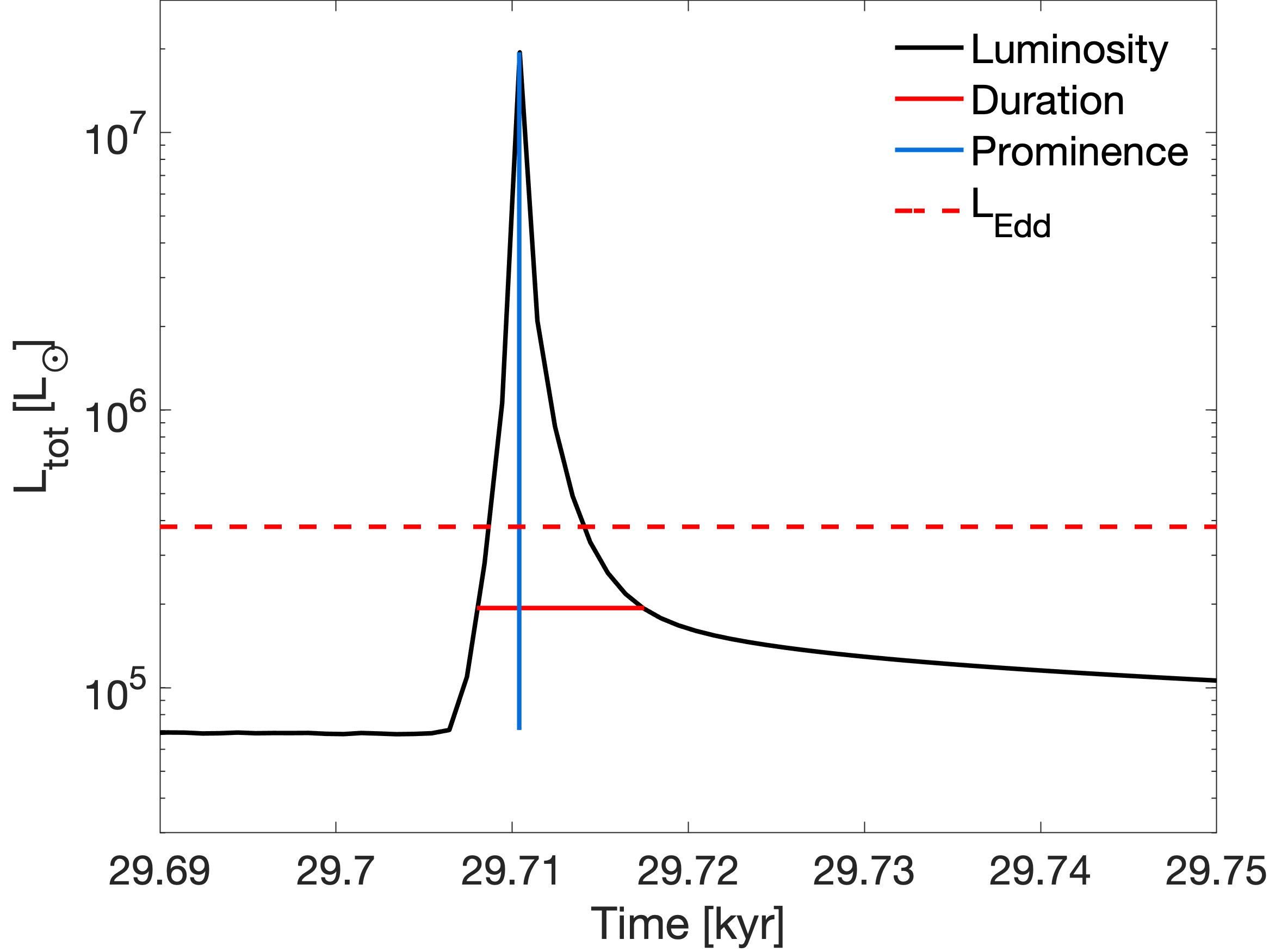}
\par\end{centering}
\caption{\label{fig:3} Zoom-in on the bottom panel of Fig. \ref{fig:2}, showing the total luminosity of the star vs time around tidal disruption of the object. The vertical blue line shows the prominence of the burst, and the horizontal red line indicates the duration of the burst measured at a vertical distance equal to 1 \bref{per cent} of the burst prominence. The dashed line shows Eddington luminosity for the central star.}
\end{figure}

Figure~\ref{fig:3} further zooms in on the burst shown in Fig. \ref{fig:2}, showing the total luminosity of the star. The vertical blue line shows the prominence of the burst. The duration of the burst is defined as the time period when $L_{\rm tot}$ exceeds  1 \bref{per cent} of the burst prominence (cf. the solid red horizontal line). According to this somewhat arbitrary definition, the burst duration is 9.5~years, with the rise time of about 4 years. The decay time, during which the total luminosity recovers its pre-burst value, is about 125 years. Such a long decay time is due to the outward viscous spreading of some of the material of the disrupted object. It is interesting to note that the observed outburst from S255IR-NIRS3 \citep{2017Caratti} had a duration of about 2 years, although it is much less luminous than the one in Fig. \ref{fig:2}. A few other accretion outbursts in HMYSOs have been observed in recent years: NGC 6334I MM1 \citep{2017Hunter, 2018MacLeod}, G358.93-0.03 MM1 \citep{2019Brogan, 2021Stecklum}, G323.46-0.08 \citep{2019Proven-Adzri}, M17 MIR \citep{2021Chen}, having duration from a few months to a decade.

The horizontal dashed line in Fig.~\ref{fig:3} shows the Eddington luminosity for the central star, $L_{\rm Edd} = 3.2\times10^4~(M_*/M_{\odot})~L_{\odot} \approx 3.8\times10^5~L_{\odot}$. During the peak times, the outburst significantly exceeds $L_{\rm Edd}$.  At this point in the calculation, the hydrostatic balance assumption of our 1D model breaks down in the inner part of our disc (cf. Fig.~\ref{fig:4} and \S\ref{sec:disc_profiles} below). It is likely that the radiation pressure generated during super-Eddington outbursts similar to this one would blow off a significant fraction of the disc that in our present calculation accretes onto the star.  Disruption of objects of smaller masses, such as massive gas giant planets or brown dwarfs, could yield outbursts of much smaller luminosity and hence not violate the Eddington limit. We note that for HMYSO bursts observed so far the gas mass accreted onto the stars during the bursts are estimated at between  0.3 to 30 $M_{\rm Jup}$, much lower than  $\sim 550 M_{\rm Jup}$ considered in this example in Fig.~\ref{fig:2}.

Finally, we note that we assumed for simplicity that the stellar radius $R_*$ is constant, $R_*=10~R_{\odot}$, during the entire simulation. A model with a larger stellar radius is presented in \S\ref{sec:stellar_rad}. We leave an analysis of models with a dynamical disc inner edge and dynamical stellar radius for a follow-up study.




\subsection{Influence of the migrating object on the disc} \label{sec:disc_profiles}

\begin{figure}
\begin{centering}
\includegraphics[width=1\columnwidth]{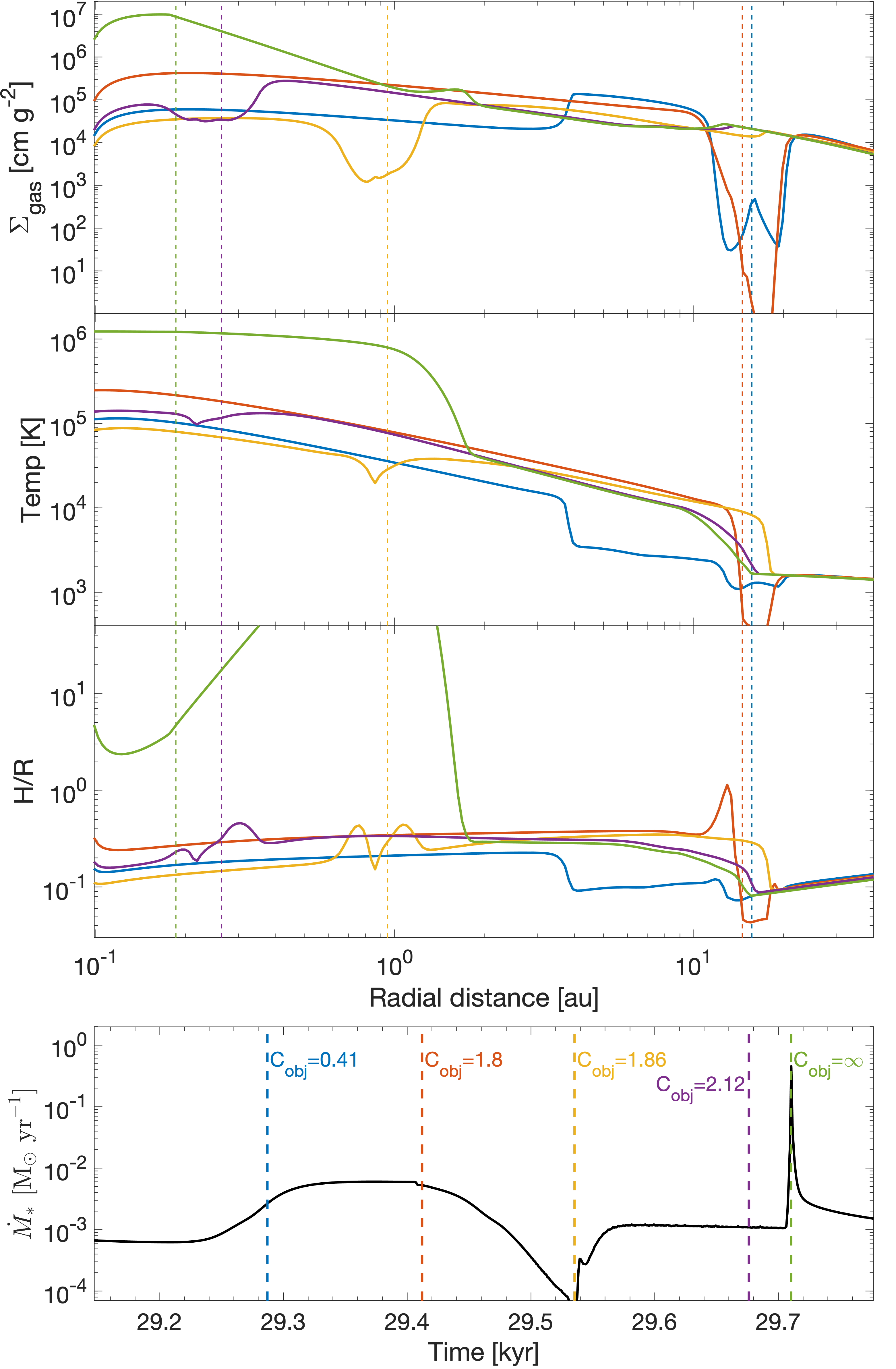}
\par\end{centering}
\caption{\label{fig:4} \textbf{Top panels:} The radial profiles of gas surface density (top panel), midplane disc temperature (second panel), disc aspect ratio (third panel) for five distinct time moments. Vertical dashed lines mark the radial distance of the object at the corresponding time moment. Note that at the very peak of the burst (the green curves), the disc is so hot as to formally make $H/R >10$ in the context of our 1D model. As discussed in \S \ref{sec:disc_profiles}, at this point the disc luminosity exceeds the Eddington limit. Strong outflows are likely to be launched but we do not model them here. \textbf{Bottom panel:} Mass accretion rate history onto the HMYSO during the inward migration and disruption of the object. Vertical dashed lines indicate the time moments for which the radial profiles with corresponding colours are shown on the top panels. The value of parameter $C_{\rm obj}$ for each time moment is shown near the vertical dashed lines.}
\label{fig:Disc_structure}
\end{figure}

We now shift focus onto the disc properties during the object migration and disruption. We plot in Figure~\ref{fig:4} the time evolution of disc surface density profile (top panel), midplane temperature profile (second panel), 
and disc aspect ratio profile (third panel) during the time period of object migration and disruption in our model. The bottom panel shows the stellar mass accretion rate, with coloured vertical dashed lines marking the time moment of the corresponding profile in the top panels. The value of gap opening parameter $C_{\rm obj}$ is shown near each vertical line.

The embedded object is injected in the 1D disc at $t=29150$~yr and shortly after a deep gap is opened in it. This is evident in the blue curve in Figure~\ref{fig:4} that corresponds to $t=29287$~yr. The disc temperature at the position of the object at that time is about 1200~K. After the object is deposited into the 1D disc, diffuse mass deposition from the 3D disc into the 1D disc continues at high rates until the end of the mass injection episode (cf. the drop in the black curve in Fig. \ref{fig:2}). As a result, the disc heats up, increasing the aspect ratio $h$. When the disc temperature at the position of the object reaches a few thousand Kelvin ($t=29412$~yr, the orange line, 262 years after the injection of the object),  the gap around the object closes due to the higher local value of $h$. As a result, the object switches its migration from Type II to Type I and migrates inwards more rapidly. A period of runaway-like inward migration results (the region shaded blue in the middle panel in Fig. \ref{fig:2}).

During this migration, the accretion rate onto the protostar drops by almost two orders of magnitude, although then recovering somewhat to $\dot M_* \sim 10^{-3}\msun$~yr$^{-1}$. This unexpected behaviour is due to the high mass (angular momentum) that the object possesses.  As its angular momentum is transferred to the disc in its vicinity, the inward flow of gas through its orbit is reduced. However, at later times, (e.g., see the yellow line, corresponding to  $t=29535$~yr, 385 years after the injection) the depth of the gap opened by the object drops, and the gas passes through planetary orbit more readily. This allows the gas accretion rate onto the protostar to recover somewhat.


Due to the high masses of both the object and the surrounding disc, the migration of the former is extremely rapid; the object migrates from $r=0.94$~\bref{au} to $r=0.134$~\bref{au} in just 175 years. At the radial distance $r=0.134$~\bref{au} the object is tidally disrupted, triggering a strong accretion burst onto the protostar. The material lost by the object is injected in the very inner disc. Some of this mass is accreted by the star very rapidly, some moves to the higher orbit due to the diffusive nature of the angular momentum transport in our model \citep[recall that in the][model the outer edge of the disc spreads outwards due to conservation of angular momentum]{1973ShakuraSunyaev}. As a result of object disruption, the surface density of the inner $r\lesssim0.4$~\bref{au} disc increases by more than two order of magnitudes, reaching $10^7$~g~cm$^{-2}$. This can be seen at $t=29710$~yr, the green curves, 560 years after the object injection. With time, this "fresh" material accretes onto the HMYSO and the disc surface density eventually recovers its pre-burst profile. The temperature and aspect ratio of the disc undergo a strong increase in the inner 1~\bref{au} of the disc. The temperature in the inner 1~\bref{au} disc reaches $10^6$~K, which in turn makes the aspect ratio of the inner 1~\bref{au} disc to {\em formally} exceed 10. As discussed in \S \ref{sec:fiducial}, this occurs because the disc luminosity exceeds Eddington limit during the peak of the burst. The thin disc approximation that we use in this paper actually breaks down at this point. 3D radiation hydrodynamics of the innermost disc region is necessary to simulate this. This is outside the scope of our paper. Luckily, the period where the Eddington limit is exceeded is brief, and therefore one may hope that our 1D simulation does capture the main properties of the outburst except at the very peak.

\subsection{Dynamics of objects with different masses} \label{sec:masses}

\begin{figure}
\begin{centering}
\includegraphics[width=1\columnwidth]{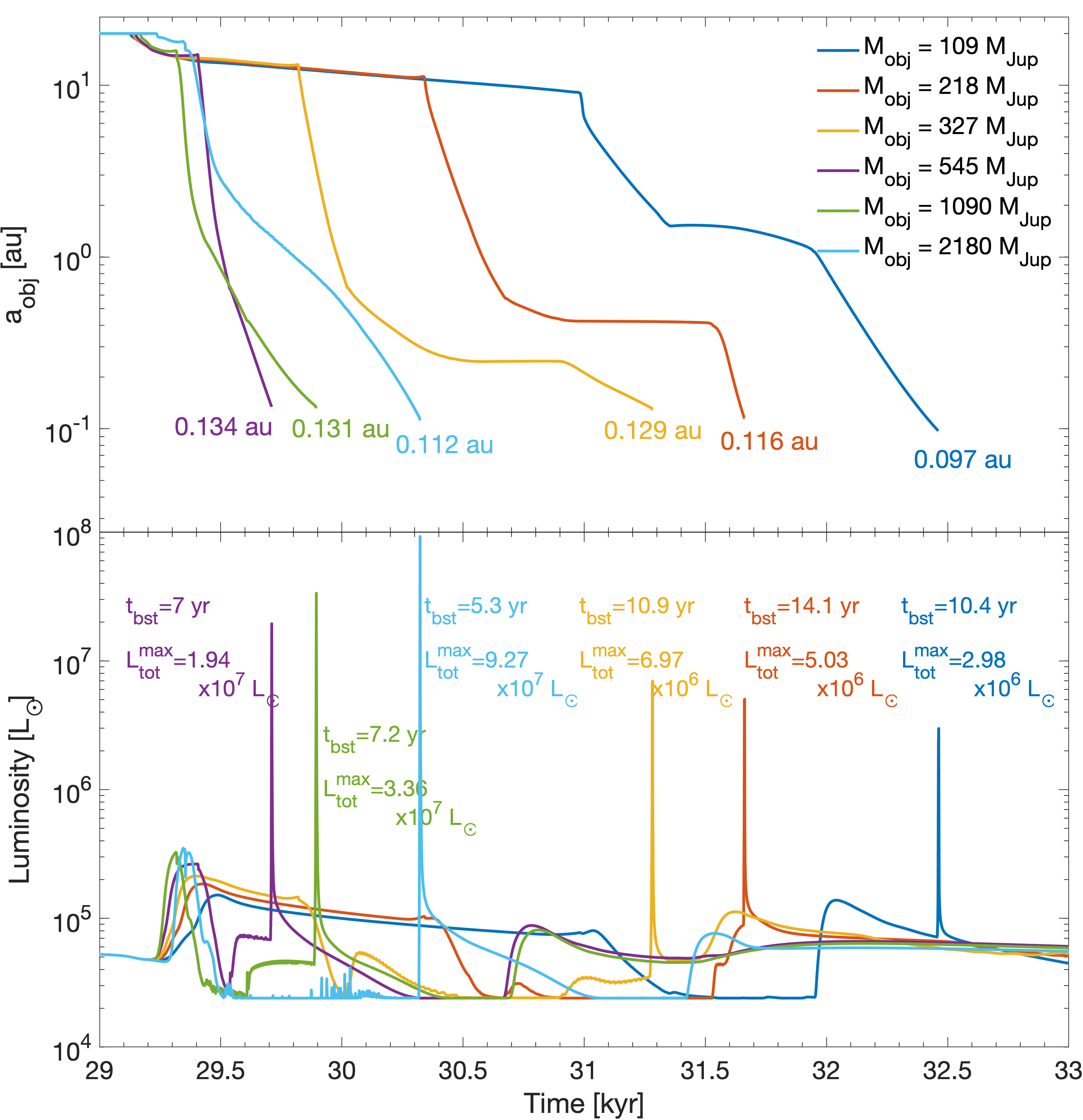}
\par\end{centering}
\caption{\label{fig:6} \textbf{Top panel:} The time evolution of radial distance of the objects with different masses. The value near each line shows the radial distance of the object at the moment of its disruption. \textbf{Bottom panel:} Time evolution of total luminosity of central HMYSO during the burst event. The peak value of total luminosity and the duration of the burst are shown near each line.}
\end{figure}

In our fiducial model, we assumed that the mass of the embedded object formed inside the clump is a 0.25 fraction (25 \bref{per cent}) of the total clump mass. However, the mass of the embedded object strongly depends on the characteristics of the clump, such as rotational-to-gravitation energy ratio, temperature, and density profiles.
Here we present calculations of the disc-object evolution entirely analogous to that presented in \S \ref{sec:fiducial}, but for a range of the mass of the embedded object, that is, 5, 10, 15, 50, 100 \bref{per cent} of the total clump mass. As was described earlier, the moment of embedded object insertion depends on its mass -- the more massive is the object, the later it is inserted during the clump crossing the 3D disc inner boundary. For example, in the model where we assume that the mass of the embedded object is equal to the mass of the clump, the object is inserted at the moment when all of the clump fully passes through the inner boundary of the 3D disc.

In Figure~\ref{fig:6} we show the time evolution of the orbital distance of the objects (top panel) and the total luminosity of the central HMYSO (bottom panel) in each model. As in the fiducial model, the objects open a gap in the disc after the insertion and migrate inward in the Type II regime before they reach the high temperature region in the inner disc. In these hot inner disc regions, $h$ is significantly larger than just outside, so this usually leads to gap closure. The objects then migrate inwards much faster in the Type I regime. The lower the mass of the object, the longer it takes them to migrate inwards. At the very high object masses, there is, however, an exception to this behaviour. Migration of objects with mass exceeding 1~$M_{\odot}$ takes slightly longer than for the $\sim$0.5~$M_{\odot}$ counterpart. This occurs because at this mass scale the objects become comparable with the disc mass in their vicinity. 

There is further complexity to some of the migration tracks. In some cases, the objects manage to open gaps in the disc for the second time. All the objects are tidally disrupted at $r$ $\approx$0.1~\bref{au}. The exact radial distance at which the objects are disrupted is listed near the end of the corresponding migration track in the top panel of the figure. We note that this is a consequence of the toy model for the internal structure of the object that we accepted in the present paper. It is likely that more sophisticated treatment of object contraction may lead to different disruption locations. This important question is left to a future study.


For each model, the duration of the burst, $t_{\rm bst}$ (calculated as described earlier in \S\ref{sect:obj_migration}), and the peak total luminosity during the burst, $L_{\rm tot}^{\rm max}$, are shown near the corresponding luminosity curves in the bottom panel of Fig.~\ref{fig:6}. As can be expected, the more massive is the object, the more luminous is the burst. Burst durations vary from 5.3 to 14.1~years, being close to the viscous timescale at the radial distance $r\approx0.1$~\bref{au}, $t_{\rm visc}\approx10$~yr.


\section{Low mass objects}\label{sec:Low_M_both_models}

\subsection{Standard discs: stalled migration at the ionisation front}\label{sect:low_mass_objects}

\begin{figure}
\begin{centering}
\includegraphics[width=1\columnwidth]{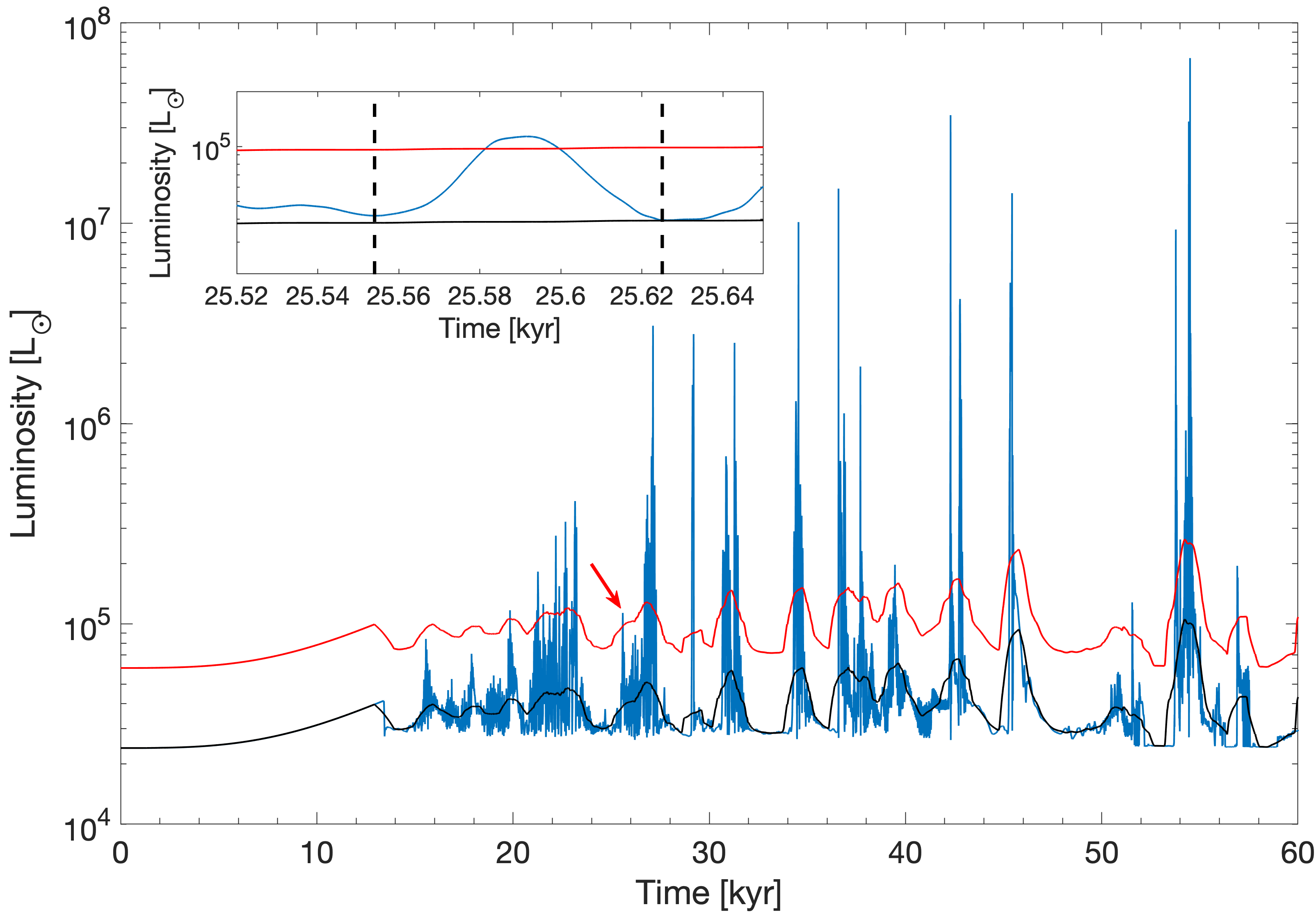}
\par\end{centering}
\caption{\label{fig:7} The blue line shows the time evolution of the total luminosity of central HMYSO calculated using Eq.~\ref{eq:Lacc} with $\dot{M}_{\rm dep}$ instead of $\dot{M}_*$, and $L_{\rm ph}=2.4\times10^4~L_{\odot}$ which is equal to the pre-burst bolometric luminosity of S255IR-NIRS3 from \citet{2017Caratti}. The background luminosity calculated with Eq.~\ref{eq:Lbg} is shown with the black line, while the 1-magnitude cut-off used for the burst filtering is shown with the red line. The arrow points at the burst event triggered by the passage of the clump through the outer edge of the 1D disc. The inset presents the magnified view on the burst pointed by the arrow. The vertical dashed lines in the inset show the beginning and the end of the burst event. }
\end{figure}

The masses of clumps formed in numerical simulations of discs around HMYSOs may vary in a broad range of values, from a few Jovian masses to a few solar masses \citep{2021Meyer}.  For the fiducial model presented in \S \ref{sec:fiducial}, we chose a burst associated with a passage of a very massive clump through the inner boundary of the 3D disc. In this section, we study the inward migration and disruption of embedded objects with masses below 100 $M_{\rm Jup}$.


The passage of low mass clumps through the inner boundary of 3D simulations are much harder to distinguish from the background variability of the disc. The latter can be caused, for example, by dynamical perturbations from large-scale spiral arms rather than clump arrival at the inner boundary.
Following the method used in \citet{2018VorobyovElbakyanPlunkett} and \citet{2021Meyer}, we filter out regular low-amplitude variability to distinguish that from clump accretion bursts.

First, using the mass deposition rate onto the 1D disc ($\dot{M}_{\rm dep}$), we calculated the total luminosity curve with Eq.~\ref{eq:Lacc} and assuming that the photospheric luminosity of central HMYSO is equal to the pre-burst bolometric luminosity of S255IR-NIRS3 from \citet{2017Caratti}, $L_{\rm bol}=2.4\times10^4~L_{\odot}$. The total luminosity curve is shown in Figure~\ref{fig:7} with the blue line. Then, we introduce the background luminosity, $L_{\rm bg}$, which removes strong luminosity bursts as follows
\begin{equation}
\label{eq:Lbg}
    L_{\mathrm{bg}}(t)=
    \begin{cases}
      \langle L_{\mathrm{ph}}(t)+L_{\mathrm{acc}}(t)\rangle, & \text{ if } \dot{M}_{\mathrm{dep}}\leq\dot{M}_{\mathrm{crit}}    \\
      \langle L_{\mathrm{ph}}(t)+(\dot{M}_{\mathrm{crit}}/\dot{M}_{\mathrm{dep}})L_{\mathrm{acc}}(t)\rangle, & \text{ if } \dot{M}_{\mathrm{dep}}>\dot{M}_{\mathrm{crit}}
    \end{cases}\,.
\end{equation}
where $\dot{M}_{\mathrm{crit}}=10^{-3}~M_{\odot}\mathrm{yr}^{-1}$ and angle brackets indicate time averaging over a period of 1000 years. The background luminosity is shown in Fig.~\ref{fig:7} with the black line. Next, we make an assumption that the peak luminosity during the outburst associated with the clump must be at least 1~magnitude (2.5~times) higher than the background luminosity. The 1-magnitude cut-off is shown with the red line in Fig.~\ref{fig:7}. Finally, we choose an isolated luminosity burst triggered due to the passage of a clump, which satisfies the peak luminosity criteria, and make an assumption that an embedded object is already formed inside the clump. The luminosity burst associated with the selected clump is marked with the red arrow in Fig.~\ref{fig:7} and the zoomed in view on the burst is shown in the inset. The vertical dashed lines in the inset show the beginning and the end of the burst. These time moments correspond to the clump beginning to enter, and then fully passing through the inner edge of the 3D disc. The total mass of the clump in this event is 86~$M_{\rm Jup}$. 

\begin{figure}
\begin{centering}
\includegraphics[width=1\columnwidth]{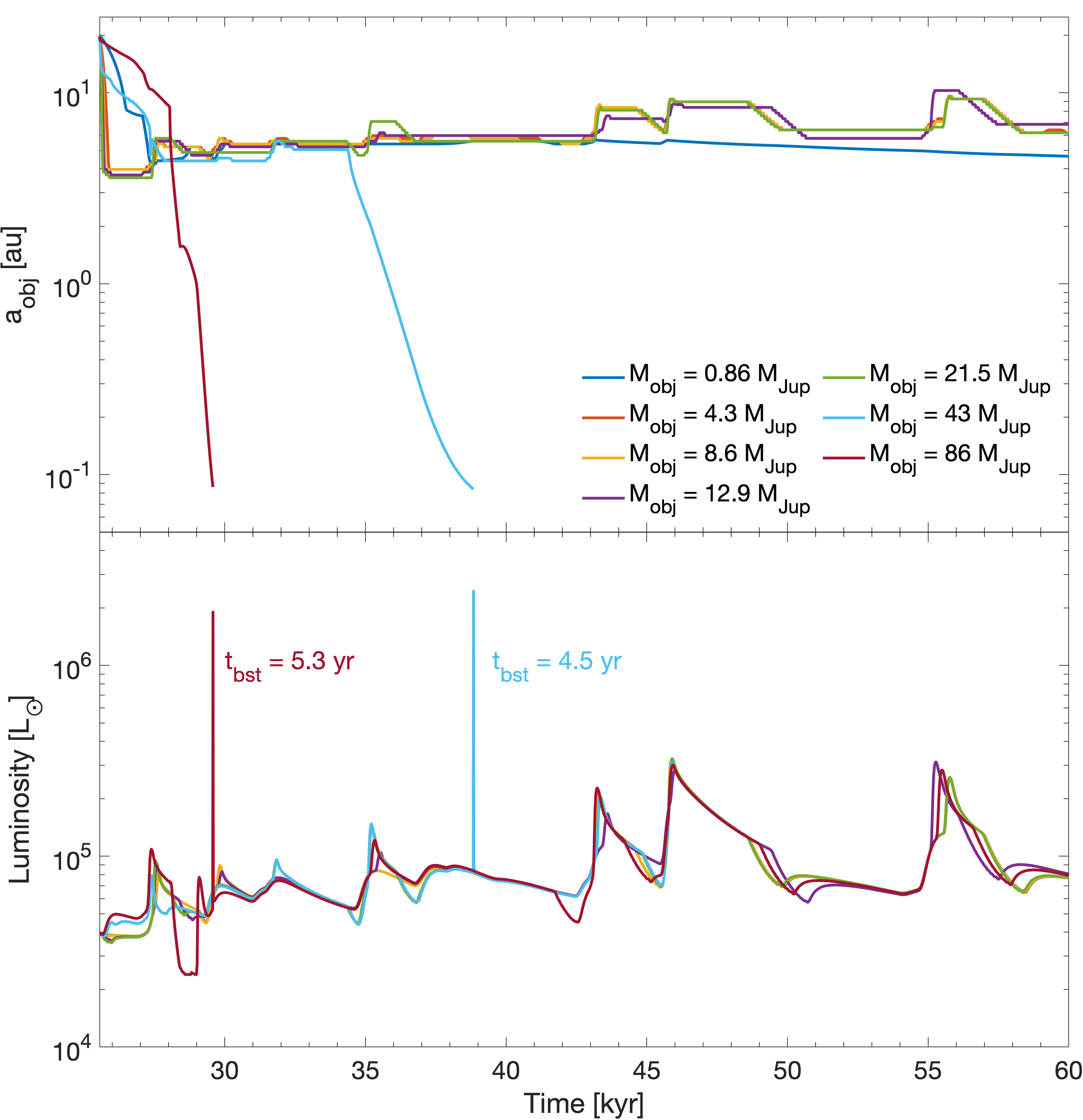}
\par\end{centering}
\caption{\label{fig:8}  Temporal evolution of the radial distance of the objects with different masses (top panel) and temporal evolution of the total luminosity of central HMYSO in all models (bottom panel). Only the two most massive objects migrate into the inner disc and get disrupted at $\sim$0.85~\bref{au}, triggering a luminous outburst on HMYSO. Lower mass objects stall at the migration trap due to the hydrogen ionisation front. The up and down jumps in object orbits occur when the front makes periodic up or down excursions. See \S \ref{sect:low_mass_objects} for more detail.}
\end{figure}

In order to have a better understanding how the objects with planetary/brown dwarf masses migrate inside the disc of HMYSO, we vary the mass of the embedded object, assuming that its mass is 1, 5, 10, 15, 25, 50, 100 \bref{per cent} of the total clump mass. The top panel of Figure~\ref{fig:8} shows the time evolution of radial distance of the objects with different masses, while the bottom panel shows the corresponding total luminosity curve for the central HMYSO in the models. Only two most massive objects, that is, those with $M_{\rm obj}= 43 M_{\rm Jup}$ and 86~$M_{\rm Jup}$, get disrupted in these experiments. This occurs at $r=0.085$~\bref{au} from the HMYSO, causing a high luminosity burst. The total luminosity of HMYSO reaches about $2\times10^6~L_{\odot}$ during the bursts. The burst duration for both of the objects is about 5 years.

All of the less massive objects ($M_{\rm obj} < 43 M_{\rm Jup}$) migrate inwards to the radial distance $\sim3 \text{--} 4$~\bref{au} and then stop, orbiting at that radial distance for more than 10 kyrs. Some of the objects even show gradual outward migration, punctuated by short periods of rapid outward migration that coincide with the moments when HMYSO  luminosity increases. The latter is driven by accretion rate increases onto HMYSO. 

\begin{figure}
\begin{centering}
\includegraphics[width=1\columnwidth]{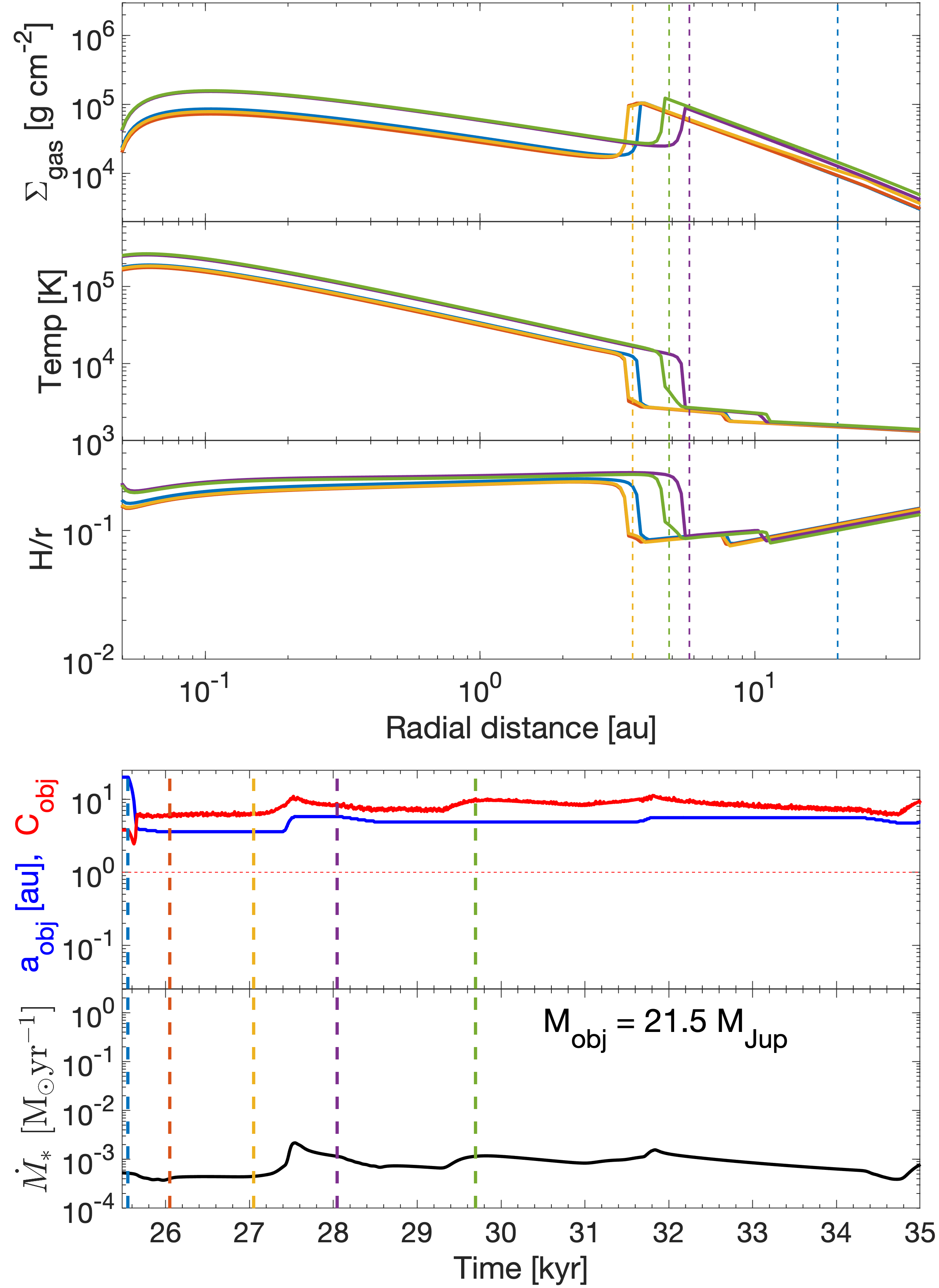}
\par\end{centering}
\caption{\label{fig:9}{\bf Top panel:} several snapshots of disc surface density, midplane temperature, and the aspect ratio $H/r$ at notable moments during the inward migration of the object. The vertical dashed lines show the radial position of the object at respective times. {\bf Bottom panel:} the radial distance of the object ($a_{\rm obj}$), gap opening parameter ($C_{\rm obj}$) and mass accretion rate onto the star ($\dot{M}_*$) vs time. Vertical dashed lines mark the time moments for which the disc properties are shown in the top panel. For more details, see \S \ref{sect:low_mass_objects}.
}
\end{figure}

The key to this stalled and outward migration of lower mass objects is the fact that they open no gap in the disc, and so migrate in the Type I regime. While the Type I migration in isothermal discs is usually directed inwards  \citep[e.g.][]{1997Ward, 2002TanakaWard}, migration in radiative discs can be directed outwards \citep{2008KleyCrida, 2008BaruteauMasset, 2009KleyBitsch, 2011Paardekooper, 2014Baruteau-PPVI, 2017JimenezMasset, 2021Guilera}. To understand this further, we plot in the top three rows of Figure~\ref{fig:9}, respectively, the surface density, temperature, and aspect ratio profiles of the disc at different time moments during the inward migration of the object with mass $M_{\rm obj} = 21.5~M_{\rm Jup}$. The vertical dashed lines in the top three rows show the position of the object at each time moment. In the bottom two panels of the figure, we plot the time evolution of the radial distance of the object (blue line), gap opening parameter (red line), and mass accretion rate onto the star (black line). The vertical dashed lines mark the time moments for which the radial profiles with the corresponding colours are shown on the top panels.

Note a step-like change in the radial profiles of the surface density, temperature, and the aspect ratio at about 4~\bref{au}. This is caused by the dip and then a rapid increase in the gas opacity in the temperature range from $T \sim 2\times 10^3$~K to $T\gtrsim 10^4$~K. Portions of the disc between these temperatures are thermally unstable, so the disc is forced to avoid this range of temperatures \citep[see \S 6 in ][]{2015Armitage}. The key difference between the cold stable and the hot stable solutions to the disc vertical structure equations is the fact that hydrogen is neutral and molecular on the former and completely ionised on the latter \citep[][]{1994BellLin}. At low enough mass accretion rates in the disc, this opacity/ionisation transition results in the well known hydrogen ionisation instability of discs, including cataclysmic variables and dwarf novae systems \citep{Lin85-CVs}. The inner region of the disc keeps switching between the low and the high stable branches, resulting in the significant accretion variability onto the star \citep[e.g.,][]{2004LodatoClarke,2021ElbakyanNayakshin}. At higher accretion rates, the innermost region of the disc is permanently on the hot stable solution branch. The ionisation fronts similar to the ones seen in Fig. \ref{fig:9} are at increasingly larger radii for larger $\dot M$ in the disc; however, even at a fixed mean accretion rate the fronts are quasi-static rather than completely stationary, performing small excursions around a mean location \citep[e.g.,][]{Clarke89-TI-FUOR}.

Analysis of the different terms in the \cite{2011Paardekooper} expressions for Type I planet migration torques shows that the (near) discontinuities in the disc due to hydrogen ionisation form a very strong migration trap. That is, to the right of the ionisation front, the migration torques push the objects inwards. However, as it reaches the ionisation front, the torques are positive and hence push the object outward; thus the object sits just outside the temperature jump. 

Further, the size of the inner $T \gtrsim 10^4 K$ region, where hydrogen is completely ionised, varies between $\sim$~3 and 5~\bref{au} in our simulations in response to the variability in the mass supply rate into the inner disc. The planets follow this motion of the ionisation front, which leads to the surprising and irregular see-saw pattern for some of the planets in the top panel of Fig.~\ref{fig:8}.

In conclusion to this section, we find that planets less massive than a few tens of $M_{\rm Jup}$ may fail to migrate into the inner few au of the disc. This is in disagreement with results of Paper I, where we used the isothermal \cite{2002TanakaWard} torque expressions for the migrating planets. While the approach of \cite{2011Paardekooper} appears more physically relevant to the problem at hand, we note that their calculations assume the disc entropy function of the form $P/\Sigma^{\gamma}$. At the ionisation front, this entropy form breaks down. Our disc equation of state (EOS) takes into account $\mathrm{H}_2$ molecule rotational and vibrational transitions, $\mathrm{H}_2$ dissociation, and $\mathrm{H}$ atom ionisation. This leads to significant variations in $\gamma$, the mean molecular weight, and the entropy function across the ionisation front. Planet migration studies specifically designed to capture disc ionisation front transitions like those we find here are needed to confirm the existence of the migration traps for lower mass objects. There is also an additional important uncertainty in the disc physics that we discuss in the next section.



\subsection{Magnetised disc winds: object migration through the ionisation front}\label{sec:low_mass_dw}

ALMA observations indicate that the turbulent velocities in protoplanetary discs are quite low, so that for most of the discs $\alpha\lesssim 10^{-3}$ \citep{2017Lodato, DSHARP-6, 2020Rosotti, 2021DoiKataoka}. Such a low level of turbulence in the disc is insufficient to ensure the observed mass accretion rates onto the stars \citep[cf. the review and references in][]{2022MiotelloPPVII}. It is widely believed that the likely candidate mechanism responsible for the majority of mass and angular momentum transport in protoplanetary discs could be the MHD driven winds \citep[e.g.,][]{2013BaiStone,Armitage13-MHD-wind,Suzuki16-MHD-winds,Lesur21-MHD-winds}.

In this section, we attempt to evaluate the potential effects of magnetised disc winds on the outcome of object migration. \bref{We note that MHD wind is not present in the 3D model, and we study the influence of MHD wind in our 1D model.}
In order to include the winds in a simple form we follow \cite{2022Tabone}, and add an advective term on the r.h.s. of Eq.~(\ref{dSigma_dt}):
\begin{equation}
\begin{split}
    \frac{\partial\Sigma}{\partial t} = \frac{3}{r} \frac{\partial}{\partial r} \left[ r^{1/2} \frac{\partial}{\partial r} \left(r^{1/2}\nu \Sigma\right) \right] - \frac{1}{r} \frac{\partial}{\partial r} \left(2\Omega^{-1} \lambda \Sigma\right) + \qquad\qquad \\
    \frac{\dot{M}_{\mathrm{ obj}}}{2\pi r}\delta (r - r_{\mathrm{obj}}) + \frac{\dot{M}_{\mathrm{dep}}}{2\pi^{3/2} r \sigma} \exp\left[-\left(\frac{r-r_{\mathrm{dep}}}{\sigma}\right)^2\right] - \frac{1}{r} \frac{\partial}{\partial r} \left( v_{\mathrm{dw} }\Sigma r \right),
\end{split}
\label{dSigma_dt_dw}
\end{equation}
where the gas flow velocity due to the disc wind is given by
\begin{equation}
        v_{\rm dw} = -\frac{3}{2}\alpha_{\rm dw} \left(\frac{H}{r}\right)^2 v_K\;,
    \label{v_dw}
\end{equation}
and $\alpha_{\rm dw}$ is a dimensionless positive parameter, defined by analogy with the $\alpha$ disc viscosity parameter.
For more detail on disc wind implementation in our model, the reader is referred to \S3.1 in \citet{2022ElbakyanWu}. Numerical experiments in the quoted paper showed that gap opening in discs with magnetised winds may be parameterised with the following modified form of \cite{2006Crida} parameter:
\begin{equation}
    C_{\rm dw} =  \frac{3H}{4R_{\rm H}} + \frac{50 \alpha_{\rm v}}{q} \left({\frac{H}{R}}\right)^2  + \frac{70\alpha_{\rm dw}^{3/2}}{q} \left({\frac{H}{R}}\right)^3\;.
\label{Crida_mod}
\end{equation}

The last term in Eq.~\ref{Crida_mod} describes the effects of laminar disc flows on the gap opening for a given $\alpha_{\rm dw}$. As with \cite{2006Crida} parameter, deep gaps in the disc are opened when $C_{\rm dw} < 1$. The substitution of $C_{\rm obj}$ by $C_{\rm dw}$ is the only modification to our object-disc angular momentum exchange prescription.

\begin{figure}
\begin{centering}
\includegraphics[width=1\columnwidth]{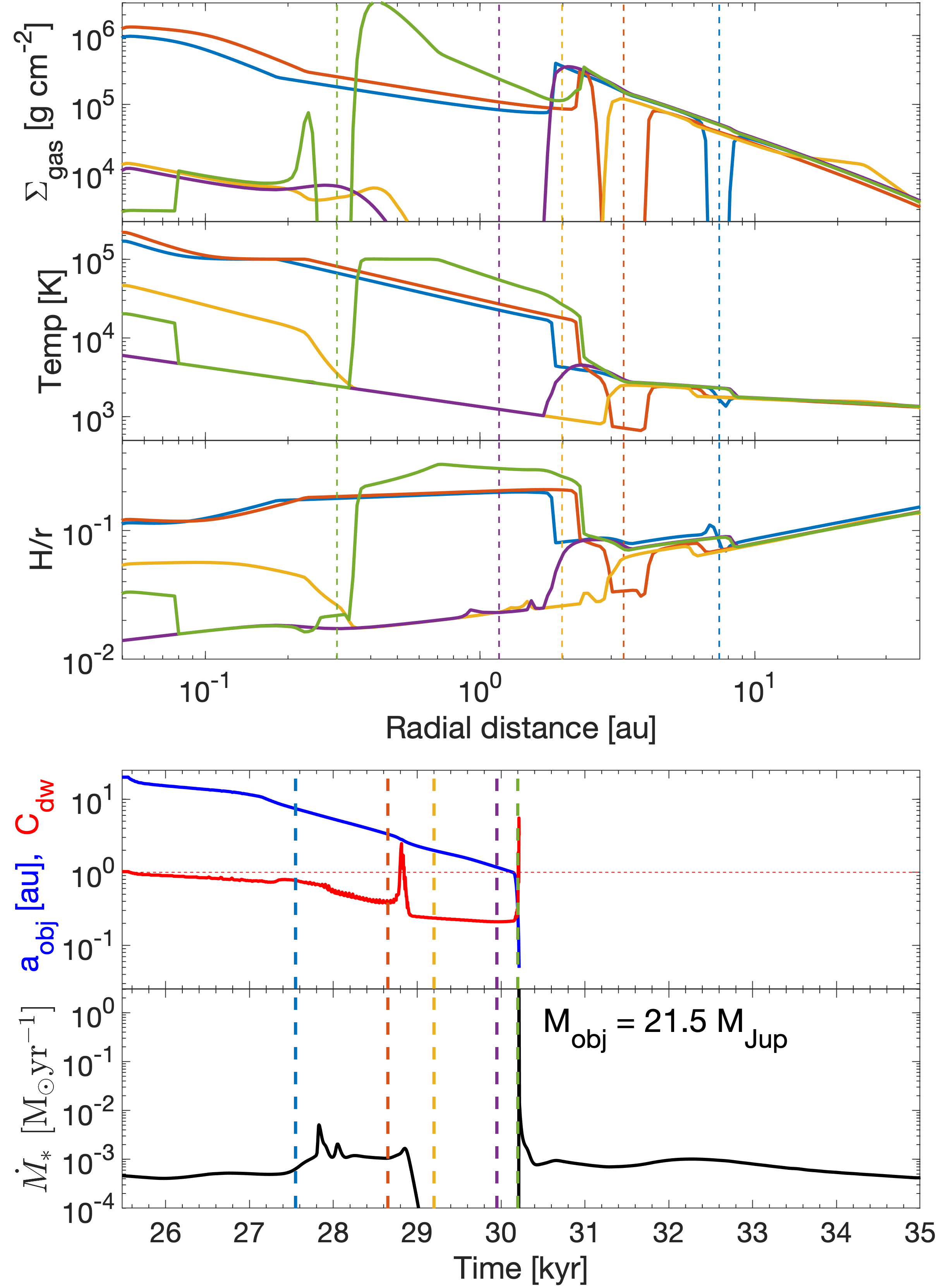}
\par\end{centering}
\caption{\label{fig:profiles_wind} Same as Fig.~\ref{fig:9} but for the model where the angular momentum in the disc is transported mainly due to the MHD disc wind with $\alpha_{\rm dw}=10^{-2}$, whereas the turbulent viscosity is set at $\alpha=10^{-4}$.
}
\end{figure}

We now experiment with a limiting case where magnetised disc winds dominate angular momentum transport, setting $\alpha_{\rm dw}=10^{-2}$, and a very low turbulent viscosity, $\alpha=10^{-4}$. Figure \ref{fig:profiles_wind} shows the results of a simulation entirely analogous to that presented in Fig.~\ref{fig:9} save for the magnetised disc wind. Since low turbulence laminar discs are much more prone to gap opening by low mass planets \citep{2022ElbakyanWu}, the object opens a gap shortly after its injection in the disc and migrates inwards in the Type II regime. The ionisation front is thus not efficient in preventing the planet from breaking the Type I migration trap at $r\approx 2$~\bref{au} \citep[which we note is closer to the star because the discs with magnetised winds are in general cooler than their turbulent counterparts][]{Suzuki16-MHD-winds}. 

The key to this outcome is the fact that the object continues to migrate in the Type II regime, even in the hot ionised region of the disc at $r\lesssim 2$~au. This is because the object is able to destroy the ionisation front by bringing it down from the hot stable branch of the S-curve \citep{1994BellLin} to the stable cold one. Consider the top three panels of Fig.~\ref{fig:profiles_wind}. Just before the object enters the hot ionised region, the disc inside the gap it opens is passive, i.e., heated mainly by the irradiation from the star rather than the local viscous dissipation. When the object enters the ionisation front, it interrupts the supply of fresh matter to it. As the inner disc drains onto the star, the region cools down and "falls" onto the cold branch of the S-curve. This is why the excursion of the modified gap opening parameter $C_{\rm dw}$ into the Type I territory when it enters the hot zone is only temporary. It then continues to migrate in the Type II regime, which is relatively slow. However, the disc material is backed up behind the object. When enough material accumulates, the region behind the object suddenly transitions back onto the stable branch. The disc $H/r$ increases, and the gap finally closes. The object then migrates in the faster Type I regime, rapidly arriving at the point where it is tidally disrupted. This sequence of events is reminiscent of the model by \cite{2004LodatoClarke} in which a massive planet modulates the inner thermally unstable disc region, except here the object is tidally disrupted.

\begin{figure}
\begin{centering}
\includegraphics[width=1\columnwidth]{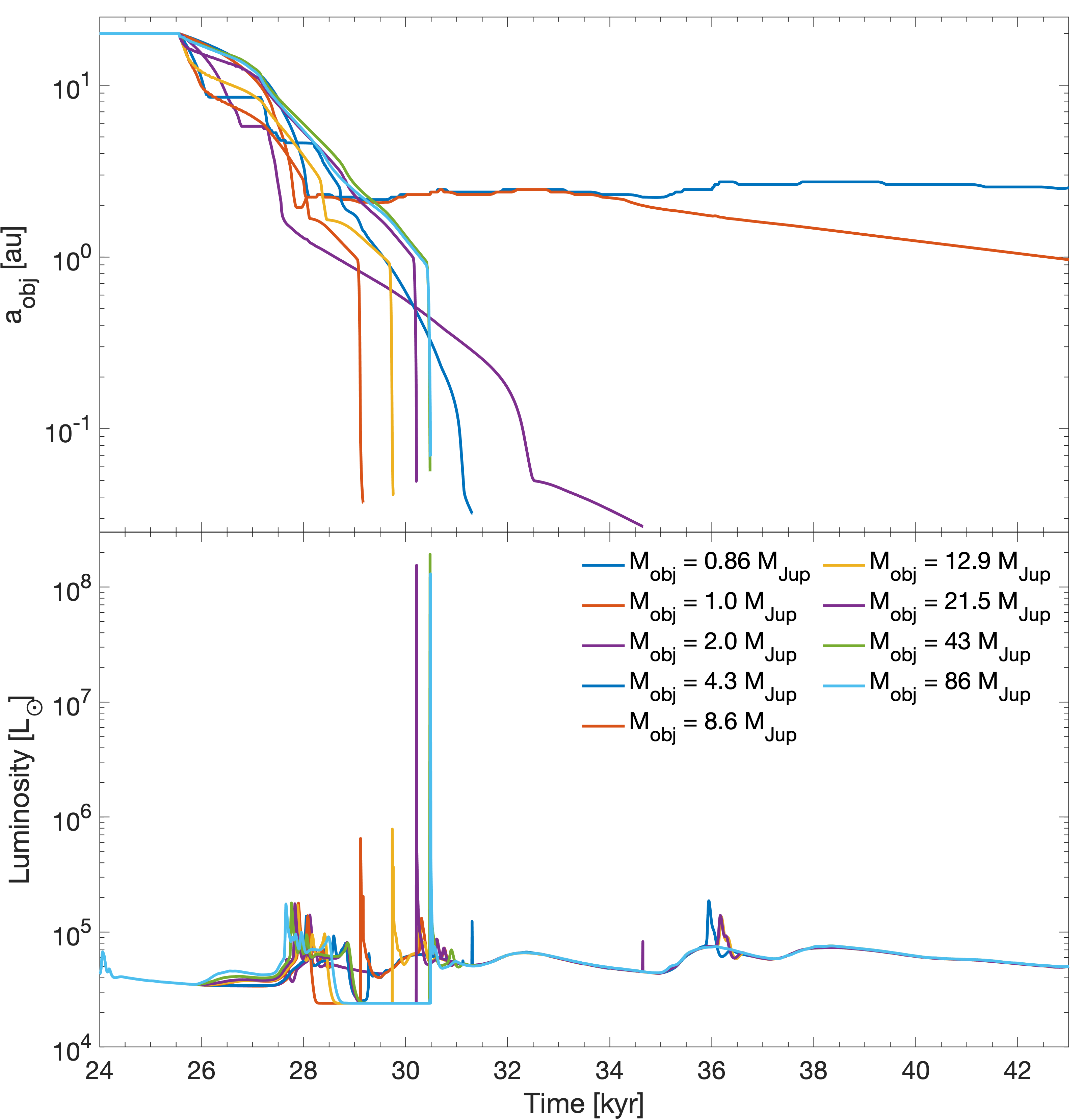}
\par\end{centering}
\caption{\label{fig:aL_MHD_winds} Same as Fig.~\ref{fig:6} but for the models where the angular momentum in the disc is transported mainly due to the MHD disc wind. Note that except for the two lowest mass objects, all of them manage to migrate inward and be tidally disrupted, producing intense accretion bursts. This is in stark contrast to the discs dominated by turbulent viscosity transport, where only the two most massive objects were disrupted.}
\end{figure}

We find that this non-linear object-disc interaction allows objects as low mass as $M_{\rm obj} = 2\mj$ to pass through the migration trap due to the ionisation front.  Figure~\ref{fig:aL_MHD_winds} shows the object separation (top panel) and the central star luminosity (bottom panel) for a wide range of object masses. We see that only objects with mass $M_{\rm obj} \leq 1\mj$ avoid tidal disruption by the virtue of their migration stalled at 1-2 \bref{au}.

\section{Some model uncertainties}\label{sec:future}
\subsection{Object internal structure} \label{sect:cooling_heating}

\begin{figure}
\begin{centering}
\includegraphics[width=1\columnwidth]{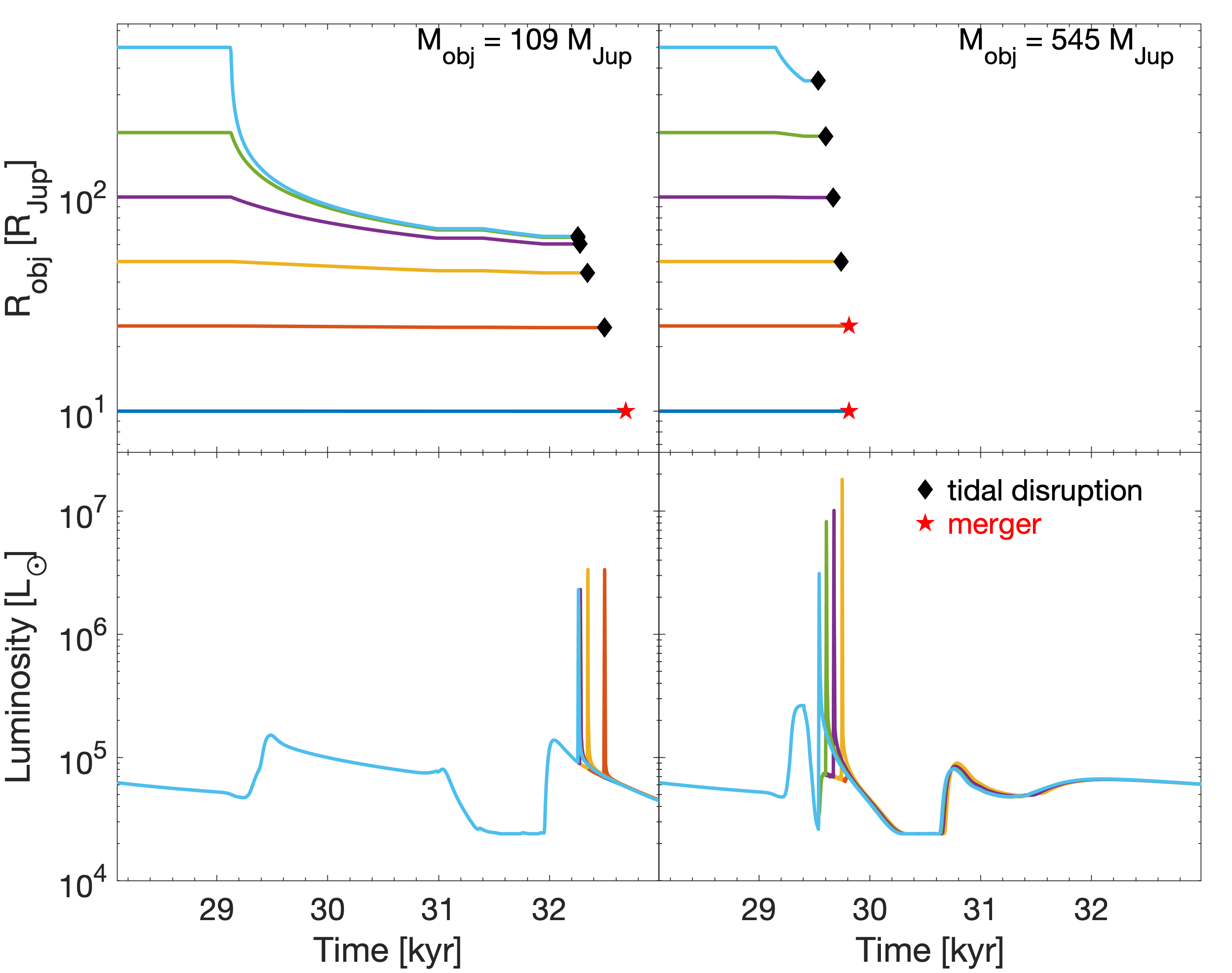}
\par\end{centering}
\caption{\label{fig:R_init} Temporal evolution of object radius ($R_{\rm obj}$) and total luminosity of central star during the inward migration of the object and its disruption/merging. The left column corresponds to the object with 109 Jovian masses, while the right column corresponds to the object with 545 Jovian masses. The objects that are merging with the central star, without being disrupted, are marked with a red star symbol on the top panels. The objects that are disrupted tidally are shown with the black diamonds.}
\end{figure}

As explained in the Introduction and in \S 2, it is numerically challenging to resolve the internal structure of the gaseous clumps in 3D simulations of circumstellar discs. So far in this paper we opted to explore a range of object masses, but for simplicity always assumed their initial radius to be $70 R_J \approx 7 \rsun$. In this section, we explore how our results depend on the initial value of $R_{\rm obj}$. In the top row of Figure~\ref{fig:R_init} the time evolution of the size of embedded objects with various initial radius, while in the bottom row the time evolution of total luminosity of the system. Each column corresponds to six objects with the same fixed mass presented in the top right corner of the top panel. The initial radius of the object varies between 10 and 500 Jovian radii.


In the top row of Fig.~\ref{fig:R_init}, the tracks of migrating objects end with a  black diamond if they are tidally disrupted, or with a red star if they merge with the parent star (that is, reach the inner radius of our computational domain). The left panels of the figure are for objects  with initial mass of $M_{\rm obj} = 109~M_{\rm Jup}$, whereas the right panels are for initial mass of 545~$M_{\rm Jup}$.

For $M_{\rm obj} = 109~M_{\rm Jup}$, only the object with initial radius of $R_{\rm init} = 10~R_{\rm Jup}$ merges with the protostar, and the rest of the objects with larger initial radii are disrupted in the disc on the radial distances from 0.08 to $0.23$~\bref{au}. The corresponding accretion bursts due to the object disruption are clearly seen in the bottom panel of Fig.~\ref{fig:R_init}. The duration of bursts for all the objects is around 10 years. Note that we do not model the outcome of the star-object merger here, so there is no corresponding accretion spike for the merger cases in Fig.~\ref{fig:R_init}.

The $M_{\rm obj} = 545~M_{\rm Jup}$ objects migrate inwards faster than their lighter cousins. Objects with $R_{\rm init} = 10 R_{\rm Jup}$ and 25~$R_{\rm Jup}$ merge with the central star, while the rest are tidally disrupted when reaching radial distances $\sim 0.1\text{--}0.5$~\bref{au}, initiating a strong accretion burst. The duration of the bursts vary from slightly less than 10 years to a few tens of years. 
Objects with smaller $R_{\rm init}$ are disrupted closer to the star, where the viscous times are shorter, and so their bursts are shorter and the maximum luminosity higher. 

These simple experiments demonstrate the importance of modelling embedded object contraction in greater detail to provide robust predictions on burst properties. We plan to undertake such study in the future.


\subsection{Central star radius uncertainties}\label{sec:stellar_rad}

\begin{figure}
\begin{centering}
\includegraphics[width=1\columnwidth]{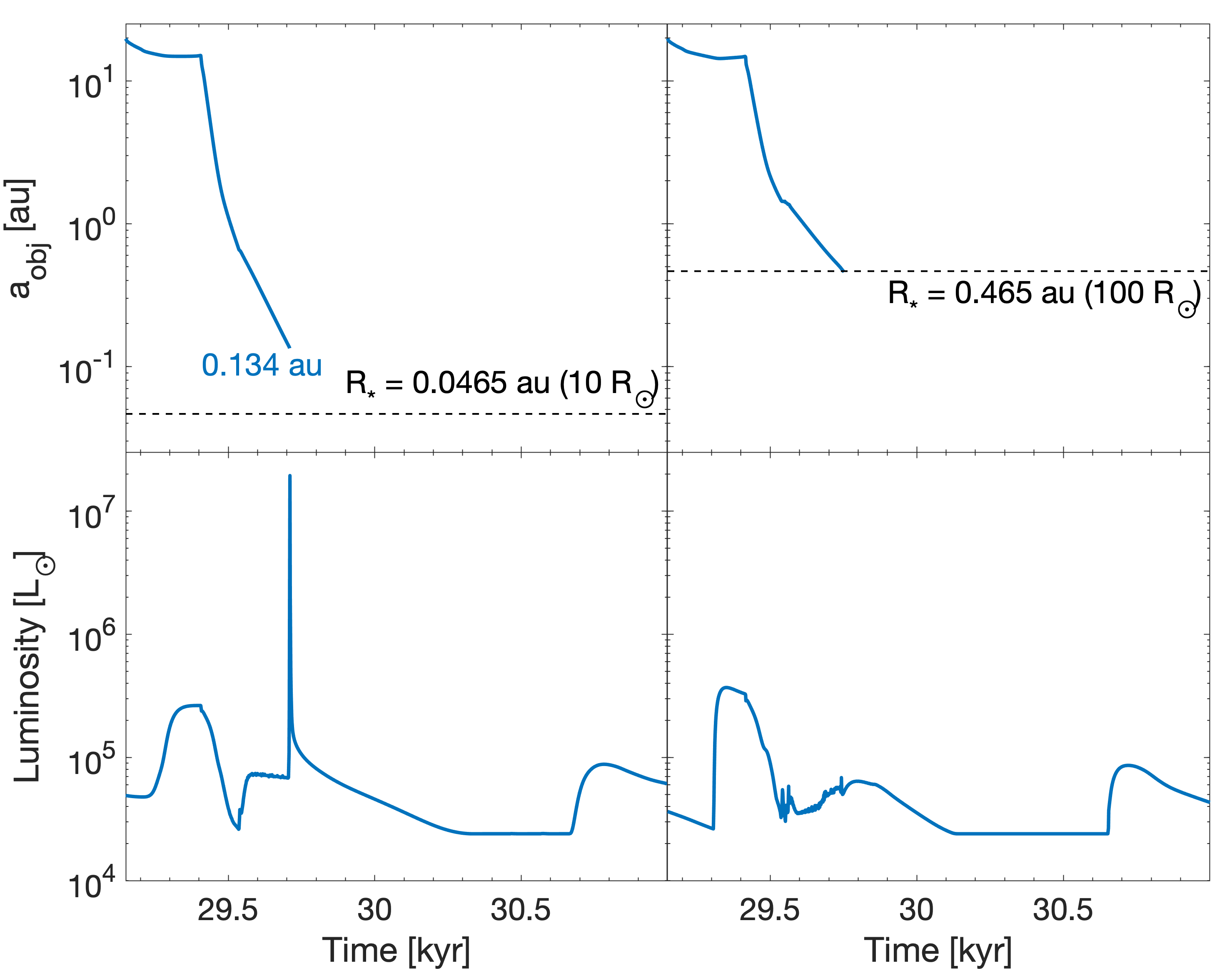}
\par\end{centering}
\caption{\label{fig:Rstar} Radial distance of the object (top row) and luminosity curve of the HMYSO (bottom row) zoomed in on the object disruption event in the model with $R_*=r_{\rm in}=10 \ R_{\odot}$ (left column) and model with $R_*=r_{\rm in}=100 \ R_{\odot}$ (right column). Horizontal dashed lines show the radial distance of the inner disc edge $r_{\rm in}$ in each model. The value presented near the blue line in the top-left panel shows the radial distance at which the object is disrupted in the model.}
\end{figure}

Episodes of very rapid gas accretion onto young stars may have lasting effects on the central stars themselves, even after their main accretion phase has ended \citep[e.g.][]{2011HosokawaOffner,2012BaraffeVorobyov,2018Jensen,2019ElbakyanVorobyov}. If accretion energy is deposited below the photosphere, then the radius of HMYSOs can bloat to as much as 100~$R_{\odot}$ \citep{2009HosokawaOmukai, 2017TanakaTan, 2019MeyerVorobyov} due to bursts of accretion.


To study how the value of $R_*$ may impact our results, here we repeat the fiducial calculation (\S \ref{sec:fiducial}) but set $R_* =100 \ R_{\odot}$ instead of $R_* = 10 \rsun$. The time evolution of radial distance of the object is shown in the top row of Fig. \ref{fig:Rstar}, while the time evolution of the total luminosity of the system in the bottom row. The horizontal dashed lines in the top row show the radial position of the disc inner edge. 
We remind the reader that in our models we assume that the inner radius of the disc, $r_{\rm in}$, is equal to the stellar radius, $R_*$, and stays constant during the entire simulation.

The top row of Figure~\ref{fig:Rstar} shows that the object separation $a_{\rm obj}$ evolves nearly identically in these two simulations, except for the inner 1 \bref{au}. This is expected because the objects in the two models has the same mass, $M_{\rm obj}=545 M_{\rm Jup}$.  However, unlike the fiducial model with $R_*=10 \ R_{\odot}$ (left panels), in the case  $R_* = 100 \rsun$ (right panels), the object is not disrupted. In our 1D model, it passes through the inner edge of the disc and merges with the star. No accretion burst due to this merger event is shown in the luminosity curves in the bottom right panel because we do not model the merger. However, usage of a stellar evolution model will most probably show an increase in stellar luminosity and an outburst event. The nature of such a burst is likely to be significantly different from the disc bursts we study here.

These two experiments indicate that solving self-consistently for the radius evolution of the central star in crucial in determining the exact outcome of the embedded object migrating towards it.




\subsection{Disc viscosity}\label{sec:param}



\begin{figure*}
\begin{centering}
\includegraphics[width=2\columnwidth]{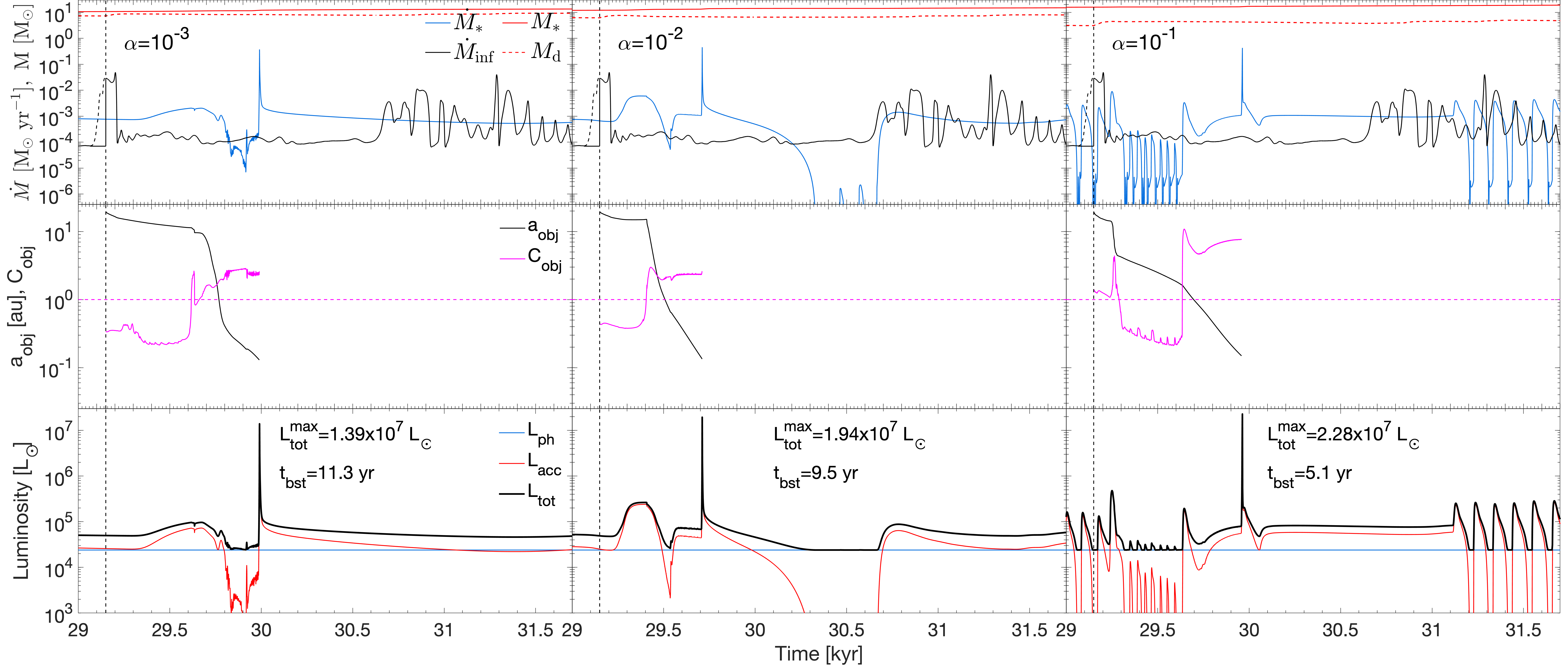}
\par\end{centering}
\caption{\label{fig:5} Comparison of the disc and central HMYSO properties in the low ($\alpha=10^{-3}$, {\bf left column}) and high ($\alpha=10^{-1}$, {\bf right column}) viscosity models with the fiducial ($\alpha=10^{-2}$) model ({\bf middle}). Notations are similar to the ones in Figure~\ref{fig:2}. The peak value of total luminosity during the burst ($L_{\rm tot}^{\rm max}$) and the burst duration ($t_{\rm bst}$) for each model are shown in the bottom panels. The mass of the migrating object is 545~$M_{\rm Jup}$.}
\end{figure*}

In our 1D disc model,  angular momentum transfer  in the disc is via turbulence, except for \S \ref{sec:low_mass_dw}, and 
is parametrized through the dimensionless $\alpha$ parameter \citep{1973ShakuraSunyaev}. Different authors argue for a wide range of values for $\alpha$ in astrophysical discs, from $10^{-4}$ to 0.1 \citep{2012MuldersDominik, 2016Pinte, 2017Lodato, 2017Rafikov, 2018AnsdellWilliams, 2018NajitaBergin, 2018Dullemond, 2020Flaherty, 2020Rosotti}. In this section, we study how the value of $\alpha$-parameter affects our results.

In the top row of Figure~\ref{fig:5} we show the accretion rate history in three different models, one of which is our fiducial model with $\alpha=10^{-2}$ discussed in \S\ref{sect:obj_migration} and two other models, which differ from the fiducial one only by the $\alpha$ value. Notations are similar to the ones in Fig.~\ref{fig:2}. As expected, due to the viscous time scaling as $\alpha^{-1}$, at any given time the disc masses are lower for the high $\alpha$ discs, while the masses of central protostars are increasing with the increasing $\alpha$ value. As an example, at the moment of object injection, the disc masses in the $\alpha=10^{-3}$, $\alpha=10^{-2}$ and $\alpha=10^{-1}$~models are 7.3, 6.2, and 3.1 $M_{\odot}$, respectively, while the protostellar masses are 10.8, 12.1, and 15.9 $M_{\odot}$. In all the three models, the object with the same mass (545~$M_{\rm Jup}$) is injected at the same time moment (vertical dashed line). In the $\alpha=10^{-3}$ and $\alpha=10^{-1}$~model the objects are disrupted about 200~years later than in our fiducial $\alpha=10^{-2}$~model. 

The middle row in Fig.~\ref{fig:5} shows orbital distance of the object ($a_{\rm obj}$) along with the gap opening parameter ($C_{\rm obj}$). In the $\alpha=10^{-3}$~model, the evolution of both the object and the accretion luminosity of the star are rather similar to the fiducial $\alpha=10^{-2}$~model. The most notable difference is the further smoothing out of the accretion rate variability $\dot M_*$ (the blue curves in the top row) compared with the mass deposition rate into the 1D disc, $\dot M_{\rm dep}$, in the $\alpha=10^{-3}$ model. This occurs due to a longer viscous time of the disc. As with the fiducial case, the only significant short scale variability event in $\dot M_*$ is due to the embedded object disruption in the inner disc.


The behaviour of the high viscosity case, $\alpha=10^{-1}$, is however qualitatively different from the low viscosity models. Due to the high $\alpha$ value, the object is not able to open a gap in the disc of $\alpha=10^{-1}$~model at the moment of its injection and starts migrating inwards in Type I regime. However, since viscosity is low, the structure of the disc evolves relatively rapidly. When the object reaches the radius of about 4 \bref{au}, the disc cools down. The object opens a gap, slowing down its migration by switching to the Type II regime. Since the gap opened by the object prevents the free interaction between the disc interior and exterior to the object, only the matter inside the inner disc is accreted onto the HMYSO during the thermal unstable periods of the disc, triggering shorter bursts with lower amplitude \citep[this is similar to TI bursts found by][in discs around Solar mass stars]{2004LodatoClarke}.
However, when the object migrates to the radial distance $r\approx$1.6~\bref{au}, it is no longer able to restrain the matter accumulated in the outer disc, and a strong TI burst is triggered. Henceforth, the disc at the location of the object heats up, closing the gap. The object switches to the Type I regime again, starts to migrate inwards faster and gets tidally disrupted at the radial distance $r=0.146$~\bref{au}.

Note that in the high $\alpha$ case there is accretion variability in the disc long after the object is disrupted. This is due to TI of the inner disc. These are relatively long duration (tens of years) periodic bursts, quite distinct in their characteristics from those of object disruption bursts \citep{2021ElbakyanNayakshin}.

The luminosity bursts due to the object disruption is shown in the bottom row of Fig.~\ref{fig:5}. By analogy with the $\alpha=10^{-2}$~model, we use the pre-burst bolometric luminosity of S255IR-NIRS3 HMSYO as a photospheric luminosity of central HMYSO for the other two models. The peak total luminosity during the burst in the $\alpha=10^{-3}$ and the $\alpha=10^{-1}$~model are, respectively, $L_{\rm tot}^{\rm max}=1.39\times10^7 \ L_{\odot}$ and $L_{\rm tot}^{\rm max}=2.28\times10^7 \ L_{\odot}$.
The burst duration, $t_{\rm bst}$, is, respectively, equal to 11.3 and 5.1 years.
The viscous timescale is longer in the lower $\alpha$ models, thus making the burst duration in these models longer. The peak luminosity during the burst is higher for the higher $\alpha$ models, due to the higher stellar masses at the moment of the burst.

\subsection{Object migration model}\label{sec:migr_model}

The prescription of the total torque acting on the object plays an important role in its migration. If the total torque is calculated with the non-isothermal disc assumption following \citet{2011Paardekooper}, then the objects with masses <40~$M_{\rm Jup}$ can stop their inward migration at the radial distance of a few au from the star. However, \citet{2011Paardekooper} Type I migration formulae were derived for a fixed $\mu$, the mean molecular weight for a gas, and an entropy function for an ideal gas with a fixed number of degrees of freedom. This is a good approximation in a broad range of conditions, e.g., for regions dominated by completely ionised Hydrogen, or for discs composed of molecular Hydrogen.
Due to significant accretion rate variability, our discs do not entirely belong to just one of these regimes (cf. Fig.~\ref{fig:Disc_structure}). In the innermost regions, midplane disc temperatures may exceed $10^6$~K but at outer regions this falls to $\sim 10^3$ K.  More work is needed (C. Baruteau, private communication) to understand if \citet{2011Paardekooper} results apply in the vicinity of regions where discs undergo an abrupt phase transition from mainly molecular to mainly ionised Hydrogen, e.g., at $r\approx 4$~\bref{au} for the blue curve in the second from the top panel of Fig.~\ref{fig:Disc_structure}.


\bref{
\subsection{The effect of stellar motion}
Massive non-axisymmetric structures in a gravitationally unstable disc, such as clumps or spirals, can move the star and the barycentre (the centre of mass) of the star-disc system from the coordinate system origin. It has been shown by \citet{2010MichaelDurisen} that stellar motion due to non-axisymmetry developed in a gravitationally unstable disc affects the overall disc evolution only slightly. They concluded that GI in the disc is weakened by about at most tens of per cent if the stellar motion is taken into account. However, \citet{2017RegalyVorobyova} studied the effect of stellar motion, including the effect of the infalling envelope. They found that the stellar motion has a notable effect on the evolution of GI discs, reducing their mass and total angular momentum. Particularly, the radial migration of giant planets (in the range of $1.0\text{--}2.5~\mj$) is significantly slower if the effect of stellar motion is taken into account. \citet{2019MeyerVorobyovElbakyan} found that inclusion of stellar motion results in a minor acceleration of disc fragmentation and yields somewhat more frequent accretion bursts.  On the other hand, a more recent high-resolution study by \citet{2022Meyer} finds that the stellar motion slows down the development of GI in the disc, reducing the number and magnitude of the accretion-driven bursts on HMYSOs. 
In the present paper, we used 3D disc simulations that did not include stellar motion.
While we do not expect our main conclusions to be affected by this simplification, this is an important caveat warrents further study. 
}

\bref{
\subsection{The effect of eccentric orbits of the embedded object}
\label{sect:ecc_orbit}
The object is inserted into the 1D disc, assuming a Keplerian-like speed and no eccentricity. However, as it is found in multidimensional numerical simulations \citep[e.g.,][]{2020OlivaKuiper, 2021VorobyovElbakyanLiu} due to the interaction with each other and spiral structures inside the GI disc, the clumps can be ejected from the disc, sling-thrown towards the central star and tend to have chaotic orbits with relatively high eccentricities. If no embedded second core object is formed inside the high-eccentricity clump when it approaches the central star, then the clump will experience shearing and be distrupted \citep{2020OlivaKuiper}, before reaching the radial distance $r_{\rm dis}$ at which the disruption will produce an observable accretion burst. On the other hand, if an embedded second core object is present inside the clump with high eccentricity, it may reach the radial distance $r_{\rm dis}$, get disrupted and produce an observable accretion burst. However, these disruptions occur at a fraction of an au, i.e., almost two orders of magnitude closer to that star than the 1D-3D disc boundary of our simulations. To get that close to the star in one single passage, the object must possess an extreme eccentricity. It is much more likely that eccentricity of the object would be damped rapidly in the inner dense disc, much more rapidly than the object migrates in \citep[e.g,][]{1980GoldreichTremaine}. Having said that, we of course acknowledge that if there are multiple objects in the inner disc then their eccentricity may be maintained by their interactions. More sophisticated multidimensional and multi-clump numerical simulations are necessary to follow these situations.
}

\bref{
\subsection{Impact on and connection with previous studies}
Here, we showed that short and powerful accretion bursts result from object disruption only when the object is in a very dense (post-collapse) configuration. Deposition of diffuse gas or pre-collapse low density clumps into the inner sink cell do not lead to powerful accretion flares because the inner 1D disc smooths this variability out very efficiently.
}

\bref{
What are the implication of our results on accretion burst statistics from published 3D simulations? On the one hand, we expect a reduction in the number of such events, since not all of the accretion spikes seen in 3D simulations contain post-collapse objects. 
We note that it is shown by \citet{2018MeyerKuiper} that central temperature at the centre of a clump inside GI disc of HMYSO can reach the dissociation temperature for the hydrogen 2000K. Thus, an embedded second core object could be formed inside the clump. However, their study did not focus on the statistics of such events. In a more recent study, \citet{2020OlivaKuiper} showed that 10 out of 19 surviving clumps form a second core object inside them. This gives us a statistics that about half of the clumps inside a GI disc of HMYSO will end up forming second core objects. Such objects may result in intense bursts of accretion.}

\bref{On the other hand, in general, we expect the number of the clumps to increase with ever-increasing resolution of the simulations. For example, \cite{2020OlivaKuiper} demonstrate that the number of second cores increases with increasing spacial resolution. It is therefore unclear at the present whether future simulations -- that would converge in terms of the number and inner structure of self-bound fragments and at the same time model the inner au discs in sufficient detail -- would result in fewer or more numerous accretion bursts.}


\bref{
There are other significant challenges ahead. Due to current numerical limitations, the only currently feasible approach to modelling post-collapse objects is the sink particle method \citep[e.g.,][]{1995Bate}. However, it is known that the sink particle prescription tends to over-simplify the process of gas accretion onto the objects, usually resulting in an over-estimate of the accretion rate. For example, \cite{2017Nayakshin} investigated gas accretion onto first cores -- objects that are much easier to resolve -- without the use of the sink particle technique in protoplanetary discs in 3D simulations. It was found that gas accretion efficiency on the first cores depends strongly on radiative cooling efficiency, and hence on dust opacity in the disc, which is currently poorly constrained. The uncertainties in gas accretion due to radiative cooling are also well known in 3D simulations of gas accretion onto solid cores in the context of the Core Accretion theory for planet formation \citep[e.g.,][]{2015OrmelShiKuiper, 2015OrmelKuiperShi}. \cite{2017Nayakshin} also compared gas accretion rates onto the first cores with that obtained with the sink particle prescription with different sink particle radii, and found that the larger the latter, the more gas accretion onto the object is over-estimated (see his Appendix). Additionally, the inner structure of the objects is not tackled by most authors in the sink particle approach.  As we have seen in \S \ref{sect:cooling_heating} and \ref{sec:stellar_rad}, until such modelling is done we will not be entirely clear on whether formation of post-collapse objects in 3D simulations leads to accretion bursts or star-object mergers.}

\bref{Furthermore, neglect of gas accretion onto embedded objects in the current study may cause not just quantitative, but also qualitative changes in the final configuration of the system. If gas accretion onto a migrating object is especially efficient, then in fact the object may accrete more gas than the primary star, which could result in formation of a roughly equal mass binary system. Indeed, simulations often show that secondary stars in a binary system accrete gas more rapidly than the primary due to their lower velocity relative to the disc material, and due to orbiting closer to the disc inner edge \citep[e.g,][]{1995Bate, 2022CeppiCuelloLodato}.}


\section{Discussion and conclusions}\label{sec:discuss}

In this paper, we studied episodic accretion onto HMYSOs using a combination of 3D and 1D approaches. We used the results of 3D simulations of protostellar discs around HMYSOs  by \citet{2019MeyerVorobyovElbakyan} to simulate evolution of the disc in the inner 20~\bref{au}, unresolved in the 3D simulations, with a 1D code. We injected a dense embedded object into the inner disc during two representative elevated accretion rate episodes of 3D disc. In our 1D method, the object interacts with the disc via gravity (tidal torques), so it can migrate self-consistently, and can be tidally disrupted, depositing its mass worth of material back into the disc. 
Our main conclusions are the following:

\begin{enumerate}

    \item In 3D simulations, an accretion burst is \bref{usually modelled} as an instance of a \bref{significantly} elevated mass flow rate through the inner boundary. \bref{Our main result is the need to pay closer attention to the physical state of the matter crossing the inner boundary of the computation domain of 3D simulations. In \S \ref{sec:pre-collapse} we have shown that matter entering the inner tens of au in the form of pre-collapse clumps \citep[first cores defined by][]{1969Larson} cannot give short duration high luminosity accretion bursts. Such objects are tidally disrupted $\sim 10$ au from the star. Similarly, diffuse matter deposition into the inner disc through, e.g., a spiral arm inflow, even at high rates, is unlikely to result in observable flares due to viscous reprocessing in the inner disc.} This results in outbursts smoothed over timescales of $\sim 10^3$ years. Such outbursts are too weak and not observationally classifiable as outbursts on human time scales. \bref{Only post-collapse (second Larson cores) are able to approach the central star to distances of a fraction of an au. If these objects are disrupted as they migrate towards the star, powerful accretion flares may result. }
    
    \item The physics of the inner accretion disc has a strong impact on the character of the outbursts, and more broadly on the outcome of object migration towards the host protostar. For example, we found that thermal instability (TI) of the inner disc can produce accretion variability even if no embedded objects are present in the disc (or it is long destroyed, as in the right panels of Fig.~\ref{fig:5}). These bursts tend to be of lower amplitude and longer duration, however. 
    \item In the standard viscous/turbulent discs \citep{1973ShakuraSunyaev,1981Pringle}, only the more massive objects are able to reach the inner few au. For the system parameters studied here, objects less massive than $\sim 40\mj$ are held at a migration trap formed by the temperature jump at the hydrogen ionisation front. In this case, only outbursts with mass budget exceeding tens of $\mj$ could be explained by GI fragmentation of the outer massive disc. However, the robustness of this conclusion requires a broader parameter range study, and a further investigation of migration physics in the so far unexplored regime of very hot protoplanetary discs (\S \ref{sec:migr_model}).
    \item If the discs are powered by magnetised disc winds, with only a minor contribution from MRI angular momentum transfer, then the discs are cooler, and so the objects are more likely to migrate in via Type II, opening a deep gap. The hydrogen ionisation front in this case is closer to the star and can actually be erased temporarily by the object migrating through it. In such discs, objects as low mass as $\sim 2\mj$ are able to reach inside the inner au of the disc. Tidal disruption of such objects produce short accretion bursts with luminosities comparable to the bursts observed in HMYSOs.
    \item Disruption of high mass objects ($M_{\rm obj} \gtrsim $ tens of $\mj$ or more) may result in super-Eddington accretion episodes onto the star. We cannot model such events reliably in our 1D disc model, however one can expect powerful jets and outflows launched from the inner regions of HMYSO discs during these. Both the high luminosity output and the outflows are likely to be very important for feedback produced by massive protostars on their gaseous surroundings while they grow.
    \item Mergers of the migrating objects with the central protostar is another possible outcome. This can occur if the object reaches the innermost fraction of an au while the protostar is bloated by a previous episode of high accretion rate, and/or if the object is particularly dense. Dedicated multidimensional simulations that would model the internal structure of the embedded object better are needed to address the outcome of such mergers, but they are likely to result in outbursts significantly different from the disc-powered outbursts in terms of their luminosity, spectra, and the outflows launched.
    \item The phenomenon of episodic accretion driven by massive object migration, disruption in the disc and/or merger with the central star is uniquely important to how massive protostars grow and produce feedback on their surroundings. \bref{In this paper we have shown that neither previous 3D studies nor isolated 1D disc simulations can predict reliably the properties of individual accretion bursts, their statistics, and trends with numerous parameters of the problem. For 3D simulations, the key uncertainties are the survivability and internal structure of the fragments. For the 1D disc models, the challenges are the disc physics, the object migration model, and the internal structure of the object.}   Future simulation efforts should aim to model the entire relevant domain, from the stellar surface, $r \lesssim 0.1$~\bref{au}, to the region where gravitational instability of the disc gives birth to massive gas clumps, $r\gtrsim 10^2-10^3$~\bref{au}. \bref{This is an extremely challenging goal.}
\end{enumerate}

\section*{Acknowledgements}

\bref{We thank the anonymous referee for an insightful report, which helped to improve this paper.} The authors acknowledge the funding from the UK Science and Technologies Facilities Council, grant No. ST/S000453/1. 
E.I.V. and V.E. acknowledge support of Ministry of Science and
Higher Education of the Russian Federation under the grant 075-15-2020-780 (N13.1902.21.0039).
This research used the ALICE High Performance Computing Facility at the University of Leicester, and DiRAC Data Intensive service at Leicester, operated by the University of Leicester IT Services, which forms part of the STFC DiRAC HPC Facility (www.dirac.ac.uk). 

\section*{Data Availability}

The data obtained in our simulations can be made available on reasonable request to the corresponding author.



\bibliographystyle{mnras}
\bibliography{ref_base} 

\begin{thebibliography}{}
\makeatletter
\relax
\def\mn@urlcharsother{\let\do\@makeother \do\$\do\&\do\#\do\^\do\_\do\%\do\~}
\def\mn@doi{\begingroup\mn@urlcharsother \@ifnextchar [ {\mn@doi@}
  {\mn@doi@[]}}
\def\mn@doi@[#1]#2{\def\@tempa{#1}\ifx\@tempa\@empty \href
  {http://dx.doi.org/#2} {doi:#2}\else \href {http://dx.doi.org/#2} {#1}\fi
  \endgroup}
\def\mn@eprint#1#2{\mn@eprint@#1:#2::\@nil}
\def\mn@eprint@arXiv#1{\href {http://arxiv.org/abs/#1} {{\tt arXiv:#1}}}
\def\mn@eprint@dblp#1{\href {http://dblp.uni-trier.de/rec/bibtex/#1.xml}
  {dblp:#1}}
\def\mn@eprint@#1:#2:#3:#4\@nil{\def\@tempa {#1}\def\@tempb {#2}\def\@tempc
  {#3}\ifx \@tempc \@empty \let \@tempc \@tempb \let \@tempb \@tempa \fi \ifx
  \@tempb \@empty \def\@tempb {arXiv}\fi \@ifundefined
  {mn@eprint@\@tempb}{\@tempb:\@tempc}{\expandafter \expandafter \csname
  mn@eprint@\@tempb\endcsname \expandafter{\@tempc}}}

\bibitem[\protect\citeauthoryear{{Ahmadi}, {Kuiper}  \& {Beuther}}{{Ahmadi}
  et~al.}{2019}]{2019Ahmadi}
{Ahmadi} A.,  {Kuiper} R.,   {Beuther} H.,  2019, \mn@doi [\aap]
  {10.1051/0004-6361/201935783}, \href
  {https://ui.adsabs.harvard.edu/abs/2019A&A...632A..50A} {632, A50}

\bibitem[\protect\citeauthoryear{{Alexander}, {Clarke}  \&
  {Pringle}}{{Alexander} et~al.}{2006}]{2006AlexanderClarke}
{Alexander} R.~D.,  {Clarke} C.~J.,   {Pringle} J.~E.,  2006, \mn@doi [\mnras]
  {10.1111/j.1365-2966.2006.10294.x}, \href
  {https://ui.adsabs.harvard.edu/abs/2006MNRAS.369..229A} {369, 229}

\bibitem[\protect\citeauthoryear{{Ansdell} et~al.,}{{Ansdell}
  et~al.}{2018}]{2018AnsdellWilliams}
{Ansdell} M.,  et~al., 2018, \mn@doi [\apj] {10.3847/1538-4357/aab890}, \href
  {https://ui.adsabs.harvard.edu/abs/2018ApJ...859...21A} {859, 21}

\bibitem[\protect\citeauthoryear{{Armitage}}{{Armitage}}{2015}]{2015Armitage}
{Armitage} P.~J.,  2015, arXiv e-prints, \href
  {https://ui.adsabs.harvard.edu/abs/2015arXiv150906382A} {p. arXiv:1509.06382}

\bibitem[\protect\citeauthoryear{{Armitage} \& {Bonnell}}{{Armitage} \&
  {Bonnell}}{2002}]{2002ArmitageBonnell}
{Armitage} P.~J.,  {Bonnell} I.~A.,  2002, \mn@doi [\mnras]
  {10.1046/j.1365-8711.2002.05213.x}, \href
  {https://ui.adsabs.harvard.edu/abs/2002MNRAS.330L..11A} {330, L11}

\bibitem[\protect\citeauthoryear{{Armitage}, {Livio}  \& {Pringle}}{{Armitage}
  et~al.}{2001}]{2001Armitage}
{Armitage} P.~J.,  {Livio} M.,   {Pringle} J.~E.,  2001, \mn@doi [\mnras]
  {10.1046/j.1365-8711.2001.04356.x}, \href
  {https://ui.adsabs.harvard.edu/abs/2001MNRAS.324..705A} {324, 705}

\bibitem[\protect\citeauthoryear{{Armitage}, {Simon}  \& {Martin}}{{Armitage}
  et~al.}{2013}]{Armitage13-MHD-wind}
{Armitage} P.~J.,  {Simon} J.~B.,   {Martin} R.~G.,  2013, \mn@doi [\apjl]
  {10.1088/2041-8205/778/1/L14}, \href
  {https://ui.adsabs.harvard.edu/abs/2013ApJ...778L..14A} {778, L14}

\bibitem[\protect\citeauthoryear{{Audard} et~al.,}{{Audard}
  et~al.}{2014}]{2014AudardAbraham}
{Audard} M.,  et~al., 2014, in {Beuther} H.,  {Klessen} R.~S.,  {Dullemond}
  C.~P.,   {Henning} T.,  eds, Protostars and Planets VI. p.~387 (\mn@eprint
  {arXiv} {1401.3368}), \mn@doi{10.2458/azu_uapress_9780816531240-ch017}

\bibitem[\protect\citeauthoryear{{Bae}, {Hartmann}, {Zhu}  \& {Nelson}}{{Bae}
  et~al.}{2014}]{2014BaeHartmann}
{Bae} J.,  {Hartmann} L.,  {Zhu} Z.,   {Nelson} R.~P.,  2014, \mn@doi [\apj]
  {10.1088/0004-637X/795/1/61}, \href
  {http://adsabs.harvard.edu/abs/2014ApJ...795...61B} {795, 61}

\bibitem[\protect\citeauthoryear{{Bai} \& {Stone}}{{Bai} \&
  {Stone}}{2013}]{2013BaiStone}
{Bai} X.-N.,  {Stone} J.~M.,  2013, \mn@doi [\apj]
  {10.1088/0004-637X/769/1/76}, \href
  {https://ui.adsabs.harvard.edu/abs/2013ApJ...769...76B} {769, 76}

\bibitem[\protect\citeauthoryear{Balay, Gropp, McInnes  \& Smith}{Balay
  et~al.}{1997}]{petsc-efficient}
Balay S.,  Gropp W.~D.,  McInnes L.~C.,   Smith B.~F.,  1997, in Arge E.,
  Bruaset A.~M.,   Langtangen H.~P.,  eds, Modern Software Tools in Scientific
  Computing. Birkh{\"{a}}user Press, pp 163--202

\bibitem[\protect\citeauthoryear{{Baraffe}, {Vorobyov}  \&
  {Chabrier}}{{Baraffe} et~al.}{2012}]{2012BaraffeVorobyov}
{Baraffe} I.,  {Vorobyov} E.,   {Chabrier} G.,  2012, \mn@doi [ApJ]
  {10.1088/0004-637X/756/2/118}, \href
  {http://adsabs.harvard.edu/abs/2012ApJ...756..118B} {756, 118}

\bibitem[\protect\citeauthoryear{{Baruteau} \& {Masset}}{{Baruteau} \&
  {Masset}}{2008}]{2008BaruteauMasset}
{Baruteau} C.,  {Masset} F.,  2008, \mn@doi [\apj] {10.1086/529487}, \href
  {https://ui.adsabs.harvard.edu/abs/2008ApJ...678..483B} {678, 483}

\bibitem[\protect\citeauthoryear{{Baruteau}, {Meru}  \&
  {Paardekooper}}{{Baruteau} et~al.}{2011}]{2011BaruteauMeru}
{Baruteau} C.,  {Meru} F.,   {Paardekooper} S.-J.,  2011, \mn@doi [\mnras]
  {10.1111/j.1365-2966.2011.19172.x}, \href
  {http://adsabs.harvard.edu/abs/2011MNRAS.416.1971B} {416, 1971}

\bibitem[\protect\citeauthoryear{{Baruteau} et~al.,}{{Baruteau}
  et~al.}{2014}]{2014Baruteau-PPVI}
{Baruteau} C.,  et~al., 2014, in {Beuther} H.,  {Klessen} R.~S.,  {Dullemond}
  C.~P.,   {Henning} T.,  eds, Protostars and Planets VI. p.~667 (\mn@eprint
  {arXiv} {1312.4293}), \mn@doi{10.2458/azu\_uapress\_9780816531240-ch029}

\bibitem[\protect\citeauthoryear{{Bate}, {Bonnell}  \& {Price}}{{Bate}
  et~al.}{1995}]{1995Bate}
{Bate} M.~R.,  {Bonnell} I.~A.,   {Price} N.~M.,  1995, \mn@doi [\mnras]
  {10.1093/mnras/277.2.362}, \href
  {https://ui.adsabs.harvard.edu/abs/1995MNRAS.277..362B} {277, 362}

\bibitem[\protect\citeauthoryear{{Bell} \& {Lin}}{{Bell} \&
  {Lin}}{1994}]{1994BellLin}
{Bell} K.~R.,  {Lin} D.~N.~C.,  1994, \mn@doi [ApJ] {10.1086/174206}, \href
  {http://adsabs.harvard.edu/abs/1994ApJ...427..987B} {427, 987}

\bibitem[\protect\citeauthoryear{{Bhandare}, {Kuiper}, {Henning}, {Fendt},
  {Marleau}  \& {K{\"o}lligan}}{{Bhandare} et~al.}{2018}]{2018Bhandare}
{Bhandare} A.,  {Kuiper} R.,  {Henning} T.,  {Fendt} C.,  {Marleau} G.-D.,
  {K{\"o}lligan} A.,  2018, \mn@doi [\aap] {10.1051/0004-6361/201832635}, \href
  {https://ui.adsabs.harvard.edu/abs/2018A&A...618A..95B} {618, A95}

\bibitem[\protect\citeauthoryear{{Bhandare}, {Kuiper}, {Henning}, {Fendt},
  {Flock}  \& {Marleau}}{{Bhandare} et~al.}{2020}]{2020Bhandare}
{Bhandare} A.,  {Kuiper} R.,  {Henning} T.,  {Fendt} C.,  {Flock} M.,
  {Marleau} G.-D.,  2020, \mn@doi [\aap] {10.1051/0004-6361/201937029}, \href
  {https://ui.adsabs.harvard.edu/abs/2020A&A...638A..86B} {638, A86}

\bibitem[\protect\citeauthoryear{{Boley}, {Hayfield}, {Mayer}  \&
  {Durisen}}{{Boley} et~al.}{2010}]{2010BoleyEtal}
{Boley} A.~C.,  {Hayfield} T.,  {Mayer} L.,   {Durisen} R.~H.,  2010, \mn@doi
  [\icarus] {10.1016/j.icarus.2010.01.015}, \href
  {https://ui.adsabs.harvard.edu/abs/2010Icar..207..509B} {207, 509}

\bibitem[\protect\citeauthoryear{{Boss}}{{Boss}}{2017}]{2017Boss}
{Boss} A.~P.,  2017, \mn@doi [\apj] {10.3847/1538-4357/836/1/53}, \href
  {https://ui.adsabs.harvard.edu/abs/2017ApJ...836...53B} {836, 53}

\bibitem[\protect\citeauthoryear{{Boss} \& {Black}}{{Boss} \&
  {Black}}{1982}]{1982BossBlack}
{Boss} A.~P.,  {Black} D.~C.,  1982, \mn@doi [\apj] {10.1086/160077}, \href
  {https://ui.adsabs.harvard.edu/abs/1982ApJ...258..270B} {258, 270}

\bibitem[\protect\citeauthoryear{{Brogan} et~al.,}{{Brogan}
  et~al.}{2019}]{2019Brogan}
{Brogan} C.~L.,  et~al., 2019, \mn@doi [\apjl] {10.3847/2041-8213/ab2f8a},
  \href {https://ui.adsabs.harvard.edu/abs/2019ApJ...881L..39B} {881, L39}

\bibitem[\protect\citeauthoryear{{Caratti o Garatti} et~al.,}{{Caratti o
  Garatti} et~al.}{2017}]{2017Caratti}
{Caratti o Garatti} A.,  et~al., 2017, \mn@doi [Nature Physics]
  {10.1038/nphys3942}, \href
  {https://ui.adsabs.harvard.edu/abs/2017NatPh..13..276C} {13, 276}

\bibitem[\protect\citeauthoryear{{Ceppi}, {Cuello}, {Lodato}, {Clarke}, {Toci}
  \& {Price}}{{Ceppi} et~al.}{2022}]{2022CeppiCuelloLodato}
{Ceppi} S.,  {Cuello} N.,  {Lodato} G.,  {Clarke} C.,  {Toci} C.,   {Price}
  D.~J.,  2022, \mn@doi [\mnras] {10.1093/mnras/stac1390}, \href
  {https://ui.adsabs.harvard.edu/abs/2022MNRAS.514..906C} {514, 906}

\bibitem[\protect\citeauthoryear{{Cha} \& {Nayakshin}}{{Cha} \&
  {Nayakshin}}{2011}]{2011ChaNayakshin}
{Cha} S.-H.,  {Nayakshin} S.,  2011, \mn@doi [\mnras]
  {10.1111/j.1365-2966.2011.18953.x}, \href
  {http://adsabs.harvard.edu/abs/2011MNRAS.415.3319C} {415, 3319}

\bibitem[\protect\citeauthoryear{{Chen}, {Sun}, {Chini}, {Haas}, {Jiang}  \&
  {Chen}}{{Chen} et~al.}{2021}]{2021Chen}
{Chen} Z.,  {Sun} W.,  {Chini} R.,  {Haas} M.,  {Jiang} Z.,   {Chen} X.,  2021,
  \mn@doi [\apj] {10.3847/1538-4357/ac2151}, \href
  {https://ui.adsabs.harvard.edu/abs/2021ApJ...922...90C} {922, 90}

\bibitem[\protect\citeauthoryear{{Chon} \& {Hosokawa}}{{Chon} \&
  {Hosokawa}}{2019}]{2019ChonHosokawa}
{Chon} S.,  {Hosokawa} T.,  2019, \mn@doi [\mnras] {10.1093/mnras/stz1824},
  \href {https://ui.adsabs.harvard.edu/abs/2019MNRAS.488.2658C} {488, 2658}

\bibitem[\protect\citeauthoryear{{Clarke}, {Lin}  \& {Papaloizou}}{{Clarke}
  et~al.}{1989}]{Clarke89-TI-FUOR}
{Clarke} C.~J.,  {Lin} D.~N.~C.,   {Papaloizou} J.~C.~B.,  1989, \mn@doi
  [\mnras] {10.1093/mnras/236.3.495}, \href
  {https://ui.adsabs.harvard.edu/abs/1989MNRAS.236..495C} {236, 495}

\bibitem[\protect\citeauthoryear{{Crida}, {Morbidelli}  \& {Masset}}{{Crida}
  et~al.}{2006}]{2006Crida}
{Crida} A.,  {Morbidelli} A.,   {Masset} F.,  2006, \mn@doi [\icarus]
  {10.1016/j.icarus.2005.10.007}, \href
  {https://ui.adsabs.harvard.edu/abs/2006Icar..181..587C} {181, 587}

\bibitem[\protect\citeauthoryear{{Doi} \& {Kataoka}}{{Doi} \&
  {Kataoka}}{2021}]{2021DoiKataoka}
{Doi} K.,  {Kataoka} A.,  2021, \mn@doi [\apj] {10.3847/1538-4357/abe5a6},
  \href {https://ui.adsabs.harvard.edu/abs/2021ApJ...912..164D} {912, 164}

\bibitem[\protect\citeauthoryear{{Dullemond} et~al.,}{{Dullemond}
  et~al.}{2018a}]{DSHARP-6}
{Dullemond} C.~P.,  et~al., 2018a, \mn@doi [\apjl] {10.3847/2041-8213/aaf742},
  \href {http://adsabs.harvard.edu/abs/2018ApJ...869L..46D} {869, L46}

\bibitem[\protect\citeauthoryear{{Dullemond} et~al.,}{{Dullemond}
  et~al.}{2018b}]{2018Dullemond}
{Dullemond} C.~P.,  et~al., 2018b, \mn@doi [\apj] {10.3847/2041-8213/aaf742},
  \href {https://ui.adsabs.harvard.edu/abs/2018ApJ...869L..46D} {869, L46}

\bibitem[\protect\citeauthoryear{{Dunham} \& {Vorobyov}}{{Dunham} \&
  {Vorobyov}}{2012}]{2012DunhamVorobyov}
{Dunham} M.~M.,  {Vorobyov} E.~I.,  2012, \mn@doi [ApJ]
  {10.1088/0004-637X/747/1/52}, \href
  {http://adsabs.harvard.edu/abs/2012ApJ...747...52D} {747, 52}

\bibitem[\protect\citeauthoryear{{Elbakyan}, {Vorobyov}  \&
  {Glebova}}{{Elbakyan} et~al.}{2016}]{2016Elbakyan}
{Elbakyan} V.~G.,  {Vorobyov} E.~I.,   {Glebova} G.~M.,  2016, \mn@doi
  [Astronomy Reports] {10.1134/S1063772916100012}, \href
  {http://adsabs.harvard.edu/abs/2016ARep...60..879E} {60, 879}

\bibitem[\protect\citeauthoryear{{Elbakyan}, {Vorobyov}, {Rab}, {Meyer},
  {G{\"u}del}, {Hosokawa}  \& {Yorke}}{{Elbakyan}
  et~al.}{2019}]{2019ElbakyanVorobyov}
{Elbakyan} V.~G.,  {Vorobyov} E.~I.,  {Rab} C.,  {Meyer} D.~M.-A.,  {G{\"u}del}
  M.,  {Hosokawa} T.,   {Yorke} H.,  2019, \mn@doi [\mnras]
  {10.1093/mnras/sty3517}, \href
  {http://adsabs.harvard.edu/abs/2019MNRAS.484..146E} {484, 146}

\bibitem[\protect\citeauthoryear{{Elbakyan}, {Nayakshin}, {Vorobyov}, {Caratti
  o Garatti}  \& {Eisl{\"o}ffel}}{{Elbakyan}
  et~al.}{2021}]{2021ElbakyanNayakshin}
{Elbakyan} V.~G.,  {Nayakshin} S.,  {Vorobyov} E.~I.,  {Caratti o Garatti} A.,
   {Eisl{\"o}ffel} J.,  2021, \mn@doi [\aap] {10.1051/0004-6361/202140871},
  \href {https://ui.adsabs.harvard.edu/abs/2021A&A...651L...3E} {651, L3}

\bibitem[\protect\citeauthoryear{{Elbakyan}, {Wu}, {Nayakshin}  \&
  {Rosotti}}{{Elbakyan} et~al.}{2022}]{2022ElbakyanWu}
{Elbakyan} V.,  {Wu} Y.,  {Nayakshin} S.,   {Rosotti} G.,  2022, \mn@doi
  [\mnras] {10.1093/mnras/stac1774}, \href
  {https://ui.adsabs.harvard.edu/abs/2022MNRAS.515.3113E} {515, 3113}

\bibitem[\protect\citeauthoryear{{Fischer}, {Hillenbrand}, {Herczeg},
  {Johnstone}, {K{\'o}sp{\'a}l}  \& {Dunham}}{{Fischer}
  et~al.}{2022}]{2022Fischer}
{Fischer} W.~J.,  {Hillenbrand} L.~A.,  {Herczeg} G.~J.,  {Johnstone} D.,
  {K{\'o}sp{\'a}l} {\'A}.,   {Dunham} M.~M.,  2022, arXiv e-prints, \href
  {https://ui.adsabs.harvard.edu/abs/2022arXiv220311257F} {p. arXiv:2203.11257}

\bibitem[\protect\citeauthoryear{{Flaherty} et~al.,}{{Flaherty}
  et~al.}{2020}]{2020Flaherty}
{Flaherty} K.,  et~al., 2020, \mn@doi [\apj] {10.3847/1538-4357/ab8cc5}, \href
  {https://ui.adsabs.harvard.edu/abs/2020ApJ...895..109F} {895, 109}

\bibitem[\protect\citeauthoryear{{Fletcher}, {Nayakshin}, {Stamatellos},
  {Dehnen}, {Meru}, {Mayer}, {Deng}  \& {Rice}}{{Fletcher}
  et~al.}{2019}]{2019FletcherNayakshin}
{Fletcher} M.,  {Nayakshin} S.,  {Stamatellos} D.,  {Dehnen} W.,  {Meru} F.,
  {Mayer} L.,  {Deng} H.,   {Rice} K.,  2019, \mn@doi [\mnras]
  {10.1093/mnras/stz1123}, \href
  {https://ui.adsabs.harvard.edu/abs/2019MNRAS.486.4398F} {486, 4398}

\bibitem[\protect\citeauthoryear{{Flock}, {Nelson}, {Turner}, {Bertrang},
  {Carrasco-Gonz{\'a}lez}, {Henning}, {Lyra}  \& {Teague}}{{Flock}
  et~al.}{2017}]{2017FlockNelson}
{Flock} M.,  {Nelson} R.~P.,  {Turner} N.~J.,  {Bertrang} G. H.~M.,
  {Carrasco-Gonz{\'a}lez} C.,  {Henning} T.,  {Lyra} W.,   {Teague} R.,  2017,
  \mn@doi [\apj] {10.3847/1538-4357/aa943f}, \href
  {https://ui.adsabs.harvard.edu/abs/2017ApJ...850..131F} {850, 131}

\bibitem[\protect\citeauthoryear{{Goldreich} \& {Tremaine}}{{Goldreich} \&
  {Tremaine}}{1980}]{1980GoldreichTremaine}
{Goldreich} P.,  {Tremaine} S.,  1980, \mn@doi [\apj] {10.1086/158356}, \href
  {https://ui.adsabs.harvard.edu/abs/1980ApJ...241..425G} {241, 425}

\bibitem[\protect\citeauthoryear{{Guilera}, {Miller Bertolami}, {Masset},
  {Cuadra}, {Venturini}  \& {Ronco}}{{Guilera} et~al.}{2021}]{2021Guilera}
{Guilera} O.~M.,  {Miller Bertolami} M.~M.,  {Masset} F.,  {Cuadra} J.,
  {Venturini} J.,   {Ronco} M.~P.,  2021, \mn@doi [\mnras]
  {10.1093/mnras/stab2371}, \href
  {https://ui.adsabs.harvard.edu/abs/2021MNRAS.tmp.2172G} {}

\bibitem[\protect\citeauthoryear{{Hartmann} \& {Kenyon}}{{Hartmann} \&
  {Kenyon}}{1996}]{1996HartmannKenyon}
{Hartmann} L.,  {Kenyon} S.~J.,  1996, \mn@doi [\araa]
  {10.1146/annurev.astro.34.1.207}, \href
  {http://adsabs.harvard.edu/abs/1996ARA%26A..34..207H} {34, 207}

\bibitem[\protect\citeauthoryear{{Hosokawa} \& {Omukai}}{{Hosokawa} \&
  {Omukai}}{2009}]{2009HosokawaOmukai}
{Hosokawa} T.,  {Omukai} K.,  2009, \mn@doi [\apj]
  {10.1088/0004-637X/691/1/823}, \href
  {http://adsabs.harvard.edu/abs/2009ApJ...691..823H} {691, 823}

\bibitem[\protect\citeauthoryear{{Hosokawa}, {Offner}  \&
  {Krumholz}}{{Hosokawa} et~al.}{2011}]{2011HosokawaOffner}
{Hosokawa} T.,  {Offner} S.~S.~R.,   {Krumholz} M.~R.,  2011, \mn@doi [\apj]
  {10.1088/0004-637X/738/2/140}, \href
  {http://adsabs.harvard.edu/abs/2011ApJ...738..140H} {738, 140}

\bibitem[\protect\citeauthoryear{{Hosokawa}, {Hirano}, {Kuiper}, {Yorke},
  {Omukai}  \& {Yoshida}}{{Hosokawa} et~al.}{2016}]{2016Hosokawa}
{Hosokawa} T.,  {Hirano} S.,  {Kuiper} R.,  {Yorke} H.~W.,  {Omukai} K.,
  {Yoshida} N.,  2016, \mn@doi [\apj] {10.3847/0004-637X/824/2/119}, \href
  {https://ui.adsabs.harvard.edu/abs/2016ApJ...824..119H} {824, 119}

\bibitem[\protect\citeauthoryear{{Hunter} et~al.,}{{Hunter}
  et~al.}{2017}]{2017Hunter}
{Hunter} T.~R.,  et~al., 2017, \mn@doi [\apjl] {10.3847/2041-8213/aa5d0e},
  \href {https://ui.adsabs.harvard.edu/abs/2017ApJ...837L..29H} {837, L29}

\bibitem[\protect\citeauthoryear{{Hunter} et~al.,}{{Hunter}
  et~al.}{2018}]{2018Hunter}
{Hunter} T.~R.,  et~al., 2018, \mn@doi [\apj] {10.3847/1538-4357/aaa962}, \href
  {https://ui.adsabs.harvard.edu/abs/2018ApJ...854..170H} {854, 170}

\bibitem[\protect\citeauthoryear{{Ilee}, {Cyganowski}, {Nazari}, {Hunter},
  {Brogan}, {Forgan}  \& {Zhang}}{{Ilee} et~al.}{2016}]{2016Ilee}
{Ilee} J.~D.,  {Cyganowski} C.~J.,  {Nazari} P.,  {Hunter} T.~R.,  {Brogan}
  C.~L.,  {Forgan} D.~H.,   {Zhang} Q.,  2016, \mn@doi [\mnras]
  {10.1093/mnras/stw1912}, \href
  {https://ui.adsabs.harvard.edu/abs/2016MNRAS.462.4386I} {462, 4386}

\bibitem[\protect\citeauthoryear{{Ilee}, {Cyganowski}, {Brogan}, {Hunter},
  {Forgan}, {Haworth}, {Clarke}  \& {Harries}}{{Ilee} et~al.}{2018}]{2018Ilee}
{Ilee} J.~D.,  {Cyganowski} C.~J.,  {Brogan} C.~L.,  {Hunter} T.~R.,  {Forgan}
  D.~H.,  {Haworth} T.~J.,  {Clarke} C.~J.,   {Harries} T.~J.,  2018, \mn@doi
  [\apjl] {10.3847/2041-8213/aaeffc}, \href
  {https://ui.adsabs.harvard.edu/abs/2018ApJ...869L..24I} {869, L24}

\bibitem[\protect\citeauthoryear{{Jensen} \& {Haugb{\o}lle}}{{Jensen} \&
  {Haugb{\o}lle}}{2018}]{2018Jensen}
{Jensen} S.~S.,  {Haugb{\o}lle} T.,  2018, \mn@doi [\mnras]
  {10.1093/mnras/stx2844}, \href
  {http://adsabs.harvard.edu/abs/2018MNRAS.474.1176J} {474, 1176}

\bibitem[\protect\citeauthoryear{{Jim{\'e}nez} \& {Masset}}{{Jim{\'e}nez} \&
  {Masset}}{2017}]{2017JimenezMasset}
{Jim{\'e}nez} M.~A.,  {Masset} F.~S.,  2017, \mn@doi [\mnras]
  {10.1093/mnras/stx1946}, \href
  {https://ui.adsabs.harvard.edu/abs/2017MNRAS.471.4917J} {471, 4917}

\bibitem[\protect\citeauthoryear{{Johnston} et~al.,}{{Johnston}
  et~al.}{2015}]{2015Johnston}
{Johnston} K.~G.,  et~al., 2015, \mn@doi [\apjl] {10.1088/2041-8205/813/1/L19},
  \href {https://ui.adsabs.harvard.edu/abs/2015ApJ...813L..19J} {813, L19}

\bibitem[\protect\citeauthoryear{{Johnston} et~al.,}{{Johnston}
  et~al.}{2020}]{2020Johnston}
{Johnston} K.~G.,  et~al., 2020, \mn@doi [\aap] {10.1051/0004-6361/201937154},
  \href {https://ui.adsabs.harvard.edu/abs/2020A&A...634L..11J} {634, L11}

\bibitem[\protect\citeauthoryear{{Kadam}, {Vorobyov}, {Reg{\'a}ly},
  {K{\'o}sp{\'a}l}  \& {{\'A}brah{\'a}m}}{{Kadam} et~al.}{2020}]{2020Kadam}
{Kadam} K.,  {Vorobyov} E.,  {Reg{\'a}ly} Z.,  {K{\'o}sp{\'a}l} {\'A}.,
  {{\'A}brah{\'a}m} P.,  2020, \mn@doi [\apj] {10.3847/1538-4357/ab8bd8}, \href
  {https://ui.adsabs.harvard.edu/abs/2020ApJ...895...41K} {895, 41}

\bibitem[\protect\citeauthoryear{{Kley} \& {Crida}}{{Kley} \&
  {Crida}}{2008}]{2008KleyCrida}
{Kley} W.,  {Crida} A.,  2008, \mn@doi [\aap] {10.1051/0004-6361:200810033},
  \href {https://ui.adsabs.harvard.edu/abs/2008A&A...487L...9K} {487, L9}

\bibitem[\protect\citeauthoryear{{Kley}, {Bitsch}  \& {Klahr}}{{Kley}
  et~al.}{2009}]{2009KleyBitsch}
{Kley} W.,  {Bitsch} B.,   {Klahr} H.,  2009, \mn@doi [\aap]
  {10.1051/0004-6361/200912072}, \href
  {https://ui.adsabs.harvard.edu/abs/2009A&A...506..971K} {506, 971}

\bibitem[\protect\citeauthoryear{{Kolb}, {Stute}, {Kley}  \& {Mignone}}{{Kolb}
  et~al.}{2013}]{2013Kolb}
{Kolb} S.~M.,  {Stute} M.,  {Kley} W.,   {Mignone} A.,  2013, \mn@doi [\aap]
  {10.1051/0004-6361/201321499}, \href
  {https://ui.adsabs.harvard.edu/abs/2013A&A...559A..80K} {559, A80}

\bibitem[\protect\citeauthoryear{{Kratter} \& {Lodato}}{{Kratter} \&
  {Lodato}}{2016}]{2016Kratter}
{Kratter} K.,  {Lodato} G.,  2016, \mn@doi [\araa]
  {10.1146/annurev-astro-081915-023307}, \href
  {http://adsabs.harvard.edu/abs/2016ARA%26A..54..271K} {54, 271}

\bibitem[\protect\citeauthoryear{{Kratter} \& {Matzner}}{{Kratter} \&
  {Matzner}}{2006}]{2006KratterMatzner}
{Kratter} K.~M.,  {Matzner} C.~D.,  2006, \mn@doi [\mnras]
  {10.1111/j.1365-2966.2006.11103.x}, \href
  {https://ui.adsabs.harvard.edu/abs/2006MNRAS.373.1563K} {373, 1563}

\bibitem[\protect\citeauthoryear{{Kratter}, {Matzner}  \& {Krumholz}}{{Kratter}
  et~al.}{2008}]{2008Kratter}
{Kratter} K.~M.,  {Matzner} C.~D.,   {Krumholz} M.~R.,  2008, \mn@doi [\apj]
  {10.1086/587543}, \href {http://adsabs.harvard.edu/abs/2008ApJ...681..375K}
  {681, 375}

\bibitem[\protect\citeauthoryear{{Kraus} et~al.,}{{Kraus}
  et~al.}{2017}]{2017Kraus}
{Kraus} S.,  et~al., 2017, \mn@doi [\apjl] {10.3847/2041-8213/835/1/L5}, \href
  {https://ui.adsabs.harvard.edu/abs/2017ApJ...835L...5K} {835, L5}

\bibitem[\protect\citeauthoryear{{Krumholz}, {Klein}  \& {McKee}}{{Krumholz}
  et~al.}{2007}]{2007Krumholz}
{Krumholz} M.~R.,  {Klein} R.~I.,   {McKee} C.~F.,  2007, \mn@doi [\apj]
  {10.1086/519305}, \href
  {https://ui.adsabs.harvard.edu/abs/2007ApJ...665..478K} {665, 478}

\bibitem[\protect\citeauthoryear{{Krumholz}, {Klein}, {McKee}, {Offner}  \&
  {Cunningham}}{{Krumholz} et~al.}{2009}]{2009Krumholz}
{Krumholz} M.~R.,  {Klein} R.~I.,  {McKee} C.~F.,  {Offner} S. S.~R.,
  {Cunningham} A.~J.,  2009, \mn@doi [Science] {10.1126/science.1165857}, \href
  {https://ui.adsabs.harvard.edu/abs/2009Sci...323..754K} {323, 754}

\bibitem[\protect\citeauthoryear{{Kuiper} \& {Yorke}}{{Kuiper} \&
  {Yorke}}{2013}]{2013KuiperYorke}
{Kuiper} R.,  {Yorke} H.~W.,  2013, \mn@doi [\apj]
  {10.1088/0004-637X/772/1/61}, \href
  {http://adsabs.harvard.edu/abs/2013ApJ...772...61K} {772, 61}

\bibitem[\protect\citeauthoryear{{Kuiper}, {Klahr}, {Dullemond}, {Kley}  \&
  {Henning}}{{Kuiper} et~al.}{2010}]{2010Kuiper}
{Kuiper} R.,  {Klahr} H.,  {Dullemond} C.,  {Kley} W.,   {Henning} T.,  2010,
  \mn@doi [\aap] {10.1051/0004-6361/200912355}, \href
  {https://ui.adsabs.harvard.edu/abs/2010A&A...511A..81K} {511, A81}

\bibitem[\protect\citeauthoryear{{Kuiper}, {Klahr}, {Beuther}  \&
  {Henning}}{{Kuiper} et~al.}{2011}]{2011Kuiper}
{Kuiper} R.,  {Klahr} H.,  {Beuther} H.,   {Henning} T.,  2011, \mn@doi [\apj]
  {10.1088/0004-637X/732/1/20}, \href
  {https://ui.adsabs.harvard.edu/abs/2011ApJ...732...20K} {732, 20}

\bibitem[\protect\citeauthoryear{{Laor} \& {Draine}}{{Laor} \&
  {Draine}}{1993}]{1993Laor}
{Laor} A.,  {Draine} B.~T.,  1993, \mn@doi [\apj] {10.1086/172149}, \href
  {https://ui.adsabs.harvard.edu/abs/1993ApJ...402..441L} {402, 441}

\bibitem[\protect\citeauthoryear{{Larson}}{{Larson}}{1969}]{1969Larson}
{Larson} R.~B.,  1969, \mn@doi [\mnras] {10.1093/mnras/145.3.271}, \href
  {http://adsabs.harvard.edu/abs/1969MNRAS.145..271L} {145, 271}

\bibitem[\protect\citeauthoryear{{Lesur}}{{Lesur}}{2021}]{Lesur21-MHD-winds}
{Lesur} G. R.~J.,  2021, \mn@doi [\aap] {10.1051/0004-6361/202040109}, \href
  {https://ui.adsabs.harvard.edu/abs/2021A&A...650A..35L} {650, A35}

\bibitem[\protect\citeauthoryear{{Lin}, {Papaloizou}  \& {Faulkner}}{{Lin}
  et~al.}{1985}]{Lin85-CVs}
{Lin} D.~N.~C.,  {Papaloizou} J.,   {Faulkner} J.,  1985, \mn@doi [\mnras]
  {10.1093/mnras/212.1.105}, \href
  {https://ui.adsabs.harvard.edu/abs/1985MNRAS.212..105L} {212, 105}

\bibitem[\protect\citeauthoryear{{Liu}, {Su}, {Zinchenko}, {Wang}, {Meyer},
  {Wang}  \& {Hsieh}}{{Liu} et~al.}{2020}]{liu_apj_904_2020}
{Liu} S.-Y.,  {Su} Y.-N.,  {Zinchenko} I.,  {Wang} K.-S.,  {Meyer} D. M.~A.,
  {Wang} Y.,   {Hsieh} I.~T.,  2020, \mn@doi [\apj] {10.3847/1538-4357/abc0ec},
  \href {https://ui.adsabs.harvard.edu/abs/2020ApJ...904..181L} {904, 181}

\bibitem[\protect\citeauthoryear{{Lodato} \& {Clarke}}{{Lodato} \&
  {Clarke}}{2004}]{2004LodatoClarke}
{Lodato} G.,  {Clarke} C.~J.,  2004, \mn@doi [\mnras]
  {10.1111/j.1365-2966.2004.08112.x}, \href
  {https://ui.adsabs.harvard.edu/abs/2004MNRAS.353..841L} {353, 841}

\bibitem[\protect\citeauthoryear{{Lodato}, {Scardoni}, {Manara}  \&
  {Testi}}{{Lodato} et~al.}{2017}]{2017Lodato}
{Lodato} G.,  {Scardoni} C.~E.,  {Manara} C.~F.,   {Testi} L.,  2017, \mn@doi
  [\mnras] {10.1093/mnras/stx2273}, \href
  {https://ui.adsabs.harvard.edu/abs/2017MNRAS.472.4700L} {472, 4700}

\bibitem[\protect\citeauthoryear{{MacLeod} et~al.,}{{MacLeod}
  et~al.}{2018}]{2018MacLeod}
{MacLeod} G.~C.,  et~al., 2018, \mn@doi [\mnras] {10.1093/mnras/sty996}, \href
  {https://ui.adsabs.harvard.edu/abs/2018MNRAS.478.1077M} {478, 1077}

\bibitem[\protect\citeauthoryear{{Machida} \& {Doi}}{{Machida} \&
  {Doi}}{2013}]{2013MachidaDoi}
{Machida} M.~N.,  {Doi} K.,  2013, \mn@doi [\mnras] {10.1093/mnras/stt1524},
  \href {https://ui.adsabs.harvard.edu/abs/2013MNRAS.435.3283M} {435, 3283}

\bibitem[\protect\citeauthoryear{{Machida}, {Inutsuka}  \&
  {Matsumoto}}{{Machida} et~al.}{2011}]{2011MachidaInutsuka}
{Machida} M.~N.,  {Inutsuka} S.-i.,   {Matsumoto} T.,  2011, \mn@doi [\apj]
  {10.1088/0004-637X/729/1/42}, \href
  {https://ui.adsabs.harvard.edu/abs/2011ApJ...729...42M} {729, 42}

\bibitem[\protect\citeauthoryear{{Masunaga} \& {Inutsuka}}{{Masunaga} \&
  {Inutsuka}}{2000}]{2000MasunagaInutsuka}
{Masunaga} H.,  {Inutsuka} S.-i.,  2000, \mn@doi [\apj] {10.1086/308439}, \href
  {http://adsabs.harvard.edu/abs/2000ApJ...531..350M} {531, 350}

\bibitem[\protect\citeauthoryear{{Mayer}, {Quinn}, {Wadsley}  \&
  {Stadel}}{{Mayer} et~al.}{2004}]{2004MayerL}
{Mayer} L.,  {Quinn} T.,  {Wadsley} J.,   {Stadel} J.,  2004, \mn@doi [\apj]
  {10.1086/421288}, \href
  {https://ui.adsabs.harvard.edu/abs/2004ApJ...609.1045M} {609, 1045}

\bibitem[\protect\citeauthoryear{{McKee} \& {Tan}}{{McKee} \&
  {Tan}}{2008}]{2008McKeeTan}
{McKee} C.~F.,  {Tan} J.~C.,  2008, \mn@doi [\apj] {10.1086/587434}, \href
  {https://ui.adsabs.harvard.edu/abs/2008ApJ...681..771M} {681, 771}

\bibitem[\protect\citeauthoryear{{Meyer}, {Vorobyov}, {Kuiper}  \&
  {Kley}}{{Meyer} et~al.}{2017}]{2017MeyerVorobyov}
{Meyer} D.~M.-A.,  {Vorobyov} E.~I.,  {Kuiper} R.,   {Kley} W.,  2017, \mn@doi
  [\mnras] {10.1093/mnrasl/slw187}, \href
  {http://adsabs.harvard.edu/abs/2017MNRAS.464L..90M} {464, L90}

\bibitem[\protect\citeauthoryear{{Meyer}, {Kuiper}, {Kley}, {Johnston}  \&
  {Vorobyov}}{{Meyer} et~al.}{2018}]{2018MeyerKuiper}
{Meyer} D.~M.-A.,  {Kuiper} R.,  {Kley} W.,  {Johnston} K.~G.,   {Vorobyov} E.,
   2018, \mn@doi [\mnras] {10.1093/mnras/stx2551}, \href
  {http://adsabs.harvard.edu/abs/2018MNRAS.473.3615M} {473, 3615}

\bibitem[\protect\citeauthoryear{{Meyer}, {Vorobyov}, {Elbakyan}, {Stecklum},
  {Eisl{\"o}ffel}  \& {Sobolev}}{{Meyer}
  et~al.}{2019a}]{2019MeyerVorobyovElbakyan}
{Meyer} D.~M.~A.,  {Vorobyov} E.~I.,  {Elbakyan} V.~G.,  {Stecklum} B.,
  {Eisl{\"o}ffel} J.,   {Sobolev} A.~M.,  2019a, \mn@doi [\mnras]
  {10.1093/mnras/sty2980}, \href
  {https://ui.adsabs.harvard.edu/abs/2019MNRAS.482.5459M} {482, 5459}

\bibitem[\protect\citeauthoryear{{Meyer}, {Haemmerl{\'e}}  \&
  {Vorobyov}}{{Meyer} et~al.}{2019b}]{meyer_mnras_484_2019}
{Meyer} D.~M.~A.,  {Haemmerl{\'e}} L.,   {Vorobyov} E.~I.,  2019b, \mn@doi
  [\mnras] {10.1093/mnras/sty3527}, \href
  {https://ui.adsabs.harvard.edu/abs/2019MNRAS.484.2482M} {484, 2482}

\bibitem[\protect\citeauthoryear{{Meyer}, {Kreplin}, {Kraus}, {Vorobyov},
  {Haemmerle}  \& {Eisl{\"o}ffel}}{{Meyer} et~al.}{2019c}]{2019MeyerVorobyov}
{Meyer} D.~M.~A.,  {Kreplin} A.,  {Kraus} S.,  {Vorobyov} E.~I.,  {Haemmerle}
  L.,   {Eisl{\"o}ffel} J.,  2019c, \mn@doi [\mnras] {10.1093/mnras/stz1585},
  \href {https://ui.adsabs.harvard.edu/abs/2019MNRAS.487.4473M} {487, 4473}

\bibitem[\protect\citeauthoryear{{Meyer}, {Vorobyov}, {Elbakyan},
  {Eisl{\"o}ffel}, {Sobolev}  \& {St{\"o}hr}}{{Meyer} et~al.}{2021}]{2021Meyer}
{Meyer} D.~M.~A.,  {Vorobyov} E.~I.,  {Elbakyan} V.~G.,  {Eisl{\"o}ffel} J.,
  {Sobolev} A.~M.,   {St{\"o}hr} M.,  2021, \mn@doi [\mnras]
  {10.1093/mnras/staa3528}, \href
  {https://ui.adsabs.harvard.edu/abs/2021MNRAS.500.4448M} {500, 4448}

\bibitem[\protect\citeauthoryear{Meyer, Vorobyov, Elbakyan, Kraus, Liu,
  Nayakshin  \& Sobolev}{Meyer et~al.}{2022}]{2022Meyer}
Meyer D. M.-A.,  Vorobyov E.~I.,  Elbakyan V.~G.,  Kraus S.,  Liu S.-Y.,
  Nayakshin S.,   Sobolev A.~M.,  2022, \mn@doi [Monthly Notices of the Royal
  Astronomical Society] {10.1093/mnras/stac2956}

\bibitem[\protect\citeauthoryear{{Michael} \& {Durisen}}{{Michael} \&
  {Durisen}}{2010}]{2010MichaelDurisen}
{Michael} S.,  {Durisen} R.~H.,  2010, \mn@doi [\mnras]
  {10.1111/j.1365-2966.2010.16694.x}, \href
  {https://ui.adsabs.harvard.edu/abs/2010MNRAS.406..279M} {406, 279}

\bibitem[\protect\citeauthoryear{{Mignone}, {Bodo}, {Massaglia}, {Matsakos},
  {Tesileanu}, {Zanni}  \& {Ferrari}}{{Mignone} et~al.}{2007}]{2007Mignone}
{Mignone} A.,  {Bodo} G.,  {Massaglia} S.,  {Matsakos} T.,  {Tesileanu} O.,
  {Zanni} C.,   {Ferrari} A.,  2007, \mn@doi [\apjs] {10.1086/513316}, \href
  {https://ui.adsabs.harvard.edu/abs/2007ApJS..170..228M} {170, 228}

\bibitem[\protect\citeauthoryear{{Miotello}, {Kamp}, {Birnstiel}, {Cleeves}  \&
  {Kataoka}}{{Miotello} et~al.}{2022}]{2022MiotelloPPVII}
{Miotello} A.,  {Kamp} I.,  {Birnstiel} T.,  {Cleeves} L.~I.,   {Kataoka} A.,
  2022, arXiv e-prints, \href
  {https://ui.adsabs.harvard.edu/abs/2022arXiv220309818M} {p. arXiv:2203.09818}

\bibitem[\protect\citeauthoryear{{Motte}, {Bontemps}  \& {Louvet}}{{Motte}
  et~al.}{2018}]{2018Motte}
{Motte} F.,  {Bontemps} S.,   {Louvet} F.,  2018, \mn@doi [\araa]
  {10.1146/annurev-astro-091916-055235}, \href
  {https://ui.adsabs.harvard.edu/abs/2018ARA&A..56...41M} {56, 41}

\bibitem[\protect\citeauthoryear{{Mulders} \& {Dominik}}{{Mulders} \&
  {Dominik}}{2012}]{2012MuldersDominik}
{Mulders} G.~D.,  {Dominik} C.,  2012, \mn@doi [\aap]
  {10.1051/0004-6361/201118127}, \href
  {https://ui.adsabs.harvard.edu/abs/2012A&A...539A...9M} {539, A9}

\bibitem[\protect\citeauthoryear{{Najita} \& {Bergin}}{{Najita} \&
  {Bergin}}{2018}]{2018NajitaBergin}
{Najita} J.~R.,  {Bergin} E.~A.,  2018, \mn@doi [\apj]
  {10.3847/1538-4357/aad80c}, \href
  {https://ui.adsabs.harvard.edu/abs/2018ApJ...864..168N} {864, 168}

\bibitem[\protect\citeauthoryear{{Nayakshin}}{{Nayakshin}}{2015}]{2015Nayakshin}
{Nayakshin} S.,  2015, \mn@doi [\mnras] {10.1093/mnras/stv1915}, \href
  {https://ui.adsabs.harvard.edu/abs/2015MNRAS.454...64N} {454, 64}

\bibitem[\protect\citeauthoryear{{Nayakshin}}{{Nayakshin}}{2017a}]{2017Nayakshin}
{Nayakshin} S.,  2017a, \mn@doi [\pasa] {10.1017/pasa.2016.55}, \href
  {http://adsabs.harvard.edu/abs/2017PASA...34....2N} {34, e002}

\bibitem[\protect\citeauthoryear{{Nayakshin}}{{Nayakshin}}{2017b}]{2017NayakshinDesert}
{Nayakshin} S.,  2017b, \mn@doi [\mnras] {10.1093/mnras/stx1351}, \href
  {http://adsabs.harvard.edu/abs/2017MNRAS.470.2387N} {470, 2387}

\bibitem[\protect\citeauthoryear{{Nayakshin} \& {Lodato}}{{Nayakshin} \&
  {Lodato}}{2012}]{2012NayakshinLodato}
{Nayakshin} S.,  {Lodato} G.,  2012, \mn@doi [\mnras]
  {10.1111/j.1365-2966.2012.21612.x}, \href
  {http://adsabs.harvard.edu/abs/2012MNRAS.426...70N} {426, 70}

\bibitem[\protect\citeauthoryear{{Ohtani}, {Kimura}, {Tsuribe}  \&
  {Vorobyov}}{{Ohtani} et~al.}{2014}]{2014OhtaniVorobyov}
{Ohtani} T.,  {Kimura} S.~S.,  {Tsuribe} T.,   {Vorobyov} E.~I.,  2014, \mn@doi
  [\pasj] {10.1093/pasj/psu098}, \href
  {https://ui.adsabs.harvard.edu/abs/2014PASJ...66..112O} {66, 112}

\bibitem[\protect\citeauthoryear{{Oliva} \& {Kuiper}}{{Oliva} \&
  {Kuiper}}{2020}]{2020OlivaKuiper}
{Oliva} G.~A.,  {Kuiper} R.,  2020, \mn@doi [\aap]
  {10.1051/0004-6361/202038103}, \href
  {https://ui.adsabs.harvard.edu/abs/2020A&A...644A..41O} {644, A41}

\bibitem[\protect\citeauthoryear{{Ormel}, {Kuiper}  \& {Shi}}{{Ormel}
  et~al.}{2015a}]{2015OrmelKuiperShi}
{Ormel} C.~W.,  {Kuiper} R.,   {Shi} J.-M.,  2015a, \mn@doi [\mnras]
  {10.1093/mnras/stu2101}, \href
  {https://ui.adsabs.harvard.edu/abs/2015MNRAS.446.1026O} {446, 1026}

\bibitem[\protect\citeauthoryear{{Ormel}, {Shi}  \& {Kuiper}}{{Ormel}
  et~al.}{2015b}]{2015OrmelShiKuiper}
{Ormel} C.~W.,  {Shi} J.-M.,   {Kuiper} R.,  2015b, \mn@doi [\mnras]
  {10.1093/mnras/stu2704}, \href
  {https://ui.adsabs.harvard.edu/abs/2015MNRAS.447.3512O} {447, 3512}

\bibitem[\protect\citeauthoryear{{Paardekooper}, {Baruteau}  \&
  {Kley}}{{Paardekooper} et~al.}{2011}]{2011Paardekooper}
{Paardekooper} S.~J.,  {Baruteau} C.,   {Kley} W.,  2011, \mn@doi [\mnras]
  {10.1111/j.1365-2966.2010.17442.x}, \href
  {https://ui.adsabs.harvard.edu/abs/2011MNRAS.410..293P} {410, 293}

\bibitem[\protect\citeauthoryear{{Palla}}{{Palla}}{2012}]{Palla12-Hayashi-Track}
{Palla} F.,  2012, in {Umemura} M.,  {Omukai} K.,  eds,  American Institute of
  Physics Conference Series Vol. 1480, First Stars IV - from Hayashi to the
  Future -. pp 22--29, \mn@doi{10.1063/1.4754323}

\bibitem[\protect\citeauthoryear{{Pinte}, {Dent}, {M{\'e}nard}, {Hales},
  {Hill}, {Cortes}  \& {de Gregorio-Monsalvo}}{{Pinte}
  et~al.}{2016}]{2016Pinte}
{Pinte} C.,  {Dent} W.~R.~F.,  {M{\'e}nard} F.,  {Hales} A.,  {Hill} T.,
  {Cortes} P.,   {de Gregorio-Monsalvo} I.,  2016, \mn@doi [\apj]
  {10.3847/0004-637X/816/1/25}, \href
  {https://ui.adsabs.harvard.edu/abs/2016ApJ...816...25P} {816, 25}

\bibitem[\protect\citeauthoryear{{Pringle}}{{Pringle}}{1981}]{1981Pringle}
{Pringle} J.~E.,  1981, \mn@doi [\araa] {10.1146/annurev.aa.19.090181.001033},
  \href {http://adsabs.harvard.edu/abs/1981ARA%26A..19..137P} {19, 137}

\bibitem[\protect\citeauthoryear{{Proven-Adzri}, {MacLeod}, {Heever}, {Hoare},
  {Kuditcher}  \& {Goedhart}}{{Proven-Adzri} et~al.}{2019}]{2019Proven-Adzri}
{Proven-Adzri} E.,  {MacLeod} G.~C.,  {Heever} S.~P. v.~d.,  {Hoare} M.~G.,
  {Kuditcher} A.,   {Goedhart} S.,  2019, \mn@doi [\mnras]
  {10.1093/mnras/stz1458}, \href
  {https://ui.adsabs.harvard.edu/abs/2019MNRAS.487.2407P} {487, 2407}

\bibitem[\protect\citeauthoryear{{Rafikov}}{{Rafikov}}{2017}]{2017Rafikov}
{Rafikov} R.~R.,  2017, \mn@doi [\apj] {10.3847/1538-4357/aa6249}, \href
  {https://ui.adsabs.harvard.edu/abs/2017ApJ...837..163R} {837, 163}

\bibitem[\protect\citeauthoryear{{Reg{\'a}ly} \& {Vorobyov}}{{Reg{\'a}ly} \&
  {Vorobyov}}{2017}]{2017RegalyVorobyova}
{Reg{\'a}ly} Z.,  {Vorobyov} E.,  2017, \mn@doi [\aap]
  {10.1051/0004-6361/201629154}, \href
  {http://adsabs.harvard.edu/abs/2017A%26A...601A..24R} {601, A24}

\bibitem[\protect\citeauthoryear{{Rosotti}, {Teague}, {Dullemond}, {Booth}  \&
  {Clarke}}{{Rosotti} et~al.}{2020}]{2020Rosotti}
{Rosotti} G.~P.,  {Teague} R.,  {Dullemond} C.,  {Booth} R.~A.,   {Clarke}
  C.~J.,  2020, \mn@doi [\mnras] {10.1093/mnras/staa1170}, \href
  {https://ui.adsabs.harvard.edu/abs/2020MNRAS.495..173R} {495, 173}

\bibitem[\protect\citeauthoryear{{Sana} \& {Evans}}{{Sana} \&
  {Evans}}{2011}]{2011SanaEvans}
{Sana} H.,  {Evans} C.~J.,  2011, in {Neiner} C.,  {Wade} G.,  {Meynet} G.,
  {Peters} G.,  eds,  Active OB Stars: Structure, Evolution, Mass Loss, and
  Critical Limits Vol. 272, {The multiplicity of massive stars}. pp 474--485
  (\mn@eprint {arXiv} {1009.4197}), \mn@doi{10.1017/S1743921311011124}

\bibitem[\protect\citeauthoryear{{Shakura} \& {Sunyaev}}{{Shakura} \&
  {Sunyaev}}{1973}]{1973ShakuraSunyaev}
{Shakura} N.~I.,  {Sunyaev} R.~A.,  1973, \aap, \href
  {http://adsabs.harvard.edu/abs/1973A%26A....24..337S} {24, 337}

\bibitem[\protect\citeauthoryear{{Stamatellos} \& {Whitworth}}{{Stamatellos} \&
  {Whitworth}}{2009}]{2009StamatellosWhitworth}
{Stamatellos} D.,  {Whitworth} A.~P.,  2009, \mn@doi [\mnras]
  {10.1111/j.1365-2966.2008.14069.x}, \href
  {http://adsabs.harvard.edu/abs/2009MNRAS.392..413S} {392, 413}

\bibitem[\protect\citeauthoryear{{Stecklum} et~al.,}{{Stecklum}
  et~al.}{2021}]{2021Stecklum}
{Stecklum} B.,  et~al., 2021, \mn@doi [\aap] {10.1051/0004-6361/202039645},
  \href {https://ui.adsabs.harvard.edu/abs/2021A&A...646A.161S} {646, A161}

\bibitem[\protect\citeauthoryear{{Suri} et~al.,}{{Suri}
  et~al.}{2021}]{2021Suri}
{Suri} S.,  et~al., 2021, \mn@doi [\aap] {10.1051/0004-6361/202140963}, \href
  {https://ui.adsabs.harvard.edu/abs/2021A&A...655A..84S} {655, A84}

\bibitem[\protect\citeauthoryear{{Suzuki}, {Ogihara}, {Morbidelli}, {Crida}  \&
  {Guillot}}{{Suzuki} et~al.}{2016}]{Suzuki16-MHD-winds}
{Suzuki} T.~K.,  {Ogihara} M.,  {Morbidelli} A.,  {Crida} A.,   {Guillot} T.,
  2016, \mn@doi [\aap] {10.1051/0004-6361/201628955}, \href
  {https://ui.adsabs.harvard.edu/abs/2016A&A...596A..74S} {596, A74}

\bibitem[\protect\citeauthoryear{{Syer} \& {Clarke}}{{Syer} \&
  {Clarke}}{1995}]{1995SyerClarke}
{Syer} D.,  {Clarke} C.~J.,  1995, \mn@doi [\mnras] {10.1093/mnras/277.3.758},
  \href {https://ui.adsabs.harvard.edu/abs/1995MNRAS.277..758S} {277, 758}

\bibitem[\protect\citeauthoryear{{Tabone}, {Rosotti}, {Lodato}, {Armitage},
  {Cridland}  \& {van Dishoeck}}{{Tabone} et~al.}{2022}]{2022Tabone}
{Tabone} B.,  {Rosotti} G.~P.,  {Lodato} G.,  {Armitage} P.~J.,  {Cridland}
  A.~J.,   {van Dishoeck} E.~F.,  2022, \mn@doi [\mnras]
  {10.1093/mnrasl/slab124}, \href
  {https://ui.adsabs.harvard.edu/abs/2022MNRAS.512L..74T} {512, L74}

\bibitem[\protect\citeauthoryear{{Tan}, {Beltr{\'a}n}, {Caselli}, {Fontani},
  {Fuente}, {Krumholz}, {McKee}  \& {Stolte}}{{Tan} et~al.}{2014}]{2014Tan}
{Tan} J.~C.,  {Beltr{\'a}n} M.~T.,  {Caselli} P.,  {Fontani} F.,  {Fuente} A.,
  {Krumholz} M.~R.,  {McKee} C.~F.,   {Stolte} A.,  2014, in {Beuther} H.,
  {Klessen} R.~S.,  {Dullemond} C.~P.,   {Henning} T.,  eds, Protostars and
  Planets VI. p.~149 (\mn@eprint {arXiv} {1402.0919}),
  \mn@doi{10.2458/azu\_uapress\_9780816531240-ch007}

\bibitem[\protect\citeauthoryear{{Tanaka}, {Takeuchi}  \& {Ward}}{{Tanaka}
  et~al.}{2002}]{2002TanakaWard}
{Tanaka} H.,  {Takeuchi} T.,   {Ward} W.~R.,  2002, \mn@doi [\apj]
  {10.1086/324713}, \href
  {https://ui.adsabs.harvard.edu/abs/2002ApJ...565.1257T} {565, 1257}

\bibitem[\protect\citeauthoryear{{Tanaka}, {Tan}  \& {Zhang}}{{Tanaka}
  et~al.}{2017}]{2017TanakaTan}
{Tanaka} K. E.~I.,  {Tan} J.~C.,   {Zhang} Y.,  2017, \mn@doi [\apj]
  {10.3847/1538-4357/835/1/32}, \href
  {https://ui.adsabs.harvard.edu/abs/2017ApJ...835...32T} {835, 32}

\bibitem[\protect\citeauthoryear{{Vazan} \& {Helled}}{{Vazan} \&
  {Helled}}{2012}]{2012Vazan}
{Vazan} A.,  {Helled} R.,  2012, \mn@doi [\apj] {10.1088/0004-637X/756/1/90},
  \href {https://ui.adsabs.harvard.edu/abs/2012ApJ...756...90V} {756, 90}

\bibitem[\protect\citeauthoryear{{Vorobyov}}{{Vorobyov}}{2013}]{2013Vorobyov}
{Vorobyov} E.~I.,  2013, \mn@doi [\aap] {10.1051/0004-6361/201220601}, \href
  {http://adsabs.harvard.edu/abs/2013A%26A...552A.129V} {552, A129}

\bibitem[\protect\citeauthoryear{{Vorobyov} \& {Basu}}{{Vorobyov} \&
  {Basu}}{2005}]{2005VorobyovBasu}
{Vorobyov} E.~I.,  {Basu} S.,  2005, \mn@doi [\apjl] {10.1086/498303}, \href
  {https://ui.adsabs.harvard.edu/abs/2005ApJ...633L.137V} {633, L137}

\bibitem[\protect\citeauthoryear{{Vorobyov} \& {Basu}}{{Vorobyov} \&
  {Basu}}{2006}]{2006VorobyovBasu}
{Vorobyov} E.~I.,  {Basu} S.,  2006, \mn@doi [ApJ] {10.1086/507320}, \href
  {http://adsabs.harvard.edu/abs/2006ApJ...650..956V} {650, 956}

\bibitem[\protect\citeauthoryear{{Vorobyov} \& {Basu}}{{Vorobyov} \&
  {Basu}}{2009}]{2009VorobyovBasu}
{Vorobyov} E.~I.,  {Basu} S.,  2009, \mn@doi [\mnras]
  {10.1111/j.1365-2966.2008.14376.x}, \href
  {http://adsabs.harvard.edu/abs/2009MNRAS.393..822V} {393, 822}

\bibitem[\protect\citeauthoryear{{Vorobyov} \& {Elbakyan}}{{Vorobyov} \&
  {Elbakyan}}{2018}]{2018VorobyovElbakyan}
{Vorobyov} E.~I.,  {Elbakyan} V.~G.,  2018, \mn@doi [\aap]
  {10.1051/0004-6361/201833226}, \href
  {http://adsabs.harvard.edu/abs/2018A%26A...618A...7V} {618, A7}

\bibitem[\protect\citeauthoryear{{Vorobyov} \& {Elbakyan}}{{Vorobyov} \&
  {Elbakyan}}{2019}]{2019VorobyovElbakyan}
{Vorobyov} E.~I.,  {Elbakyan} V.~G.,  2019, \mn@doi [\aap]
  {10.1051/0004-6361/201936132}, \href
  {https://ui.adsabs.harvard.edu/abs/2019A&A...631A...1V} {631, A1}

\bibitem[\protect\citeauthoryear{{Vorobyov}, {Elbakyan}, {Hosokawa}, {Sakurai},
  {Guedel}  \& {Yorke}}{{Vorobyov} et~al.}{2017}]{2017VorobyovElbakyan}
{Vorobyov} E.~I.,  {Elbakyan} V.,  {Hosokawa} T.,  {Sakurai} Y.,  {Guedel} M.,
   {Yorke} H.,  2017, \mn@doi [\aap] {10.1051/0004-6361/201630356}, \href
  {http://adsabs.harvard.edu/abs/2017A%26A...605A..77V} {605, A77}

\bibitem[\protect\citeauthoryear{{Vorobyov}, {Elbakyan}, {Plunkett}, {Dunham},
  {Audard}, {Guedel}  \& {Dionatos}}{{Vorobyov}
  et~al.}{2018}]{2018VorobyovElbakyanPlunkett}
{Vorobyov} E.~I.,  {Elbakyan} V.~G.,  {Plunkett} A.~L.,  {Dunham} M.~M.,
  {Audard} M.,  {Guedel} M.,   {Dionatos} O.,  2018, \mn@doi [\aap]
  {10.1051/0004-6361/201732253}, \href
  {http://adsabs.harvard.edu/abs/2018A%26A...613A..18V} {613, A18}

\bibitem[\protect\citeauthoryear{{Vorobyov}, {Skliarevskii}, {Elbakyan},
  {Pavlyuchenkov}, {Akimkin}  \& {Guedel}}{{Vorobyov}
  et~al.}{2019}]{2019VorobyovSkliarevskii}
{Vorobyov} E.~I.,  {Skliarevskii} A.~M.,  {Elbakyan} V.~G.,  {Pavlyuchenkov}
  Y.,  {Akimkin} V.,   {Guedel} M.,  2019, \mn@doi [\aap]
  {10.1051/0004-6361/201935438}, \href
  {https://ui.adsabs.harvard.edu/abs/2019A&A...627A.154V} {627, A154}

\bibitem[\protect\citeauthoryear{{Vorobyov}, {Elbakyan}, {Liu}  \&
  {Takami}}{{Vorobyov} et~al.}{2021}]{2021VorobyovElbakyanLiu}
{Vorobyov} E.~I.,  {Elbakyan} V.~G.,  {Liu} H.~B.,   {Takami} M.,  2021,
  \mn@doi [\aap] {10.1051/0004-6361/202039391}, \href
  {https://ui.adsabs.harvard.edu/abs/2021A&A...647A..44V} {647, A44}

\bibitem[\protect\citeauthoryear{{Ward}}{{Ward}}{1997}]{1997Ward}
{Ward} W.~R.,  1997, \mn@doi [\icarus] {10.1006/icar.1996.5647}, \href
  {https://ui.adsabs.harvard.edu/abs/1997Icar..126..261W} {126, 261}

\bibitem[\protect\citeauthoryear{{Zhao}, {Caselli}, {Li}  \&
  {Krasnopolsky}}{{Zhao} et~al.}{2018}]{2018ZhaoCaselli}
{Zhao} B.,  {Caselli} P.,  {Li} Z.-Y.,   {Krasnopolsky} R.,  2018, \mn@doi
  [\mnras] {10.1093/mnras/stx2617}, \href
  {http://adsabs.harvard.edu/abs/2018MNRAS.473.4868Z} {473, 4868}

\bibitem[\protect\citeauthoryear{{Zhu}, {Hartmann}  \& {Gammie}}{{Zhu}
  et~al.}{2010}]{2010ZhuHartmann}
{Zhu} Z.,  {Hartmann} L.,   {Gammie} C.,  2010, \mn@doi [\apj]
  {10.1088/0004-637X/713/2/1143}, \href
  {https://ui.adsabs.harvard.edu/abs/2010ApJ...713.1143Z} {713, 1143}

\bibitem[\protect\citeauthoryear{{Zhu}, {Jiang}  \& {Stone}}{{Zhu}
  et~al.}{2020}]{2020Zhu}
{Zhu} Z.,  {Jiang} Y.-F.,   {Stone} J.~M.,  2020, \mn@doi [\mnras]
  {10.1093/mnras/staa952}, \href
  {https://ui.adsabs.harvard.edu/abs/2020MNRAS.495.3494Z} {495, 3494}

\bibitem[\protect\citeauthoryear{{Zinchenko} et~al.,}{{Zinchenko}
  et~al.}{2015}]{2015Zinchenko}
{Zinchenko} I.,  et~al., 2015, \mn@doi [\apj] {10.1088/0004-637X/810/1/10},
  \href {https://ui.adsabs.harvard.edu/abs/2015ApJ...810...10Z} {810, 10}

\makeatother
\end{thebibliography}




\appendix

\section{The normalized specific torque for Type I and Type II} \label{sect:app}

The normalized specific torque for Type II migration is defined as \citep{2002ArmitageBonnell, 2006AlexanderClarke, 2015Nayakshin}
\begin{equation}
\lambda_{\mathrm{II}}= 
\begin{cases}
      \frac{q^2}{2} \left(\frac{a}{\Delta r}\right)^4 \frac{G M_*}{a}, & \text{ if } r>a, \\
      -\frac{q^2}{2}\left(\frac{R}{\Delta r}\right)^4 \frac{G M_*}{a}, & \text{ if } r<a,
\end{cases}
\end{equation}
where $\Delta r = |r-a|$. To avoid a discontinuity at $r=a$, we smooth the torque at $|\Delta r| <\max[H_{\mathrm{obj}},R_{\mathrm{Hill}}]$ following the method described in \citet{1995SyerClarke}.

The normalized specific torque for Type I is defined as
\begin{equation}
    \lambda_{\mathrm{I}} = \lambda'_{\mathrm{I}} \left[ -\frac{|\Delta r|}{\Delta r_{\mathrm{I}}} \right]
\end{equation}
where $\Delta r_{\mathrm{I}} = H_{\mathrm{obj}} + a(q/3)^{1/3} + a/3$. We find the value of $\lambda'_{\mathrm{I}}$ by equating the absolute total torques acting from the disc onto the planet and from the planet onto the disc. Doing so ensures that we conserve the angular momentum of the planet-disc system manifestly. The total torque from the disc onto the planet in the Type I regime is given by Eq.~\ref{eq:gamma_tot}. 


\bsp	
\label{lastpage}
\end{document}